\documentclass[a4paper,11pt]{article}
\pdfoutput=1 
\usepackage{jheppub}
\usepackage{tikz}
\usepackage{subfig}
\usepackage{enumitem}
\usepackage[T1]{fontenc} 
\usepackage{floatrow}
\usepackage{appendix}
\usepackage{mathabx}
\usepackage{slashed}
\usepackage{dsfont}
\usepackage{epsf}
\usepackage{amsmath}
\usepackage{amsfonts}
\usepackage{amssymb}
\usepackage{psfrag,epsfig,graphicx,graphics}
\usepackage{braket}
\usepackage{cases}
\usepackage{xargs} 
\newcommand{\der}{\mathrm{d}}

\usepackage[colorinlistoftodos,prependcaption,textsize=tiny]{todonotes}
\newcommandx{\Q}[2][1=]{\todo[linecolor=blue,backgroundcolor=blue!25,bordercolor=blue,#1]{#2}}
\newcommandx{\FS}[2][1=]{\todo[linecolor=red,backgroundcolor=red!25,bordercolor=red,#1]{#2}}
\newcommandx{\EL}[2][1=]{\todo[linecolor=orange,backgroundcolor=orange!25,bordercolor=orange,#1]{#2}}

\newcommand\numberthis[1][]{%
    \refstepcounter{equation}%
    \ifx#1\empty\else\label{#1}\fi%
    \tag{\theequation}%
}

\DeclareMathOperator{\Tr}{Tr}

\usepackage{bm}
\usepackage{scalerel}
\usepackage{aligned-overset}

\allowdisplaybreaks

\title{Direct quarkonium-plus-gluon production in DIS in the Color Glass Condensate}
\author[a,b,c]{Zhong-Bo Kang,}
\author[d]{Emilie Li,}
\author[e,f,a,b]{and Farid Salazar}
\affiliation[a]{Department of Physics and Astronomy, University of California, Los Angeles, California 90095, USA}
\affiliation[b]{Mani L. Bhaumik Institute for Theoretical Physics, University of California, Los Angeles, California 90095, USA}
\affiliation[c]{Center for Frontiers in Nuclear Science, Stony Brook University, Stony Brook, NY 11794, USA}
\affiliation[d]{Université Paris-Saclay, CNRS/IN2P3, IJCLab, 91405, Orsay, France}
\affiliation[e]{Nuclear Science Division, Lawrence Berkeley National Laboratory, Berkeley, California 94720, USA}
\affiliation[f]{Physics Department, University of California, Berkeley, California 94720, USA}

\emailAdd{zkang@ucla.edu}
\emailAdd{emilie.li@ijclab.in2p3.fr}
\emailAdd{salazar@lbl.gov}

\abstract{We compute the differential cross-section for direct quarkonium production accompanied by a gluon in high-energy deep inelastic scattering (DIS) at small-$x$. We employ the Non-Relativistic
QCD factorization framework, focusing on the $S$-wave contribution to the formation of the quarkonium, and including both color singlet and octet contributions. Our short distance coefficients for the production of the heavy quark pair are obtained within the Color Glass Condensate effective field theory. Our results pave the way towards the next-to-leading order computation of direct quarkonium in DIS, as well as the study of azimuthal correlations of direct quarkonium and jet.}

\begin{document} 
\maketitle

\listoftodos[Notes]

\section{Introduction}

Due to their rapid proliferation at small $x$, it is well-known that gluons constitute the dominant partonic component of the content of hadronic and nuclear matter in the high-energy limit. It is expected that this fast growth cannot go unchecked and that the gluon density must eventually saturate due to the non-linear interactions of quantum chromodynamics \cite{Gribov:1984tu,Mueller:1985wy}. The experimental observation for gluon saturation in collider experiments is one of the outstanding topics in high-energy nuclear physics \cite{Boer:2011fh,Accardi:2012qut,Aschenauer:2017jsk,AbdulKhalek:2021gbh,AbdulKhalek:2022hcn,ALICE:2012mj,ATLAS:2016xpn,LHCb:2021vww,LHCb:2022tjh}. The Color Glass Condensate (CGC) effective theory provides a systematic framework to study this novel regime of matter of high parton densities \cite{Iancu:2003xm,Gelis:2010nm,Kovchegov:2012mbw,Albacete:2014fwa,Blaizot:2016qgz,Morreale:2021pnn}. A hallmark consequence of the CGC is the suppression of particle production when the typical transverse momenta or invariant mass is commensurate to the energy-dependent momentum saturation scale \cite{Kharzeev:2004yx,Marquet:2007vb,Kowalski:2007rw,Lappi:2012nh,Albacete:2018ruq,Zheng:2014vka,Lappi:2013zma,JalilianMarian:2012bd,Ducloue:2017kkq,Ducloue:2015gfa,Shi:2021hwx,Tong:2022zwp,Benic:2022ixp,Al-mashad:2022zbq,Liu:2022ijp,Liu:2023aqb,Caucal:2023fsf}. Heavy-quarkonium production in high-energy proton-proton and proton-nucleus collisions at RHIC and the LHC \cite{Kang:2013hta,Ma:2014mri,Ma:2015sia,Ma:2017rsu,Ma:2018bax,Levin:2019fvb,Salazar:2021mpv,Qiu:2013qka,Watanabe:2015yca,Ma:2018qvc,Stebel:2021bbn,Boussarie:2018zwg} (see also \cite{Gelis:2003vh,Blaizot:2004wv,Tuchin:2004rb,Fujii:2005vj,Kovchegov:2006qn,Fujii:2006ab,Altinoluk:2015vax,Marquet:2017xwy} for open heavy flavor studies) is a particularly compelling observable as the mass of the heavy quarkonium provides a semi-hard scale comparable to the expected saturation scales reached on these collision systems.  Analogous studies have been proposed in deep inelastic scattering (DIS) at the future Electron-Ion Collider (EIC), albeit the majority of them focusing on diffractive production \cite{Munier:2001nr,Kowalski:2003hm,Marquet:2007qa,Ivanov:2004ax,Kowalski:2006hc,Goncalves:2009za,Toll:2012mb,Rezaeian:2013tka,Armesto:2014sma,Mantysaari:2016ykx,Mantysaari:2020lhf, Mantysaari:2021ryb,Mantysaari:2022kdm}. Recently, direct quarkonium production in DIS has been studied at small-$x$ within the transverse momentum dependent framework in \cite{Bacchetta:2018ivt,Boer:2020bbd,Boer:2023zit}, and within the CGC effective theory in \cite{Kang:2023xxx}. In this work, we further extend these studies by computing direct quarkonium production accompanied by a gluon in DIS and within the joint framework of the CGC and the Non-relativistic QCD (NRQCD) formalism~\cite{Bodwin:1994jh}. Our calculation paves the way for the computation of direct quarkonium production at the next-to-leading order in the CGC, as well as the study of azimuthal correlations of direct quarkonium + jet/hadron production in the saturated regime.

This paper is organized as follows. In Sec.\,\ref{sec:theoretical_framework} we begin by setting up the kinematics of the process under consideration, as well as the notations and conventions to be followed throughout the manuscript. We then briefly review the basic theoretical tools of our computation: the NRQCD formalism and the CGC Effective Field Theory. In particular, we introduce the momentum space Feynman rules with CGC effective vertices that encode the multiple scattering of quarks and gluons with the strong gluon field of the nucleus.
We then review the computation of heavy quark pair + gluon production in virtual photon nucleus collision in Sec.\,\ref{sec:QQbar_gluon_CGC}. We follow the strategy in \cite{Caucal:2021ent} by performing our computation using covariant perturbation theory. The projection of the heavy quark pair into the $S$-wave states is performed in Sec.\,\ref{sec:projectionQQbar}, where we include both color octet and singlet contributions. The complete results for short distance coefficients for the differential cross-section are gathered in Sec.\,\ref{sec:cross-section}. We present a brief outlook of the potential application of our results in Sec.\,\ref{sec:outlook}. 

Our manuscript is supplemented by multiple appendices. First, we remind the readers of the usual Feynman rules and set conventions in appendix\,\ref{sec:Feynmanrules}. We then present several useful identities for the traces of gamma matrices, simplification of the Dirac structures, and Lorentz contractions in appendices\,\ref{sec:GammaTrace}\,, \ref{sec:Dirac-identities} and \ref{sec:Lorentz-Contractions} respectively. We close the manuscript by presenting integrals over transverse components of the loop momenta that are present in scattering amplitudes for diagrams where the gluon is emitted before the shockwave.

\section{Theoretical framework}
\label{sec:theoretical_framework}
\begin{figure}
    \centering
    \includegraphics[scale=0.5]{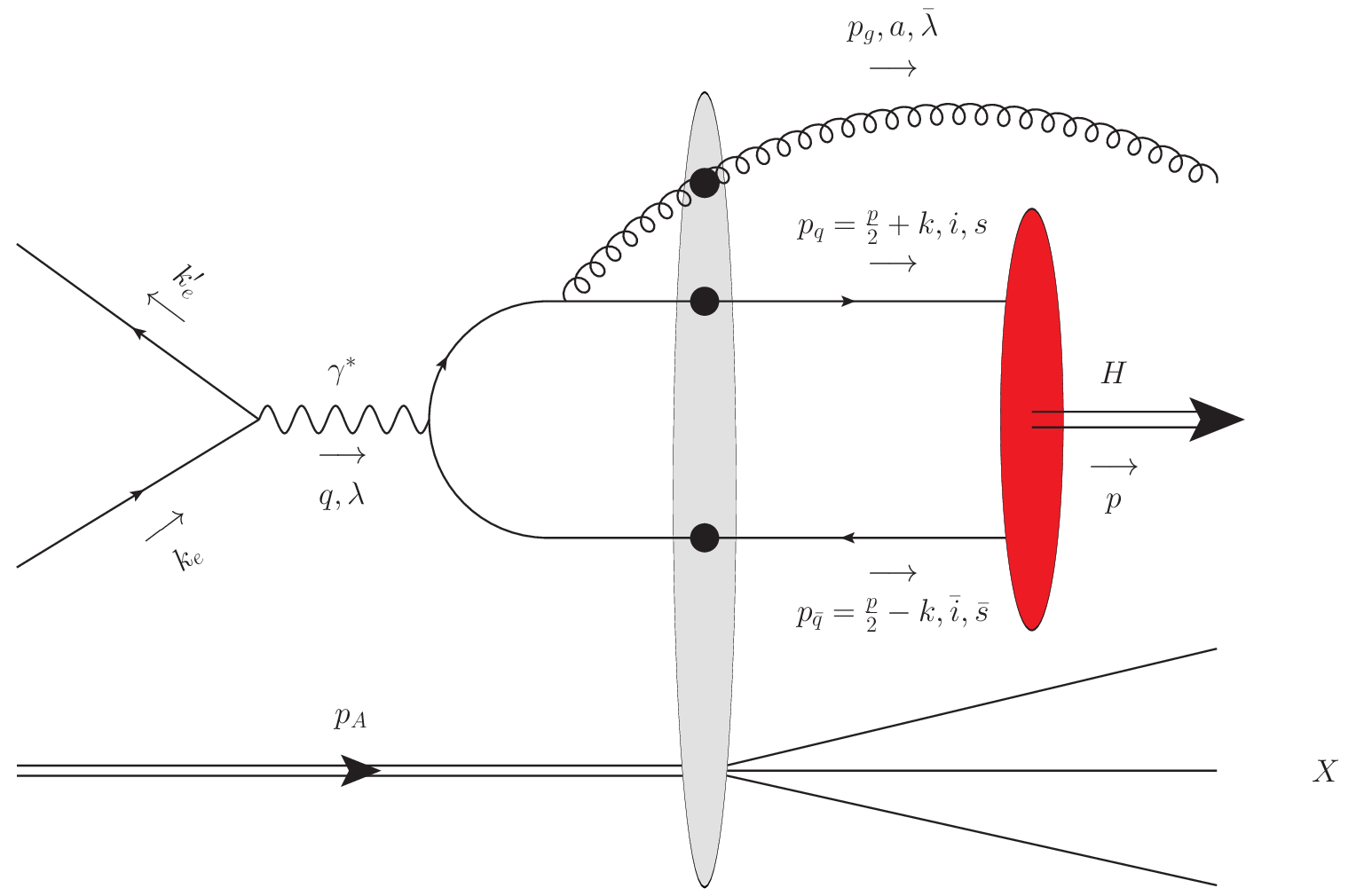}
    \caption{Representative diagram for direct quarkonium + gluon production in electron-nucleus collisions within the joint CGC + NRQCD framework. The gray oval with the black circles represents the interaction of the heavy quark pair and gluon with the gluon shockwave of the nucleus. The red oval represents the hadronization of the heavy quark pair to the quarkonium $H$ with momentum $p$.  }
    \label{fig:schematic-quarkonium+gluon-prod}
\end{figure}

We study direct quarkonium production $H$ accompanied by a gluon $g$ in deep inelastic lepton-nucleus scattering at small-$x$,
\begin{align}
e(k_e) + A(A\,p_n)\to e(k_e') + H(p) + g(p_g) + X\,,
\end{align}
as shown in Fig.~\ref{fig:schematic-quarkonium+gluon-prod}. In this section, we introduce the basic conventions and kinematic variables to be used throughout the manuscript. We then briefly review the basic elements of Color Glass Condensate effective field theory and the Non-relativistic QCD formalism needed for our computation.

\subsection{Kinematics and notations}

We define two light-cone vectors $n_1^\mu = (1, 0, 0_\perp)$ and $n_2^\mu = (0,1, 0_\perp)$ giving the $+$ and $-$ directions respectively. 
The Sudakov decomposition of any vector has the form 
\begin{equation}
p^\mu = p^+ n_1^\mu + p^- n_2^\mu + p_\perp^\mu \; .
\end{equation}
The scalar product of two vectors is given by 
\begin{align*}
a \cdot b &= a^+ b^- + a^- b^+ + a_\perp \cdot b_\perp  \; , \\ 
&=  a^+ b^- + a^- b^+ -\Vec{a} \cdot \Vec{b} \; .  \numberthis 
\end{align*}
We work in a dipole frame where the target moves ultra-relativistically in the $-$ light-cone direction while the photon flies close to the $+$ light-cone direction, as shown in Fig.~\ref{fig:schematic-quarkonium+gluon-prod}. 
The average four-momentum of the nucleon in the nucleus is
\begin{align}
    p_n^\mu = \left(0, p_n^-, 0_\perp  \right) \,,
\end{align}
where we ignored the mass of the nucleon. The four-momentum of the virtual photon is given by
\begin{equation}
q^\mu = \left(q^+, - \frac{Q^2}{2 q^+}, 0_\perp  \right) \,.
\end{equation}
We will denote $p_q, p_{\Bar{q}}, p_g$ the momenta of the on-shell massive quark, antiquark, and gluon respectively.
The momenta $p_q, p_{\Bar{q}}$ will be parameterized as
\begin{align*}
p_q & = \frac{p}{2} + k \; , \\
p_{\bar{q}} & = \frac{p}{2} - k \; . \numberthis[ParamNRQCD]
\end{align*}
where $p$ is the total momentum of the heavy quark pair, and $k$ is their relative momentum. The on-shell conditions for the quark (antiquark) are
\begin{align*}
p_{q}^2 & = \left(\frac{p}{2} + k\right)^2 = m_Q^2 \; , \\ 
p_{\bar{q}}^2 & = \left(\frac{p}{2} - k \right)^2 = m_Q^2 \; ,\numberthis 
\end{align*}
with $m_Q$ the heavy quark mass. These two conditions imply 
\begin{align*}
p \cdot k & = 0 \; , \\ 
p^2 & = 4(m_Q^2 - k^2) \; .  \numberthis[psquared]
\end{align*} 

We define the longitudinal momentum fraction of the quark, antiquark, and gluon with respect to the photon as:
\begin{equation}
    x_q = \frac{p_q^+}{q^+}, \quad x_{\bar{q}} =  \frac{p_{\bar{q}}^+}{q^+}, \quad x_g = \frac{p_g^+}{q^+} \;  .
    \label{xqxqbar}
\end{equation}
Furthermore, we introduce $\xi = k^+/q^+$. Then we find the following relations between the momentum fractions, using the longitudinal momentum conservation $q^+ = p^+ + p_g^+$:
\begin{equation}
\begin{split}
p_q & = \frac{p}{2} + k \Rightarrow x_q = \frac{1}{2} -\frac{x_g}{2} + \xi \,, \\ 
p_{\Bar{q}} & = \frac{p}{2} - k \Rightarrow x_{\Bar{q}} = \frac{1}{2} - \frac{x_g}{2} - \xi \; . 
\end{split}
\label{LongFractNot}
\end{equation}
We denote the polarizations of the virtual photon and the radiated gluon as $\lambda$ and $\bar{\lambda}$ respectively. The two transverse polarizations are denoted by $\lambda, \bar{\lambda} = \pm 1$, and for the virtual photon the longitudinal polarization is denoted by $\lambda=0$. The spin indices for the quark and antiquark are indicated as $s$ and $\bar{s}$ respectively. The color indices of the quark, antiquark, and gluon are $i$, $\bar{i}$ and $a$ respectively.

For convenience, we summarize the kinematics variables and quantum numbers in Table.\,\ref{table:kinematic_variables}.

\begin{table}[H]
\caption{Summary of kinematic variables (see Fig.\,\ref{fig:schematic-quarkonium+gluon-prod}).}
\begin{tabular}{|c | c  |}
\hline
$p_A$ & four-momentum of the nucleus \\
$p_n = p_A/A$ & average four-momentum of a nucleon in the nucleus \\
$k_e$ & incoming electron four-momentum\\
$k_e'$ & outgoing electron four-momentum\\
$q= k_e- k_e'$ &  virtual photon four-momentum \\
$s=(p_n+k_e)^2$ & center of mass energy squared  per nucleon \\
& of the electron-nucleus system\\
$W^2=(p_n+q)^2$ & center of mass energy squared  per nucleon \\
& of the virtual photon-nucleus system\\
$Q^2=-q^2$ & virtuality squared of incoming photon \\
$p_q, p_{\bar{q}}, p_g $ & quark, antiquark, gluon four-momenta   \\
$m_Q^2$ & quark (antiquark) mass squared\\
$p = p_q  + p_{\bar{q}} $ & total momentum of the heavy quark pair\\
$k = \frac{1}{2}(p_q  - p_{\bar{q}}) $ & relative momentum of the heavy quark pair\\
$x_q, x_{\bar{q}}, x_g $ & longitudinal momentum fractions of the quark, antiquark and gluon   \\
 & relative to the virtual photon \\
$\lambda (\bar{\lambda})$ & polarization of virtual photon (gluon)\\
$s, \bar{s}$ & spin of quark and antiquark \\
$i, \bar{i}, a $ & color indices of quark, antiquark and gluon\\
\hline 
\end{tabular}
\label{table:kinematic_variables}
\end{table}

\subsection{NRQCD factorization}

The differential cross-section for quarkonium production can be factorized using the NRQCD factorization formalism \cite{Bodwin:1994jh,Zwirner:2003kw} : 
\begin{equation}
\der \sigma^H=\sum_\kappa \der \hat{\sigma}_\kappa \braket{\mathcal{O}_\kappa^H}\,,
\label{eq:NRQCD_decomp}
\end{equation} 
where $\kappa = {}^{2S+1}L_J^{[c]}$ and the symbols $S$, $L$ $J$, $[c]$ stand for spin, orbital angular momentum, total angular momentum, and color state of the heavy quark pair. In this work, we consider both color singlet $[1]$ and octet contributions $[8]$. $\der \hat{\sigma}_\kappa $ are the short-distance coefficients for the production of the heavy quark pair in the quantum state $\kappa$, and $\braket{\mathcal{O}_\kappa^H}$ are the Long-Distance matrix Element (LDME), which are typically fitted and encode the non-perturbative mechanism for the hadronization of the heavy-quark pair $Q\bar{Q}[\kappa]$ to the quarkonium. The short-distance coefficient functions $\der \hat{\sigma}^\kappa $ are averaged over the degenerate quantum states of the heavy-quark pair
\begin{equation}
\der \hat{\sigma}_\kappa = \frac{1}{N_{\rm color}} \frac{1}{2 J + 1} \sum_{J_z} \der \hat{\sigma}^{\kappa, J_z}  \; ,
\label{eq:SDC_NRQCD}
\end{equation}
where $ N_{\rm color} = 1$ for the color singlet state while $N_{\rm color} = N_c^2 - 1$ for the color octet state. Eq.\,\eqref{eq:SDC_NRQCD} is obtained from the scattering amplitude projected onto a specific $\kappa$ quantum state of the heavy-quark pair. 

In this work, we focus on the $S$-wave contribution to the heavy quarkonium production, for which the projection reads:
\begin{align*}
\mathcal{M}^{\kappa, J_z}(p,p_g) 
& = \frac{1}{\sqrt{m_Q}}  \sum_{s \bar{s }} \sum_{i \bar{i}} \left\langle \frac{1}{2}s ; \frac{1}{2} \Bar{s} | J J_z \right\rangle \braket{3 i; \Bar{3} \Bar{i} | (1, 8d)} \left[ \mathcal{M}_{ s \Bar{s}, i \Bar{i}}(p,0, p_g) \right] \,, \numberthis[AmpProjected]
\end{align*}
where we used the fact that $\braket{L L_z; S S_z | J J_z} = \delta_{JS} \delta_{0 L_z} \delta_{S_z J_z}$ for $L=0$ ($S$-wave). The $P$-wave state can be computed in a similar fashion~\cite{Kang:2013hta} though more involved and will be presented in a future publication. The color projector is defined as 
\begin{equation}
\braket{3 i; \Bar{3} \Bar{i} | c}  =  C^{[c]}_{\Bar{i}i }  = \begin{cases}
& \frac{1}{\sqrt{N_c}} \delta_{\Bar{i}i} \text{ if } c = 1 \\ 
& \sqrt{2} t^d_{\Bar{i}i} \text{ if } c = 8 \; . 
\end{cases}
\label{eq:ColorProjector}
\end{equation}
To carry out the sum over spins, it is useful to introduce the spin projector:
\begin{equation}
\Pi^{J J_z}(p,k) = \frac{1}{\sqrt{m_Q}} \sum_{s \Bar{s}} \left\langle \frac{1}{2} s ; \frac{1}{2} \Bar{s} | J J_z \right \rangle v_{\Bar{s}}  \left( \frac{p}{2} - k\right) \Bar{u}_{s}\left(\frac{p}{2}+ k\right) \; \,.
\end{equation}
In particular, for the states ${}^{1} S_0$ and ${}^{3} S_1$ with $k=0$ these projectors are given by \cite{Kang:2013hta}
\begin{align}
\Pi^{00}(p,0) & = -\frac{1}{\sqrt{8m_Q}} \gamma^5 \left(2 m_Q + \slashed{p}\right) \,, \label{eq:ProjectorSpin0} \\
\Pi^{1 J_z}(p,0)  & = -\frac{\varepsilon_\rho^*(J_z)}{\sqrt{8m_Q}} \gamma^\rho \left( 2 m_Q + \slashed{p}\right)  \,, \label{eq:ProjectorSpin1}
\end{align}
where $\varepsilon(J_z)$ is the polarization vector of the quarkonium which satisfies the relation
\begin{align*}
\sum_{J_z} \varepsilon^*_\rho(J_z) \varepsilon_\alpha(J_z) 
& =  - g_{\rho \alpha} + \frac{p_\rho p_\alpha}{p^2}  \equiv \mathbb{P}_{\rho \alpha} \,.\numberthis[SumTensorSWave]
\end{align*}

\subsection{Color Glass Condensate}
\label{sec:CGC}
In the Color Glass Condensate effective field theory, the large-$x$ components of the nuclear target are treated as stochastic classical color sources, characterized by
a charge density $\rho_A$, which generates the small-$x$ background color fields~\cite{McLerran:1993ni,McLerran:1993ka,McLerran:1994vd,Ayala:1995kg,Ayala:1995hx}. For a fast-moving nucleus along the minus light-cone direction, the color sources generate
a current of the form
\begin{align}
    J^\mu(x^+,\vec{x}) = \delta^{\mu-} \rho_A(x^+,\vec{x})\,,
    \label{eq:current_CGC}
\end{align}
where the sub-eikonal components of the current are neglected. In the semi-classical approximation, the small-$x$ color field is obtained by solving the classical Yang-Mills equations $[D_{\mu}, F^{\mu\nu}] = J^{\nu}$, where the current given by Eq.\,\eqref{eq:current_CGC}. In light-cone gauge $A^+_{\mathrm{cl}}=0$, these equations have the solution
\begin{align}
    A^\mu(x^+,\vec{x}) = \delta^{\mu-} \alpha_A(x^+,\vec{x})\,,
\end{align}
where $\alpha_A(x^+,\vec{x})$ satisfied the Poisson equation
\begin{align}
    \nabla_\perp^2 \alpha_A(x^+,\vec{x}) = -\rho_A(x^+,\vec{x}) \,.
\end{align}
The expectation value of any observable $\mathcal{O}$ in the CGC is computed in perturbation
theory in the presence of the background field $A_{\mathrm{cl}}$
for a given configuration of sources $\rho_A$,
and then averaging over all possible configurations according to a gauge-invariant weight functional $W_Y[\rho_A]$:
\begin{align}
    \left \langle \mathcal{O}[\rho_A] \right \rangle_Y = \int D \rho_{A} W_Y[\rho_A] \mathcal{O}[\rho_A] \,.
\end{align}
The rapidity dependence $Y$ is acquired after absorbing large logs $Y$ in the quantum corrections to the semi-classical approximation. The evolution of $W_Y[\rho_A]$ follows the JIMWLK non-linear renormalization group evolution equations \cite{JalilianMarian:1996xn,JalilianMarian:1997dw,Kovner:2000pt,Iancu:2000hn,Iancu:2001ad,Ferreiro:2001qy}.

Light-like Wilson lines are the relevant degrees of freedom in high-energy scattering and are defined as
\begin{align}
    V\left(\vec{x}\, \right) = \mathcal{P}_+ \exp\left[ig \int dz^+ A^{-a}(z^+,\vec{x}\,) t^a\right] \,, \label{eq:V}
\end{align}
and
\begin{align}
    U\left(\vec{x}\,\right) = \mathcal{P}_+ \exp\left[ig \int dz^+ A^{- a }\left(z^+,\vec{x} \, \right)T^a\right] \; , \label{eq:U}
\end{align}
where $t^a$ and $T^a$ are the generators of $SU(3)$ in the fundamental and adjoint representation respectively.  $\mathcal{P}_+$ is the path-ordering in the $x^+$- direction. The CGC effective vertex that resums multiple eikonal interactions of the quark with the background field of the nucleus reads \cite{McLerran:1998nk}
\begin{equation}
T^q_{\sigma \sigma', ij}(l,l') = 2 \pi \delta(l^+ -l'^+ ) \gamma^+_{\sigma \sigma'} \int d^d \Vec{z} \;  e^{- i (\vec{l} - \vec{l'}) \cdot \vec{z} } V_{ij}(\Vec{z} \, ) \; . \label{eq:CGC-quark-effective-vertex}
\end{equation}
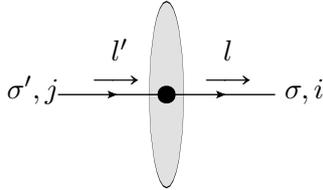
\begin{figure}[H]
\begin{picture}(100,100)
\put(0,0){\includegraphics[scale=0.5]{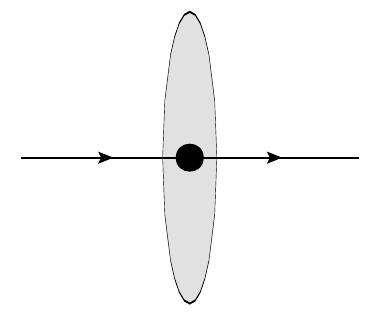}}
\put(18,40){$  \longrightarrow$} 
\put(60,40){$ \longrightarrow$} 
\put(25,50){$ l'$}
\put(67,50){$l$}
\put(90,36){$\sigma, i$}
\put(-14,36){$\sigma', j $}
\end{picture}
\caption{CGC vertex for quarks.}
\end{figure}
\noindent Similarly, for the antiquark interaction one has
\begin{equation}
    T^{\bar{q}}_{\sigma \sigma', ij}(l,l') = - 2 \pi \delta(l'^+ - l^+ ) \gamma^+_{\sigma \sigma'} \int d^d \Vec{z} \;  e^{- i (\vec{l'} - \vec{l}) \cdot \vec{z} } V_{ij}^\dag(\Vec{z} \, ) \; . 
    \label{eq:CGC-antiquark-effective-vertex}
\end{equation}
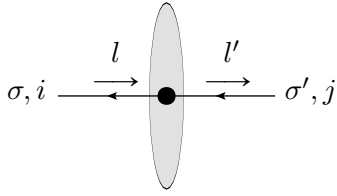
\begin{figure}[H]
\begin{picture}(100,100)
\put(0,0){\includegraphics[scale=0.5]{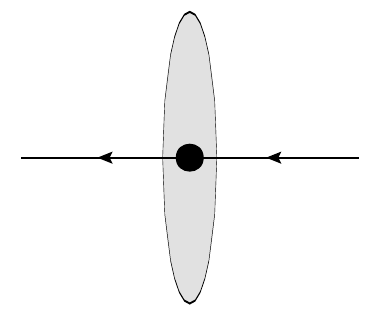}}
\put(18,40){$\longrightarrow$} 
\put(60,40){$\longrightarrow$} 
\put(25,50){$l$}
\put(67,50){$l'$}
\put(90,36){$\sigma', j$}
\put(-14,36){$\sigma, i $}
\end{picture}
\caption{CGC vertex for antiquarks.}
\end{figure}
\noindent Finally, the CGC effective vertex for the gluon reads \cite{Gelis:2005pt,Ayala:1995kg}
\begin{equation}
    T^g_{\mu\nu,ab}(l,l')= - (2 \pi) \delta(l^+ -l'^+ )  (2 l^+ ) g_{\mu \nu } \operatorname{sgn}(l^+) \int d^d \Vec{z} \;  e^{-i(\vec{l}-\vec{l'}) \cdot \vec{z}} U_{ab}^{\operatorname{sgn}(l^+)} (\Vec{z} \, ) \; . 
    \label{eq:CGC-gluon-effective-vertex}
\end{equation}
\begin{figure}[H]
\begin{picture}(100,100)
\put(0,0){\includegraphics[scale=0.5]{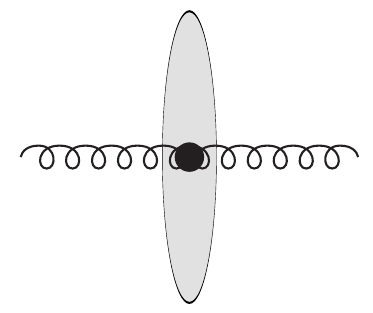}}
\put(18,40){$\longrightarrow$} 
\put(60,40){$\longrightarrow$} 
\put(25,50){$l'$}
\put(67,50){$l$}
\put(90,36){$\mu, a$}
\put(-14,36){$\nu, b $}
\end{picture}
\caption{CGC vertex for gluons.}
\end{figure}
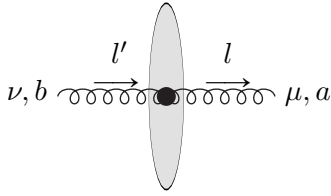
\noindent The adjoint Wilson line has a useful identity 
\begin{equation}
V^\dag (\vec{x} \, ) t^a V(\Vec{x} \, )  = U_{ab}(\Vec{x} \, ) t^b \; .  
\label{eq:WilsonLinePts}
\end{equation}
The effective CGC vertices include all possible scatterings off the target, including the possibility of no-scattering which has to be subtracted to obtain the physical amplitude. This can be done systematically by subtracting from the scattering amplitude a term in which all the Wilson lines inside the effective CGC vertices are set to unity. Furthermore, to obtain the ``reduced amplitude'' one factorizes the overall delta function from ``plus'' lightcone momentum conservation. Mathematically, this amounts to the following operation:
\begin{align}
    (2\pi) \delta\left(\sum_{i = 1}^{N} p_i^+ - q^+ \right) \mathcal{M} = \mathcal{S} - \mathcal{S}[A=0] \label{eq:def_scattering_amplitude_CGC} \,.
\end{align}

\subsection{Outline of the computation}
\label{sec:outline_computation}
Without loss of generality, we compute the differential cross-section for direct quarkonium production + gluon in virtual photon-nucleus collision. The DIS result can be readily obtained from\footnote{Here we assume the azimuthal angle of the scattered electron is integrated out, and thus we do not consider azimuthal correlations with respect to the electron. Thus, it is sufficient to consider the diagonal elements in the polarization of the virtual photon \cite{Mantysaari:2020lhf}. }
\begin{equation}
    \frac{\der\sigma^{e+A\to e'+H +g +X}}{\der W^2 \der Q^2 \der^2 \vec{p} \der^2 \vec{p}_g d x_g }=\sum_{\lambda=\mathrm{L,T}}f_{\lambda}(W^2, Q^2) \ \frac{\der \sigma_{\lambda}^{\gamma^*+A\to H+g+X}}{\der^2 \vec{p} \der^2 \vec{p}_g d x_g } \,.
\end{equation}
Here $\lambda$ denotes the polarization of the virtual photon, and $f_{\lambda}(W^2, Q^2)$ are the photon flux factors:
\begin{align}
    f_{\lambda=\mathrm{L}}(W^2, Q^2)&=\frac{\alpha_{\mathrm{em}}}{\pi Q^2 W^2}(1-y)\,,\\
    f_{\lambda=\mathrm{T}}(W^2, Q^2)&=\frac{\alpha_{\mathrm{em}}}{2\pi Q^2 W^2}[1+(1-y)^2]\,,
\end{align}
with $y$ the inelasticity satisfying $y=(W^2 + Q^2)/s$.

Following Eq.\,\eqref{eq:NRQCD_decomp}, the differential cross-section for direct quarkonium + gluon production reads
\begin{equation}
    \frac{\der \sigma_{\lambda}^{\gamma^*+A\to H+g+X}}{\der^2 \vec{p} \der^2 \vec{p}_g d x_g } = \sum_{\kappa} \frac{\der \hat{\sigma}^{\lambda}_{\kappa}}{\der^2 \vec{p} \der^2 \vec{p}_g d x_g } \braket{\mathcal{O}_\kappa^H} \,,
\end{equation} 
where the short-distance coefficients for $Q\bar{Q}[\kappa]g$ are given by 
\begin{align}
d \hat{\sigma}_{\kappa}^{\lambda} & = \frac{1}{(2q^+)^2} \frac{\der^2 \vec{p}}{(2\pi)^2} \frac{\der^2 \vec{p}_g}{(2\pi)^2} \frac{dx_g}{4 \pi x_g (1 - x_g)} \nonumber \\
& \times \frac{1}{N_{\rm color}} 
 \widebar{\sum_{J_z}} \sum_{a=1}^{N_c^2-1}\sum_{\Bar{\lambda}= \pm 1} \braket{ \mathcal{M}^{\lambda, \bar{\lambda}, a, \kappa, J_z}(p,p_g) \mathcal{M}^{\dag \lambda, \bar{\lambda}, a, \kappa, J_z}(p,p_g)}_Y \,,
\label{eq:diff-Xsec}
\end{align} 
$\braket{\cdots}_Y$ refers to the CGC average over all possible charge configurations inside the target at rapidity scale $Y$ (see Sec.\,\ref{sec:CGC}). 
We summed over the two transverse polarizations $\bar{\lambda}$ and the color $a$ of the gluon. Furthermore, following the convention in \cite{Kang:2013hta}, we denote $\widebar{\sum\limits_{J_z}} = \frac{1}{2J+1} \sum\limits_{J_z}  $. 

\begin{figure}[H]
    \begin{picture}(200,270)
  \put(-90,15){\includegraphics[scale=0.7]{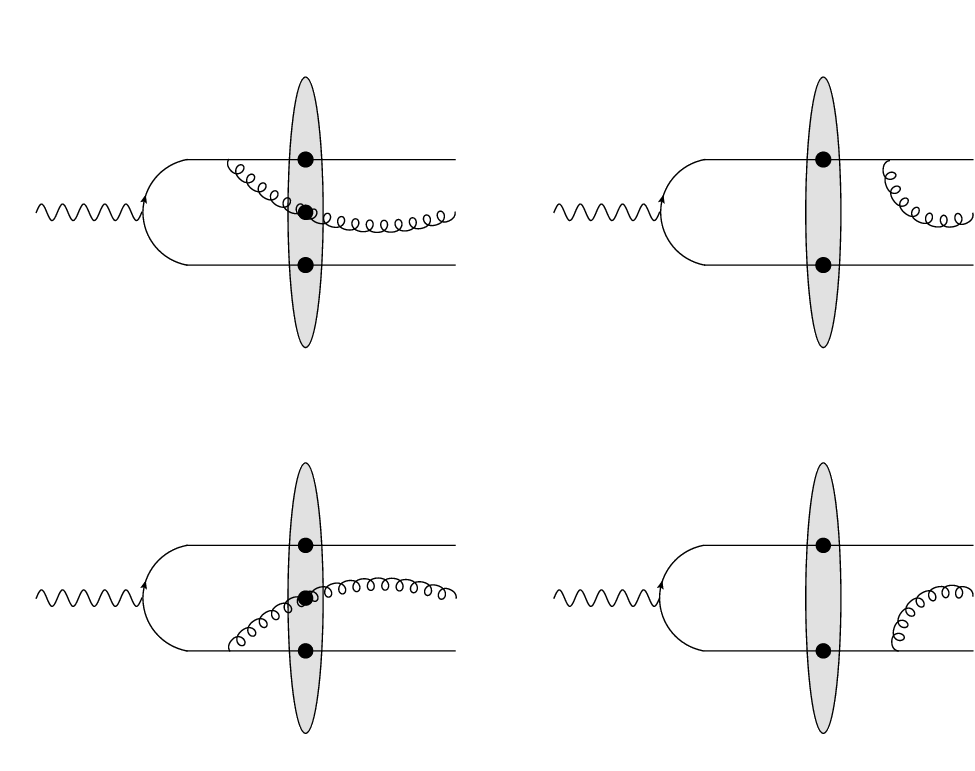}}
    \put(8,135){\small $R_1$} 
    \put(183, 135){\small $R_2$}
    \put(8,5){\small $R_{3}$}
    \put(183,5){\small $R_{4}$}
    \end{picture}
    \caption{The four diagrams for heavy quark pair + gluon production in DIS in the CGC formalism. The black circles with the gray oval represent the interaction of the partons with the background field of the nucleus.}
    \label{fig:RealEmission}
\end{figure}
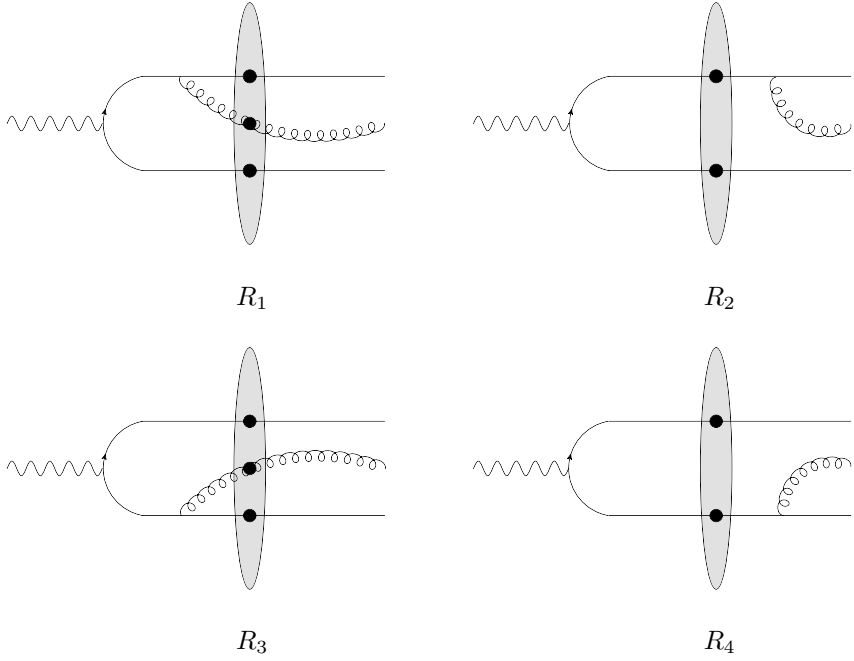

At the level of the amplitude, there are four contributions to $Q\bar{Q}g$ production in DIS:
\begin{equation}
\mathcal{M} = \mathcal{M}_{R_1} + \mathcal{M}_{R_2} + \mathcal{M}_{R_{3}} + \mathcal{M}_{R_{4}} \,,
\label{eq:amplitude_generic}
\end{equation} 
where $R_1\ (R_{3}), R_2\ (R_{4}) $ refer to the contributions where the gluon is emitted by the quark (antiquark) before and after the shockwave respectively. The diagrams associated with these contributions
are shown in Fig.~\ref{fig:RealEmission}. 
Thus, at the level of the differential cross-section in Eq.\,\eqref{eq:diff-Xsec} there are 16 contributions. The amplitudes for $Q\bar{Q}g$ production have the generic form
\begin{align*}
\mathcal{M}_{R_{1(3)}, s\bar{s}, i \Bar{i}}^{\lambda, \bar{\lambda},a}(p,k,p_g) & = \frac{e e_Q}{2 \pi} \int \der^2 \vec{r} \, \der^2 \vec{b} \, \der^2 \vec{z} \,  e^{- i \Vec{k} \cdot \Vec{r}}  e^{- i \Vec{p} \cdot \Vec{b}} e^{- i \vec{p}_g \cdot \Vec{z}} \,  \mathcal{C}^{a}_{R_{1(3)},i \Bar{i}}\left(\vec{r}, \vec{b}, \vec{z} \, \right) \\ 
& \times \Bar{u}_s\left(\frac{p}{2}+k\right)\mathcal{N}_{R_{1(3)}}^{\lambda, \Bar{\lambda}}(p,k,p_g;\vec{r}, \vec{b}, \vec{z} \, ) v_{\Bar{s}}\left(\frac{p}{2} - k \right) \,, \numberthis 
\label{eq:amplitude_unprojected_1}
\end{align*}
for the diagrams in which the gluon is emitted before the shockwave, and 
\begin{align*}
\mathcal{M}_{R_{2(4)}, s\bar{s}, i \Bar{i}}^{\lambda, \bar{\lambda},a}(p,k,p_g) & = \frac{e e_Q}{2 \pi} \int \der^2 \vec{r} \, \der^2 \vec{b} \, \, e^{- i \Vec{k} \cdot \Vec{r}}  e^{- i \Vec{p} \cdot \Vec{b}}   e^{- i \vec{p}_g \cdot \left(\vec{b} \pm  \frac{\vec{r}}{2} \,\right)}\,  \mathcal{C}^a_{R_{2(4)}i \Bar{i}}\left(\vec{b} + \frac{\vec{r}}{2}, \vec{b}  - \frac{\vec{r}}{2}\, \right) \\ 
& \times \Bar{u}_s\left(\frac{p}{2}+k\right)\mathcal{N}_{R_{2(4)}}^{\lambda, \Bar{\lambda}}(p,k,p_g;\vec{r}, \vec{b} \, ) v_{\Bar{s}}\left(\frac{p}{2} - k \right) \,, \numberthis 
\label{eq:amplitude_unprojected_2}
\end{align*}
for the diagrams in which the gluon is emitted after the shockwave. Here $s$ ($\bar{s}$) and $i$ ($\bar{i}$) are the spin and color index of the quark (antiquark) respectively.
As is customary in CGC calculations, we work with transverse spatial coordinates. The quark (antiquark) scatters with the shockwave at transverse location $\vec{x}$ $(\vec{y}\,)$, and the gluon scatters at location $\vec{z}$. We then introduce the coordinates
\begin{align}
    \vec{b} &= \frac{1}{2}\left(\vec{x} + \vec{y} \right) \,, \nonumber \\
    \vec{r} &= \vec{x} - \vec{y} \,,
    \label{eq:rb-definition}
\end{align}
which are conjugate to $\vec{p}$ and $\vec{k}$ respectively. In addition, $C^a_{R_{i},i \Bar{i}},i \Bar{i}$ and $\mathcal{N}^{\lambda,\bar{\lambda}}_{R_{i}}$ stand for the color structure and
the perturbative factor associated with the diagram $R_{i}$. Their explicit expressions will be computed in Sec.~\ref{sec:QQbar_gluon_CGC}. The short-distance coefficients for the amplitude of $Q\bar{Q}[\kappa]g$ are obtained by inserting Eq.\,\eqref{eq:amplitude_unprojected_1} into Eq.~\eqref{AmpProjected} and we have
\begin{align}
\mathcal{M}_{R_{1(3)}}^{\lambda, \Bar{\lambda},a, \kappa, J_z }(p,p_g) &= \frac{e e_Q}{2\pi} \int \der^2 \vec{r} \, \der^2 \vec{b} \, \der^2 \vec{z} \,   e^{- i \Vec{p} \cdot \Vec{b}} e^{- i \vec{p}_g \cdot \Vec{z}} \,\Tr\left[\mathcal{C}^a_{R_{1(3)}}(\vec{r}, \vec{b}, \vec{z} \, ) C^{[c]}\right] \, \nonumber \\
&\times \mathcal{F}_{R_{1(3)}}^{\lambda, \Bar{\lambda}, \kappa, J_z}\left(p,p_g,Q; \vec{r}, \vec{b}, \vec{z} \,  \right)
\end{align}
where we have specialized on the $S$-wave state and defined
\begin{align*}
& \mathcal{F}_{R_{1(3)}}^{\lambda, \Bar{\lambda}, \kappa, J_z}(p,p_g,Q; \vec{r} , \vec{b} , \vec{z} \, )  =  \Tr \left[ \Pi^{J J_z}(p,0) \mathcal{N}^{\lambda,\bar{\lambda}}_{R_{1(3)}}(p,0,p_g; \vec{x} , \vec{y} , \vec{z} \, ) \right] \,. \numberthis[eq:Ffunction]
\end{align*}
Similar expressions are obtained for $\mathcal{M}_{R_{2(4)}}^{\lambda, \Bar{\lambda}, \kappa, J_z }(p,p_g) $. The functions $\mathcal{F}$ will be computed in Sec.~\ref{sec:projectionQQbar}.

\section{$Q\bar{Q}g$ production in the CGC}
\label{sec:QQbar_gluon_CGC}

In this section, we compute the amplitude for $Q\bar{Q}g$ production in virtual photon-nucleus collision in the CGC. This calculation has been carried out in \cite{Beuf:2021qqa,Beuf:2022ndu} within light-cone perturbation theory. Here, we follow the approach in \cite{Caucal:2021ent} by performing our computation using covariant perturbation theory. We employ the standard QCD+QED Feynman rules in momentum space (see appendix \ref{sec:Feynmanrules}) with the effective CGC vertices in Eqs.\,\eqref{eq:CGC-quark-effective-vertex},\,\eqref{eq:CGC-antiquark-effective-vertex} and \eqref{eq:CGC-gluon-effective-vertex}. We work in light-cone gauge for the photon and gluon fields: $n_2 \cdot A_{\mathrm{QED}} = 0 = A_{\mathrm{QED}}^+$ and $n_2 \cdot A_{\mathrm{QCD}} = 0 = A_{\mathrm{QCD}}^+$. In this gauge the polarization vectors are
\begin{align}
\varepsilon^\mu(q,\lambda = 0) & = \left( 0, \frac{Q}{q^+}, 0_\perp \right) \; , \label{eq:polarization_vector_photonLongitudinal} \\ 
\varepsilon^\mu(q, \lambda = \pm 1) & =  (0,0, \epsilon_\perp^{\; \lambda})  \,, \label{eq:polarization_vector_photonTransverse}
\end{align}
for the virtual photon, and
\begin{align}
    \varepsilon^\mu(p_g, \bar{\lambda} = \pm 1) & =  \left(0, \frac{\vec{\epsilon}^{\; \bar{\lambda}} \cdot \vec{p}_g}{p_{g}^+}, \epsilon_\perp^{\; \bar{\lambda}}\right)  \; , 
    \label{eq:polarization_vector_gluonTransverse}
\end{align}
for the real gluon. Here we introduced the two-dimensional vector
\begin{align}
    \epsilon_\perp^{\; \lambda} =  \frac{1}{\sqrt{2}} (1, \pm i) \; . \nonumber
\end{align}
It is useful to define $ \omega_{\mu \nu} = \frac{1}{2} \left[\gamma^\mu, \gamma^\nu\right]$. 
For $i,j$ transverse component indices, we have 
\begin{equation}
\gamma^i \gamma^j = \frac{1}{2} \left\{\gamma^i, \gamma^j\right\} + \frac{1}{2} \left[ \gamma^i, \gamma^j\right]  =  g_\perp^{ij} + \omega^{ij}  \; . 
\end{equation}
We perform the explicit computation for the amplitudes corresponding to the diagrams $R_1$ (gluon emission from quark before the shockwave) and $R_2$ (gluon emission from quark after the shockwave). The computation for the contributions $R_{3}$ and $R_{4}$ are almost identical, thus we simply present their results. 

\subsection{Gluon emission from quark before the shockwave}

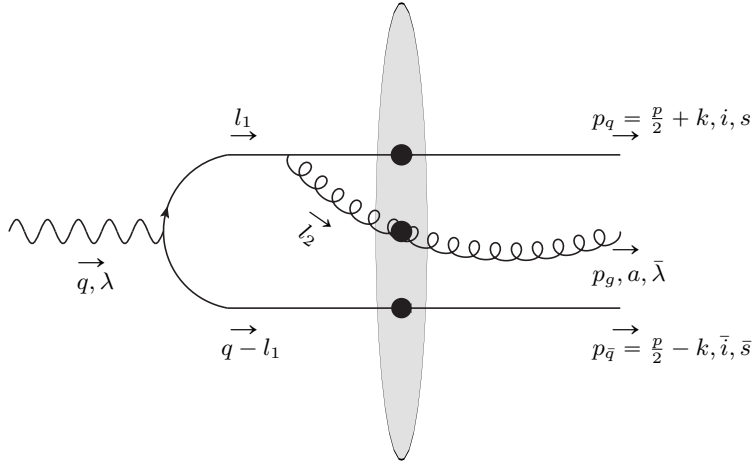
\begin{figure}[h!]
    \begin{picture}(200,200)
    \put(-150,0){\includegraphics[scale=0.6]{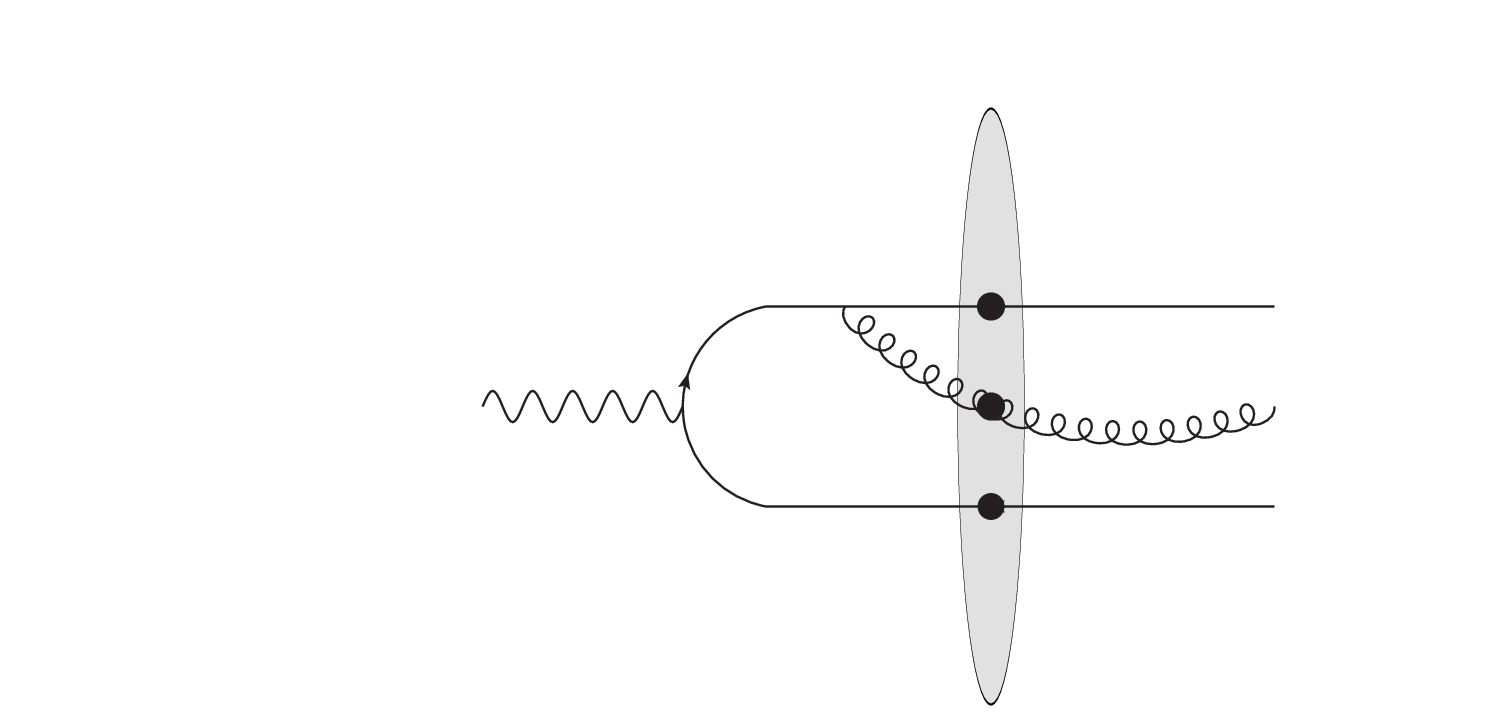}}
    \put(72,50){$\rightarrow$}
    \put(69,43){\footnotesize  $q-l_1$}
    \put(72,123){$\rightarrow$}
    \put(74,130){\footnotesize {$l_1$}}
    \put(100,95){\mbox{\rotatebox{-35}{$\rightarrow$}}}
    \put(97,88){\mbox{\rotatebox{-35}{\footnotesize $l_2$}}}
    \put(15, 75){$\rightarrow$}
    \put(15,68){\footnotesize $q,\lambda$}
    \put(215,123){$\rightarrow$}
    \put(208,130){\footnotesize $p_q = \frac{p}{2}+ k , i, s$}
    \put(215,50){$\rightarrow$}
     \put(208,43){\footnotesize $p_{\Bar{q}} = \frac{p}{2}- k , \bar{i}, \bar{s}$}
    \put(215,80){$\rightarrow$}
    \put(208,71){\footnotesize $p_g, a, \Bar{\lambda}$}
    \end{picture}
    \caption{Real gluon emission from the quark before the shockwave. Here, $l_1$ and $l_2$ are loop momenta.}
    \label{fig:R1}
\end{figure}

\subsubsection{The scattering amplitude }

Using the standard momentum space Feynman rules in appendix \ref{sec:Feynmanrules} and the CGC effective vertices presented  in Sec.~\ref{sec:CGC}, one can write the scattering amplitude for the $R_1$ diagram (Fig.~\ref{fig:R1}) as :
\begin{align*}
&  S_{R_1, s \Bar{s}, i \Bar{i}}^{\lambda, \Bar{\lambda},a}(p,k,p_g) \\
& = \int \frac{d^4 l_1}{(2\pi)^{4}} \int \frac{d^4 l_2}{(2\pi)^{4}} \Bar{u}_s(p_q) T^q_{in}(p_q,l_1 - l_2) S_0(l_1 -l_2) i g  \gamma_\mu t^b_{nm} S_0(l_1) (-ie e_Q \slashed{\varepsilon}(q,\lambda))  \\*
& \times S_0(l_1 - q) T^{\Bar{q}}_{m \Bar{i}}(q-l_1,p_{\Bar{q}}) v_{\Bar{s}}(p_{\Bar{q}}) \varepsilon^{* \alpha}(p_g,\bar{\lambda }) T^g_{\alpha \nu, ab }(p_g,l_2) G_0^{\nu \mu}(l_2) \,. \numberthis 
\end{align*}
Replacing the propagators and the vertices by their expressions and using the relations in Eq.~\eqref{eq:GluonPts} and Eq.~\eqref{eq:WilsonLinePts}, then the scattering amplitude reads
\begin{align*}
&  S_{R_1, s \Bar{s}, i \Bar{i}}^{\lambda, \Bar{\lambda},a}(p,k,p_g) \\
& = (2 \pi) \delta(p_q^+ + p_{\Bar{q}}^+ + p_g^+ - q^+ ) \frac{e e_Q}{2 \pi} \int \der^2 \vec{x} \; \der^2 \vec{y} \;   \der^2 \vec{z}  \;  \left[V(\vec{x}) V^\dag(\vec{z} \,) t^a V(\vec{z} \,)V^\dag(\vec{y} \,)  \right]_{i \Bar{i}} e^{- i \Vec{p}_q \cdot \Vec{x}} \\
& \times   e^{- i \Vec{p}_{\Bar{q}} \cdot \Vec{y}}  e^{- i \vec{p}_g \cdot \Vec{z}} (- 2 q^+) \int \frac{d^4 l_1}{(2\pi)^{2}} \int \frac{d^4 l_2}{(2\pi)^{3}} \delta(p_{\Bar{q}}^+ - q^+ + l_1^+ ) \delta(p_g^+ - l_2^+) e^{i \Vec{l}_1 \cdot (\Vec{x} - \Vec{y})} e^{i \Vec{l}_2 \cdot (\Vec{z} - \Vec{x})}  \\
& \times g (2 p_g^+) \frac{1}{2q^+}  \frac{\Bar{u}_s(p_q) \gamma^+ \left[(\slashed{l}_1 - \slashed{l}_2) + m_Q \right] \slashed{\varepsilon}^*(l_2,\Bar{\lambda}) (\slashed{l}_1 + m_Q) \slashed{\varepsilon}(q,\lambda,) \left[ (\slashed{l}_1 - \slashed{q}) + m_Q \right] \gamma^+ v_{\Bar{s}}(p_{\Bar{q}})}{[(l_1 - l_2)^2 - m_Q^2 + i \varepsilon] \left[l_1^2 - m_Q^2 + i \varepsilon\right] \left[ (l_1 - q)^2 - m_Q^2 + i \varepsilon\right] \left[l_2^2 + i \varepsilon \right]} \,.  \numberthis 
\end{align*} 
The $+$ components of the loop variables $l_1$ and $l_2$ are easily integrated over with the two delta functions thanks to the eikonal interaction of the partons in the projectile with the shockwave. Following the strategy in \cite{Caucal:2021ent} and using Eq.~\eqref{Dirac_Gluon_Q_before}, the Dirac structure can be divided into two parts: 
\begin{align*}
& \gamma^+ \left[(\slashed{l}_1- \slashed{l}_2) + m_Q\right] \slashed{\varepsilon}^* (l_2,\Bar{\lambda})(\slashed{l}_1 + m_Q) \\
& =  \Vec{\epsilon}^{\; \Bar{\lambda}*}_n 2 \frac{x_q}{x_g}   \left\{ \left( \Vec{l}_2 - \frac{x_g}{1- x_{\Bar{q}}} \Vec{l}_1\right)_m  \left[\left( 1 + \frac{x_g}{2x_q}\right) \delta^{nm} - \frac{x_g}{2x_q} \omega^{mn}\right] + \frac{x_q}{x_g + x_q} \left(\frac{x_g}{x_q}\right)^2 \frac{m_Q}{2} \gamma^n   \right \}  \\
& \times \gamma^+ (\slashed{l}_1 + m_Q) - \Vec{\epsilon}^ {\; \Bar{\lambda}*}_n \gamma^n \gamma^+ \left(\frac{x_q}{1-x_{\Bar{q}}}\right) (l_1^2 - m_Q^2 ) \; . \label{eq:reg_vs_inst} \numberthis 
 \end{align*}
The first term corresponds to a \textit{regular} term. The second one, proportional to the factor $l_1^2 - m_Q^2 $, can be identified as an \textit{instantaneous} term. It is associated with the diagram in LCPT where the quark with momentum $l_1$ is instantaneous; the factor canceling the quark propagator. 
 
Using Eq.\,\eqref{eq:def_scattering_amplitude_CGC}, the reduced amplitude for $Q\bar{Q}g$ with the gluon emitted by the quark before the shockwave reads
 \begin{align*}
\mathcal{M}_{R_1, s \Bar{s}, i \Bar{i}}^{\lambda, \Bar{\lambda},a}(p,k,p_g) & = \frac{ee_Q}{2 \pi} \int \der^2 \vec{x}  \; \der^2 \vec{y} \; \der^2 \vec{z} \;    \mathcal{C}^{a}_{R_1}(\vec{x} , \vec{y} , \vec{z} \, )_{i \Bar{i}} \; e^{- i \Vec{p}_q \cdot \Vec{x}} e^{- i \Vec{p}_{\Bar{q}} \cdot \Vec{y}} e^{- i \vec{p}_g \cdot \Vec{z}} \\
& \times  \bar{u}_s(p_q) \left( \mathcal{N}_{R_1, reg}^{\lambda, \Bar{\lambda}}(p,k,p_g; \vec{x} , \vec{y} , \vec{z} \, ) + \mathcal{N}_{R_1, inst}^{\lambda, \Bar{\lambda}}(p,k,p_g; \vec{x} , \vec{y} , \vec{z} \, )  \right) v_{\Bar{s}}(p_{\Bar{q}}) \,, \numberthis 
 \end{align*}
where we define the color structure
\begin{equation}
    \mathcal{C}^a_{R_1}(\vec{x} , \vec{y} , \vec{z} \, )  = V(\vec{x}  \, ) V^\dag(\vec{z} \,) t^a V(\vec{z} \,)V^\dag(\vec{y} \,) - t^a \mathds{1} \; ,  
    \label{eq:color-structure-R1}
\end{equation}    
and the perturbative factors
\begin{align*}
    \mathcal{N}_{R_1, reg}^{\lambda, \Bar{\lambda}}(p,k,p_g; \vec{x} , \vec{y} , \vec{z} \, ) & = \frac{g}{2\pi} \int \frac{\der^2 \Vec{l}_1}{(2\pi)^2} \frac{\der^2 \Vec{l}_2}{(2\pi)^2} e^{i \Vec{l}_1 \cdot (\Vec{x}- \Vec{y})}e^{i \Vec{l}_2\cdot (\Vec{z}- \Vec{x})} I_{R_1,reg} \mathcal{T}_{R_1,reg}^{\lambda, \Bar{\lambda}}(l_1, l_2)\bigg|_{ \begin{subarray}{l}  l_1^+ = q^+ - q^+ x_{\bar{q}}, \\  l_2^+ = q^+ x_g   \end{subarray}} \,, \numberthis[Nreg]
\end{align*}
for the regular, and 
\begin{align*}
\mathcal{N}_{R_1, inst}^{\lambda, \Bar{\lambda}}(p,k,p_g; \vec{x} , \vec{y} , \vec{z} \, ) & = \frac{g}{2\pi} \int \frac{\der^2 \Vec{l}_1}{(2\pi)^2} \frac{\der^2 \Vec{l}_2}{(2\pi)^2} e^{i \Vec{l}_1 \cdot (\Vec{x}- \Vec{y})}e^{i \Vec{l}_2\cdot (\Vec{z}- \Vec{x})} I_{R_1,inst} \mathcal{T}_{R_1,inst}^{\lambda, \Bar{\lambda}}(l_1, l_2)\bigg|_{ \begin{subarray}{l}   l_1^+ = q^+ - q^+ x_{\bar{q}}, \\  l_2^+ = q^+ x_g \,,\end{subarray}} 
\numberthis[Ninst]
\end{align*}
for instantaneous contributions.

We have decomposed the two perturbative factors into Dirac structures according to Eq.~\eqref{eq:reg_vs_inst}
\begin{align*}
\mathcal{T}_{R_1,reg}^{\lambda, \Bar{\lambda}}(l_1, l_2) & = \Vec{\epsilon}^{\; \Bar{\lambda}*}_n 4q^+ x_q \left\{ \left( \Vec{l}_2 - \frac{x_g}{1- x_{\Bar{q}}} \Vec{l}_1\right)_m  \left[\left( 1 + \frac{x_g}{2x_q}\right) \delta^{nm} - \frac{x_g}{2x_q} \omega^{mn}\right] \right. \\
& \left. + \frac{x_q}{x_g + x_q} \left(\frac{x_g}{x_q}\right)^2 \frac{m_Q}{2}\gamma^n   \right \}  \frac{1}{2q^+ } \gamma^+ (\slashed{l}_1 + m_Q) \slashed{\varepsilon}(\lambda,q) \left[ (\slashed{l}_1 - \slashed{q}) + m_Q \right] \gamma^+ \; , \numberthis[TR1Reg] \\
\mathcal{T}_{R_1,inst}^{\lambda, \Bar{\lambda}}(l_1, l_2) & = -  2q^+ x_g \frac{x_q}{1-x_{\Bar{q}}}  \Vec{\epsilon}^{\; \Bar{\lambda}*}_n \frac{1}{2q^+}\gamma^n \gamma^+ \slashed{\varepsilon}(\lambda,q) \left[ (\slashed{l}_1 - \slashed{q}) + m_Q \right] \gamma^+ \; , \numberthis[TR1Inst]
\end{align*}
and defined the corresponding pole integrals over $l_{i}^-$ 
\begin{align*}
I_{R_1, reg} &  = \int d l_1^- d l_2^- \frac{-2q^+ }{[(l_1 - l_2)^2 - m_Q^2 + i \varepsilon] \left[l_1^2 - m_Q^2 + i \varepsilon\right] \left[ (l_1 - q)^2 - m_Q^2 + i \varepsilon\right] \left[l_2^2 + i \varepsilon \right]} \,,  \numberthis \\ 
I_{R_1,inst} & = \int d l_1^- d_2^- \frac{-2q^+}{[(l_1 - l_2)^2 - m_Q^2 + i \varepsilon]  \left[ (l_1 - q)^2 - m_Q^2 + i \varepsilon\right] \left[l_2^2 + i \varepsilon \right]} \;. \numberthis 
\end{align*}
The pole structures are computed using Cauchy's residue theorem, closing the contour on the upper half-plane. We finally obtain the following results 
\begin{align*}
I_{R_1,inst}  & = -  \frac{(2\pi)^2 \theta(x_g) \theta(x_q) \theta(x_{\Bar{q}}) \theta(1- x_{\Bar{q}})}{2q^+ x_q x_{\Bar{q}} x_g } \frac{1}{\left[Q^2 + \frac{\Vec{l}_1^{\; 2} + m_Q^2}{x_{\Bar{q}}} + \frac{(\Vec{l}_2- \Vec{l}_1)^2 + m_Q^2 }{x_q} + \frac{\Vec{l}_2^{\; 2}}{x_g}\right]}  \; , \numberthis[Pole_inst] \\
I_{R_1, reg} & = \frac{(2\pi)^2 \theta(x_g) \theta(x_q)\theta(1- x_{\Bar{q}}) \theta(x_{\Bar{q}})}{2q^+ x_g x_q} \\
& \times \frac{1}{\left[Q^2x_{\Bar{q}}(1-x_{\Bar{q}}) + m_Q^2 + \vec{l}_1^{\; 2} \right] \left[ Q^2 + \frac{\Vec{l}_1^{\; 2}+ m_Q^2}{x_{\Bar{q}}} + \frac{(\Vec{l}_2 - \Vec{l}_1)^2 + m_Q^2 }{x_q} + \frac{\Vec{l}_2^{\; 2}}{x_g}  \right]}  \; . \numberthis[Pole_reg]
\end{align*}

\subsubsection{Explicit calculation of the Dirac structures and transverse momenta integrals}

We consider first the case in which the photon is longitudinally polarized. From Eq.~\eqref{eq:polarization_vector_photonLongitudinal}, we have $\slashed{\varepsilon}(q,\lambda = 0) = \frac{Q}{q^+}\gamma^+$. 
We observe easily from Eq.~\eqref{TR1Inst} that the corresponding instantaneous term vanishes as $(\gamma^+)^2 = 0$:
\begin{align*}
\mathcal{T}_{R_1,ins}^{\lambda = 0 , \Bar{\lambda}}(l_1, l_2) &  = 0 \,. \numberthis 
\label{eq:T-long-ins}
\end{align*}
For the regular term, after some elementary algebra, we obtain
\begin{align*}
\mathcal{T}_{R_1,reg}^{\lambda = 0 , \Bar{\lambda}}(l_1, l_2) &  = -8 Q q^+ x_q  x_{\Bar{q}} (1- x_{\Bar{q}})   \Vec{\epsilon}^{\; \Bar{\lambda}*}_n \left\{ \left( \Vec{l}_2 - \frac{x_g}{1- x_{\Bar{q}}} \Vec{l}_1\right)_m  \left[\left( 1 + \frac{x_g}{2x_q}\right) \delta^{nm} - \frac{x_g}{2x_q} \omega^{mn}\right] \right. \\
& \left.+ \frac{x_q}{x_g + x_q} \left(\frac{x_g}{x_q}\right)^2 \frac{m_Q}{2}\gamma^n   \right \} \gamma^+ \,. \numberthis 
\label{eq:T-long-reg}
\end{align*}

For the transversely polarized photon case, using Eq.~\eqref{eq:polarization_vector_photonTransverse}, we have $\slashed{\varepsilon}(q,\lambda = \pm 1) = - \vec{\epsilon}^{\; \lambda}_i \gamma^i$. The regular term takes the form:
\begin{align*}
\mathcal{T}_{R_1,reg}^{\lambda = \pm 1 , \Bar{\lambda}}(l_1, l_2) & =  -  4q^+ x_q  \Vec{\epsilon}^{\; \Bar{\lambda}*}_n \Vec{\epsilon}^{\; \lambda}_i \left\{ \left( \Vec{l}_2 - \frac{x_g}{1- x_{\Bar{q}}} \Vec{l}_1\right)_m  \left[\left( 1 + \frac{x_g}{2x_q}\right) \delta^{nm} - \frac{x_g}{2x_q} \omega^{mn}\right] \right.  \\
& \left. + \frac{x_q}{x_g + x_q}   \left(\frac{x_g}{x_q}\right)^2 \frac{m_Q}{2}\gamma^n   \right \}   \left\{ \Vec{l}_{1j} \left[ \delta^{ij} (1- 2 x_{\Bar{q}}) - \omega^{ij}\right] \gamma^+ + m_Q \gamma^i\gamma^+   \right\}  \;,
\numberthis
\label{eq:T-trans-red}
\end{align*}
while the instantaneous term reads
\begin{align*}
\mathcal{T}_{R_1,inst}^{\lambda = \pm 1, \Bar{\lambda}}(l_1, l_2) & =   2q^+ x_g \frac{x_q {x_{\Bar{q}}} }{1-x_{\Bar{q}}}  \Vec{\epsilon}^{\; \Bar{\lambda}*}_n \Vec{\epsilon}^{\; \lambda}_i  \gamma^n  \gamma^i \gamma^+ \;. \numberthis
\label{eq:T-trans-ins}
\end{align*}
\subsection{Results for the amplitude}
Armed with the results in Eqs.~\eqref{Pole_inst}-\eqref{Pole_reg} and Eqs.~\eqref{eq:T-long-ins}-\eqref{eq:T-trans-ins}, we are in a position to compute the perturbative factors in Eqs.\,\eqref{Nreg} and \eqref{Ninst}, they read for longitudinally polarized photon
\begin{align}
    \mathcal{N}_{R_1, inst}^{\lambda = 0 , \Bar{\lambda}}(p,k,p_g; \vec{x} , \vec{y} , \vec{z} \, ) =0 \,,
\end{align}
\begin{align*}
& \mathcal{N}_{R_1, reg}^{\lambda = 0 , \Bar{\lambda}}(p,k,p_g; \vec{x} , \vec{y} , \vec{z} \, ) 
 = -2 Q  \frac{x_{\Bar{q}} (1- x_{\Bar{q}})}{x_g }  \frac{g}{\pi}  \theta(x_g) \theta(x_q)\theta(1- x_{\Bar{q}}) \theta(x_{\Bar{q}})     \Vec{\epsilon}^{\; \Bar{\lambda}*}_n \gamma^+   \\
& \times \left\{ I_{1m}(\Vec{y}-\Vec{x}, \Vec{z}-\Vec{x}) \left[\left( 1 + \frac{x_g}{2 x_q}\right) \delta^{nm} - \frac{x_g}{2 x_q} \omega^{mn}\right]  -   m_Q I_0(\Vec{y}-\Vec{x}, \Vec{z}- \Vec{x}) \frac{x_g^2}{2 (1-x_{\bar{q}})x_q}  \gamma^n    \right\} \; , \numberthis[eq:N_R1reg_L] 
\end{align*}
and for transversely polarized photon
\begin{align*}
&\mathcal{N}_{R_1, reg}^{\lambda = \pm 1, \Bar{\lambda}}(p,k,p_g; \vec{x} , \vec{y} , \vec{z} \, )  =  \frac{1}{x_g} \frac{g}{\pi} \theta(x_g) \theta(x_q)\theta(1- x_{\Bar{q}}) \theta(x_{\Bar{q}})  \Vec{\epsilon}^{\; \Bar{\lambda}*}_n \Vec{\epsilon}^{\; \lambda}_i  \\
 & \times \left\{ I_{2mj}(\Vec{y}- \Vec{x}, \Vec{z}- \vec{x})  \left[\left( 1 + \frac{x_g}{2x_q}\right) \delta^{nm} - \frac{x_g}{2x_q} \omega^{mn}\right]\left[ \delta^{ij} (1- 2 x_{\Bar{q}}) - \omega^{ij}\right] \gamma^+ \right. \\ 
& - m_Q  I_{1m}(\Vec{y}- \Vec{x}, \Vec{z}- \Vec{x}) \left[\left( 1 + \frac{x_g}{2x_q}\right) \delta^{nm} - \frac{x_g}{2x_q} \omega^{mn}\right]  \gamma^i\gamma^+  \\
& + m_Q \Tilde{I}_{1j}(\Vec{y}- \Vec{x}, \Vec{z}- \Vec{x})  \frac{x_g^2}{2(1-x_{\bar{q}}) x_q} \gamma^n  \left[ \delta^{ij} (1- 2 x_{\Bar{q}}) - \omega^{ij}\right] \gamma^+    \\
& \left. - m_Q^2 I_0(\Vec{y}- \Vec{x}, \Vec{z}- \Vec{x} ) \frac{x_g^2}{2(1-x_{\bar{q}})x_q}\gamma^n 
 \gamma^i\gamma^+     \right\} \; ,  \numberthis[eq:N_R1reg_T] 
 \end{align*}

\begin{align*}
& \mathcal{N}_{R_1, inst}^{\lambda = \pm 1 , \Bar{\lambda}}(p,k,p_g; \vec{x} , \vec{y} , \vec{z} \, ) \\*
& =  - \frac{ x_g x_q x_{\Bar{q}}}{2(1-x_{\Bar{q}})}    \frac{g}{\pi}  \theta(x_g) \theta(x_q) \theta(x_{\Bar{q}}) \theta(1- x_{\Bar{q}})  \Vec{\epsilon}^{\; \Bar{\lambda}*}_n \Vec{\epsilon}^{\; \lambda}_i \frac{\bar{Q}_{R_1, \xi} K_1(\Bar{Q}_{R_1, \xi}X_{R_1, \xi}) }{X_{R_1, \xi}}   \gamma^n  \gamma^i \gamma^+ \; , \numberthis[eq:N_R1ins_T] 
\end{align*}
where we introduced the variables
\begin{align*}
\Bar{Q}_{R_1, \xi}^2 & = Q^2 + m_Q^2 \frac{1-x_g}{x_q x_{\Bar{q}}} \,, \\
X_{R_1,\xi}^2  & = x_g x_q (\Vec{z} - \Vec{x})^2 + x_q x_{\Bar{q}} (\Vec{y}- \Vec{x})^2 + x_{\Bar{q}} x_g (\Vec{y}- \Vec{z})^2 \; . \numberthis \label{eq:variablesR1}
\end{align*}

The transverse integrals have been absorbed in the definition of the $I_0, I_{1m},I_{2mj}, \tilde{I}_{1j}$ functions. Explicit expressions for these functions can be found in appendix \ref{sec:R1TransverseIntegral} where they have been computed using both Schwinger and Feynman parametrization. They have to be taken with $z_1 = x_{\bar{q}}, z_2 = x_g, z_3 = x_q$. In the massless limit, the perturbative factors in Eqs.\,\eqref{eq:N_R1reg_L},\,\eqref{eq:N_R1reg_T} and \eqref{eq:N_R1ins_T} match with those presented in \cite{Caucal:2021ent} (before spinor contraction) up to a factor of $1/(2q^+)$ that comes from our different convention of the perturbative factor $\mathcal{N}_{R_1}^{\lambda, \Bar{\lambda }}$. We have also checked that our results are consistent with those obtained in \cite{Beuf:2022ndu} and \cite{Beuf:2021qqa}
within the light-cone perturbation formalism.

Now, let us express our results in terms of the variables introduced in Eqs.\,\eqref{ParamNRQCD},\,\eqref{LongFractNot}, and \eqref{eq:rb-definition}. The reduced amplitude becomes
 \begin{align*}
&  \mathcal{M}_{R_1, s \Bar{s}, i \Bar{i}}^{\lambda, \Bar{\lambda},a}(p,k,p_g) \\
& = \frac{ee_Q}{2 \pi} \int \der^2 \vec{r} \; \der^2 \vec{b}  \; \der^2 \vec{z} \;    \mathcal{C}^{a}_{R_1}\left(\vec{b} + \frac{\vec{r} }{2}, \vec{b}  - \frac{\vec{r} }{2}, \vec{z} \, \right)_{i \Bar{i}} \; e^{- i \Vec{p} \cdot \Vec{b} } e^{- i \Vec{k} \cdot \Vec{r}} e^{- i \vec{p}_g \cdot \Vec{z}} \\
& \times  \bar{u}_s\left(\frac{p}{2} + k\right) \left[ \mathcal{N}_{R_1, reg}^{\lambda, \Bar{\lambda}}\left(p,k,p_g; \vec{b} + \frac{\vec{r} }{2}, \vec{b}  - \frac{\vec{r} }{2}, \vec{z} \, \right) + \mathcal{N}_{R_1, inst}^{\lambda, \Bar{\lambda}}\left(p,k,p_g; \vec{b} + \frac{\vec{r} }{2}, \vec{b}  - \frac{\vec{r} }{2}, \vec{z} \, \right)  \right] v_{\Bar{s}}\left(\frac{p}{2}-k\right)  \numberthis[AmplitudeR1QQbar] \,, 
 \end{align*}
with $ z_1 = x_{\bar{q}} = \frac{1}{2}- \frac{x_g}{2} - \xi$, $z_2 = x_g $ and $z_3 = x_q = \frac{1}{2}- \frac{x_g}{2} + \xi $.

\subsection{Gluon emission from quark after the shockwave}

\begin{figure}[h!]
    \begin{picture}(200,190)
    \put(-20,0){\includegraphics[scale=0.6]{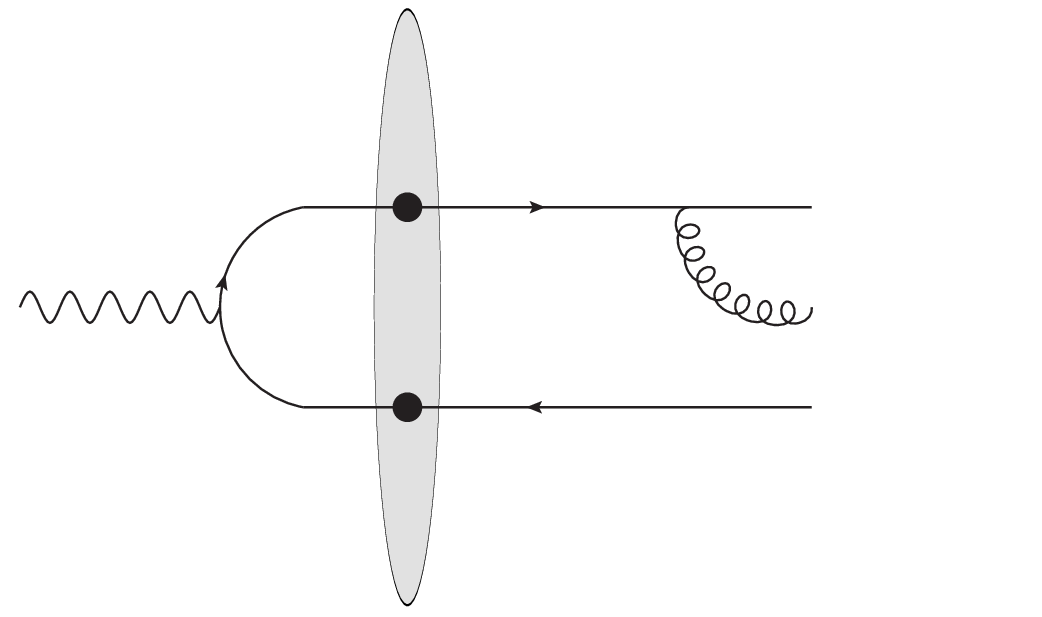}}
    \put(72,50){$\rightarrow$}
    \put(69,43){\footnotesize  {$q-l$}}
    \put(72,123){$\rightarrow$}
    \put(74,130){\footnotesize {$l$}}
    \put(15, 75){$\rightarrow$}
    \put(15,68){\footnotesize $q,\lambda$}
    \put(215,123){$\rightarrow$}
    \put(208,130){\footnotesize $p_q = \frac{p}{2}+ k , i, s$}
    \put(215,50){$\rightarrow$}
     \put(208,43){\footnotesize $p_{\Bar{q}} = \frac{p}{2}- k , \bar{i}, \bar{s}$}
    \put(215,80){$\rightarrow$}
    \put(208,71){\footnotesize $p_g,a,\Bar{\lambda}$}
    \end{picture}
    \caption{Real emission of the gluon after the shockwave and from the quark}
    \label{fig:R2}
\end{figure}
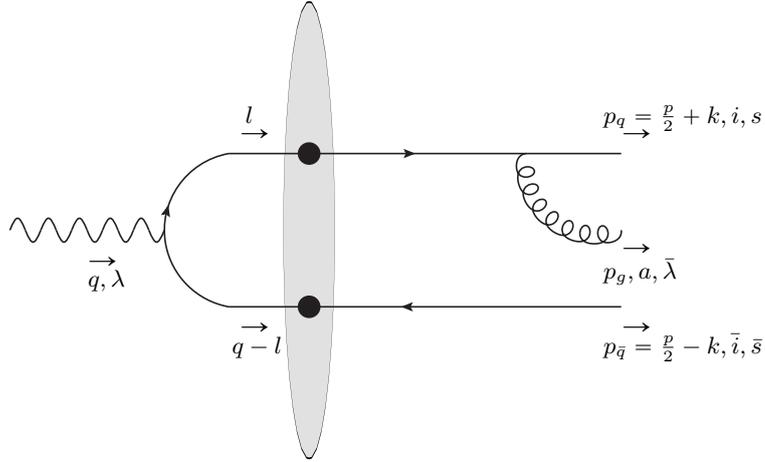
Now we discuss the scattering amplitude for the $R_2$ diagram as shown in Fig.~\ref{fig:R2}. It represents the real emission of the gluon from the quark but after the shockwave. The scattering amplitude is given by 
\begin{align*}
& S_{R_2, s\Bar{s}, i\Bar{i}}^{\lambda, \Bar{\lambda},a}(p,k,p_g) \\
& = \int \frac{d^4 l }{(2\pi)^4}\Bar{u}_s(p_q)\left(ig\slashed{\varepsilon}^*(p_g,\Bar{\lambda})t^a_{in}\right) S_0(p_q + p_g) T^q_{nm}(p_q+p_g,l)S_0(l) (-ie e_Q \slashed{\varepsilon}(q,\lambda)) \\ 
& \times S_0(l-q)T^{\Bar{q}}_{m\Bar{i}}(q-l,p_{\Bar{q}})v_{\Bar{s}}(p_{\Bar{q}}) \; .  \numberthis 
\end{align*}
Inserting the corresponding propagators and vertices, we find
\begin{align*}
& S_{R_2, s\Bar{s}, i\Bar{i}}^{\lambda, \Bar{\lambda},a}(p,k,p_g) \\
& = (2\pi)\delta(p_q^+ + p_g^+ + p_{\Bar{q}}^+-q^+ ) \frac{ee_Q}{2\pi}\int \der^2 \vec{x} \; \der^2 \vec{y} \; e^{-i(\Vec{p}_q + \vec{p}_g)\cdot \Vec{x}} e^{-i \Vec{p}_{\Bar{q}} \cdot \Vec{y}} \left[t^a V(\vec{x} \,) V^\dag(\vec{y} \,)\right]_{i\Bar{i}}\\
& \times ig  \int \frac{d^4 l }{(2\pi)^2} \delta(p_{\bar{q}}^+ - q^+ + l^+)  e^{i \Vec{l}\cdot (\Vec{x}- \Vec{y})} \frac{1}{[(p_q+p_g)^2-m_Q^2+i\varepsilon]}\\
& \times \Bar{u}_s(p_q) \frac{\slashed{\varepsilon}^*(\Bar{\lambda},p_g)(\slashed{p}_q+ \slashed{p}_{3}+m_Q)\gamma^+ (\slashed{l}+m_Q)\slashed{\varepsilon}(\lambda,q)(\slashed{l}-\slashed{q}+m_Q)\gamma^+}{[l^2-m_Q^2+i\varepsilon][(l-q)^2-m_Q^2+i\varepsilon]} v_{\Bar{s}}(p_{\Bar{q}}) \; . \numberthis
\end{align*}
The corresponding reduced amplitude takes the form 
\begin{align*}
    \mathcal{M}_{ R_2, s \Bar{s}, i \Bar{i}}^{\lambda, \Bar{\lambda},a}(p,k,p_g) & = \frac{e e_Q}{2\pi} \int \der^2 \vec{x} \; \der^2 \vec{y} \; e^{- i (\Vec{p}_q + \vec{p}_g) \cdot \Vec{x}} e^{- i \Vec{p}_{\Bar{q}} \cdot \Vec{y}} \mathcal{C}^a_{R_2}(\vec{x} , \vec{y} \,)_{i\Bar{i}} \\ 
    & \times \Bar{u}_s(p_q) \mathcal{N}_{R_2}^{\lambda, \Bar{\lambda}}(p,k,p_g;\vec{r} \,) v_{\bar{s}}(p_{\Bar{q}}) \,,
\end{align*}
with the color structure:
\begin{equation}
\mathcal{C}^a_{R_2}(\vec{x} , \vec{y} \,)  = \left[t^a V(\vec{x} \,)V^\dag(\vec{y} \,) - t^a \mathds{1} \right]  \; ,
\label{eq:color-structure-R2}
\end{equation}
and the perturbative factor: 
\begin{equation} 
\mathcal{N}_{R_2}^{\lambda, \Bar{\lambda}}(p,k,p_g; \vec{r} \,) = i g 2 q^+ \int \frac{d^4 l}{(2\pi)^2}\delta(p_{\bar{q}}^+ - q^+ + l^+) \frac{e^{i \Vec{l} \cdot \Vec{r}} \mathcal{T}_{R_2}^{\lambda, \Bar{\lambda}}(l) }{[l^2 - m_Q^2 + i \varepsilon][(l-q)^2 - m_Q^2 + i \varepsilon]} \,.
\end{equation}
The Dirac structure which has been simplified using Eq.\,\eqref{Dirac_Gluon_Q_after} reads:
\begin{align*} 
\mathcal{T}_{R_2}^{\lambda, \Bar{\lambda}}(l) & = \frac{\Vec{\epsilon}^{\; \Bar{\lambda}*}_n}{[(\vec{p}_g - x \Vec{p}_q)^2 + x^2 m_Q^2 ]} \left\{2(\vec{p}_g- x \Vec{p}_q)_m\left[\delta^{mn}\left(1 + \frac{x}{2}\right) + \frac{x}{2} \omega^{nm}\right] \right. \\
& \left.  + x^2 m_Q \gamma^n \right\} \frac{1}{2q^+ } \gamma^+ (\slashed{l}+m_Q) \slashed{\varepsilon}(\lambda,q) (\slashed{l}- \slashed{q}+m_Q) \gamma^+ \numberthis  \;,
\end{align*}
where $x = x_g/x_q$. 

As in our calculation for the $R_1$ contribution, the integration over $l^+$ is trivial due to the delta function $\delta(p_{\bar{q}}^+ - q^+ + l^+)$, and the integration over $l^-$ is performed using the residue theorem, closing the contour on the upper plane. We find then 
\begin{equation}
\mathcal{N}_{R_2}^{\lambda, \Bar{\lambda} }(p,k,p_g;\vec{r} \,) = - g  \theta(x_{\Bar{q}}) \theta(1-x_{\Bar{q}})\int \frac{\der^2 \Vec{l}}{2\pi} \frac{e^{i \Vec{l} \cdot \Vec{r}}}{\Bar{Q}_{R_2,\xi }^2 + \Vec{l}^{\; 2}}  \mathcal{T}^{\lambda, \bar{\lambda}}_{R_2}(l) \bigg|_{l^+ = q^+ - q^+ x_{\bar{q}} } \,,
\end{equation}
where 
\begin{equation}
    \bar{Q}_{R_2,\xi }^2 = Q^2 (1-x_{\bar{q}}) x_{\bar{q}} + m_Q^2\,.
    \label{eq:QR2}
\end{equation}
The explicit expressions for the perturbative factors are:
\begin{align*}
& \mathcal{N}_{R_2}^{\lambda= 0 , \Bar{\lambda}}(p,k,p_g;\vec{r} \, )  = 2Q (1-x_{\bar{q}}) x_{\Bar{q}} \; g \; \theta(x_{\Bar{q}}) \theta(1-x_{\Bar{q}})  K_0(\Bar{Q}_{R_2,\xi} |\vec{r} \, |) \frac{\Vec{\epsilon}^{\; \Bar{\lambda}*}_n}{[(\vec{p}_g - x \Vec{p}_q)^2 + x^2 m_Q^2 ]}  \\
&  \times \left\{2(\vec{p}_g- x \Vec{p}_q)_m \left[\delta^{mn}\left(1 + \frac{x}{2}\right) + \frac{x}{2} \omega^{nm}\right]  + m_Q x^2  \gamma^n \right\}   \gamma^+ \; ,  \label{eq:N_R2reg_L}\numberthis 
\end{align*}
for longitudinally polarized photons, and 
\begin{align*}
& \mathcal{N}_{R_2}^{\lambda = \pm 1, \Bar{\lambda}} (p,k,p_g; \vec{r} \, )  =  - g \theta(x_{\Bar{q}}) \theta(1-x_{\Bar{q}}) \frac{\Vec{\epsilon}^{\; \Bar{\lambda}*}_n  \Vec{\epsilon}^{\; \lambda}_i}{[(\vec{p}_g - x \Vec{p}_q)^2 + m_Q^2 x^2  ]} \\
& \times \left\{2(\vec{p}_g- x \Vec{p}_q)_m\left[\delta^{mn}\left(1 + \frac{x}{2}\right) + \frac{x}{2} \omega^{nm}\right]   + m_Q x^2  \gamma^n \right\} \nonumber \\
& \times \left\{ \frac{i \Vec{r}_j}{|\vec{r} \,|} \Bar{Q}_{R_2,\xi} K_1(\Bar{Q}_{R_2,\xi}|\vec{r} \,| )\left[\delta^{ij} (2x_{\Bar{q}} - 1) + \omega^{ij}\right]\gamma^+  - m_Q \gamma^i \gamma^+ K_0(\Bar{Q}_{R_2,\xi}|\vec{r} \,| ) \right\} \;, \label{eq:N_R2reg_T} \numberthis 
\end{align*}
for the transversely polarized photon case.

The reduced scattering amplitude, in the notations of Eqs.\,\eqref{ParamNRQCD},\,\eqref{LongFractNot} and \eqref{eq:rb-definition} becomes 
\begin{align*}
\mathcal{M}_{R_2, s \bar{s}, i \Bar{i}} ^{\lambda, \Bar{\lambda},a}(p,k,p_g) & = \frac{e e_Q}{2 \pi} \int \der^2 \vec{b} \; \der^2 \vec{r} \; e^{- i \Vec{p} \cdot \Vec{b}} e^{- i \Vec{k} \cdot \Vec{r}}  e^{- i \vec{p}_g \cdot \left(\Vec{b} + \frac{\Vec{r}}{2}\right)} \mathcal{C}^a_{R_2} \left( \vec{b}  + \frac{\vec{r} }{2}, \vec{b}  - \frac{\vec{r} }{2}\right)_{i \Bar{i}} \\
& \times \Bar{u}_s\left( \frac{p}{2} + k \right) \mathcal{N}_{R_2}^{\lambda, \Bar{\lambda} }(p,k,p_g; \vec{r} \,)v_{\Bar{s}}\left(\frac{p}{2}+k\right) \,,
\end{align*}
with $ z_1 = x_{\bar{q}} = \frac{1}{2}- \frac{x_g}{2} - \xi$, $z_2 = x_g $ and $z_3 = x_q = \frac{1}{2}- \frac{x_g}{2} + \xi $. As in the $R_1$ case, we have checked that Eqs.\,\eqref{eq:N_R2reg_L} and \eqref{eq:N_R2reg_T} match with those presented in \cite{Caucal:2021ent} (before spinor contraction) up to a factor of $1/(2q^+)$ due to our convention in the normalization.

\subsection{Gluon emission from antiquark}

The computations for the diagrams $R_3$ and $R_4$ where the gluon is emitted by the antiquark follow the same steps as those in $R_1$ and $R_2$. We omit the specific details of the derivation and simply present the results.

\paragraph{$R_{3}$ diagram: gluon emission before the shockwave}

\begin{align*}
& \mathcal{M}_{R_{3},s\bar{s}, i \bar{i}}^{\lambda, \Bar{\lambda},a} (p,k,p_g) \\
& = \frac{e e_Q }{2 \pi} \int \der^2 \vec{r} \; \der^2 \vec{b} \; \der^2 \vec{z} \;  \mathcal{C}^a_{R_{3}}\left(\vec{b}  + \frac{\vec{r} }{2}, \vec{b}  - \frac{\vec{r} }{2} ,\vec{z} \, \right)_{i\Bar{i}} \; e^{- i \Vec{p}\cdot \Vec{b}} e^{- i \Vec{k} \cdot \Vec{r}} e^{- i \vec{p}_g  \cdot \Vec{z}} \\
& \times \Bar{u}_s\left(\frac{p}{2} + k \right)\left(\mathcal{N}_{R_{3},reg}^{\lambda,\Bar{\lambda}}(p,k,p_g;\vec{r} ,\vec{b} ,\vec{z} \, ) +\mathcal{N}_{R_{3},inst}^{\lambda, \bar{\lambda}}(p,k,p_g; \vec{r} ,\vec{b} ,\vec{z} \, ) \right) v_{\Bar{s}}\left(\frac{p}{2}- k \right) \;, \numberthis 
\end{align*}
where the color factor is the identical as in the $R_1$ contribution
\begin{equation}
    \mathcal{C}^a_{R_{3}}(\vec{x} , \vec{y} , \vec{z} \, )  = V(\vec{x}  \, ) V^\dag(\vec{z} \,) t^a V(\vec{z} \,)V^\dag(\vec{y} \,) - t^a \mathds{1} \; ,  
    \label{eq:color-structure-R1c}
\end{equation} 
and the perturbative factors are 
\begin{align*}
& \mathcal{N}_{R_{3},reg}^{\lambda = 0, \bar{\lambda}}(p,k,p_g; \vec{r} , \vec{b} ,\vec{z} \, ) = 2 Q    \frac{x_q (1-x_q)}{ x_g} \frac{g}{\pi}     \theta(x_g) \theta\left(x_{\bar{q}} \right)  \theta\left(1-x_q \right) \theta\left(x_q\right)  \Vec{\epsilon}^{\; \bar{\lambda}*}_n  \gamma^+     \\
& \times \left\{   I_{1m}(\Vec{x}- \Vec{y}, \Vec{z}- \Vec{y} \, )  \left[ \left(1 + \frac{x_g  }{2 x_{\bar{q}}} \right) \delta^{nm} + \frac{x_g}{2 x_{\bar{q}}} \omega^{mn} \right] -  m_Q 
I_0(\Vec{x}- \Vec{y}, \Vec{z}- \Vec{y} \, )   \frac{x_g^2}{2(1-x_q)x_{\bar{q}}} \gamma^n \right\} \; , \numberthis \\
& \mathcal{N}_{R_{3},inst}^{\lambda = 0, \bar{\lambda}}(p,k,p_g; \vec{x} , \vec{y} ,\vec{z} \, ) = 0 \; , 
\end{align*}
for longitudinally polarized photons, and
\begin{align*}
& \mathcal{N}_{R_{3},reg}^{\lambda = \pm 1, \bar{\lambda}}(p,k,p_g; \vec{r} , \vec{b} ,\vec{z} \, ) \\
& = - \frac{1}{x_g} \frac{g}{\pi}       \theta(x_g) \theta\left(x_{\bar{q}} \right)  \theta\left(1-x_q\right) \theta\left(x_q \right)   \Vec{\epsilon}^{\; \bar{\lambda}*}_n  \vec{\epsilon}^{\; \lambda }_i \\
& \times \left\{ I_{2mj}(\Vec{x}-\Vec{y},\Vec{z}- \Vec{y} \, ) \gamma^+  \left[ \delta^{ij}(1-2x_q) + \omega^{ij}\right]     \left[\delta^{nm} \left(1 + \frac{x_g}{2 x_{\bar{q}}}\right) + \frac{x_g}{2 x_{\bar{q}}} \omega^{mn} \right] \right.\\
& + m_Q I_{1m}(\Vec{x}-\Vec{y},\Vec{z}- \Vec{y} \, )  \gamma^+  \gamma^i      \left[\delta^{nm} \left(1 +\frac{x_g}{2 x_{\bar{q}}}\right) + \frac{x_g}{2 x_{\bar{q}}} \omega^{mn} \right] \\
& - m_Q \Tilde{I}_{1j}(\Vec{x}-\Vec{y},\Vec{z}- \Vec{y}\,)  \frac{x_g^2}{2(1-x_q) x_{\bar{q}}} \gamma^+  \left[ \delta^{ij}(1-2x_q) + \omega^{ij}\right]  \gamma^n  \\
& \left. -  m_Q^2 \frac{x_g^2}{(1-x_q) x_{\bar{q}}} I_0(\Vec{x}-\Vec{y},\Vec{z}- \Vec{y}\,) \gamma^+  \gamma^i     \gamma^n \right\} \;  , \numberthis \\
&\mathcal{N}_{R_{3},inst}^{\lambda = \pm 1, \bar{\lambda}}(p,k,p_g; \vec{x} , \vec{y} ,\vec{z} \, )  \nonumber \\
& = \frac{x_g x_q x_{\bar{q}}}{2(1-x_q)}  \frac{g}{\pi}  \theta(x_g) 
\theta\left(x_{\bar{q}}\right) \theta\left(x_q\right)  \theta\left(1-x_q\right) \Vec{\epsilon}^{\; \Bar{\lambda}*}_n \Vec{\epsilon}^{\; \lambda}_i \frac{\bar{Q}_{R_{3},\xi}K_{1}(\Bar{Q}_{R_{3},\xi}X_{R_{3},\xi})  }{X_{R_{3},\xi}} \gamma^+ \gamma^i \gamma^n \; , \numberthis 
\end{align*}
for the transversely polarized photon case.

The variables $\bar{Q}_{R_{3}, \xi}^2$ and $X_{R_{3},\xi}^2 $ are identical to those in Eq.\,\eqref{eq:variablesR1}.
In this case, the $I_0, I_{1m},I_{2mj}, \tilde{I}_{1j}$ functions have to be evaluated with $z_1 = x_q, z_2 = x_g, z_3 = x_{\bar{q}}$. 

\paragraph{$R_{4}$ diagram: gluon emission after the shockwave}

\begin{align*}
    \mathcal{M}_{R_{4}, s\Bar{s}, i \Bar{i}}^{\lambda, \Bar{\lambda},a}(p,k,p_g) & = \frac{e e_Q}{2 \pi} \int \der^2 \vec{r} \; \der^2 \vec{b} \; e^{- i \Vec{p} \cdot \Vec{b}} e^{- i \vec{p}_g \cdot \left( \Vec{b}- \frac{\Vec{r}}{2} \,\right)} e^{- i \Vec{k} \cdot \Vec{r}} \mathcal{C}^a_{R_{4}}\left(\vec{b}  + \frac{\vec{r} }{2}, \vec{b}  - \frac{\vec{r} }{2} \, \right)_{i \Bar{i}} \\*
    & \times \Bar{u}_s\left(\frac{p}{2}+ k\right) \mathcal{N}_{R_{4}}^{\lambda, \Bar{\lambda}}(p,k,p_g;\vec{r} \,) v_{\bar{s}}\left(\frac{p}{2}- k \right)\,, 
\end{align*}
where the color factor is
\begin{equation}
\mathcal{C}^a_{R_{4}}(\vec{x} , \vec{y} \,)  = \left[ V(\vec{x} \,)V^\dag(\vec{y} \,)t^a - t^a \mathds{1} \right] \label{eq:color-structure-R2c} \,,
\end{equation}
and the perturbative factors are
\begin{align*}
    & \mathcal{N}_{R_{4}}^{\lambda = 0, \Bar{\lambda}}(p,k,p_g; \vec{r} \,)= -2 Q (1 -x_q) x_q \; g \; \theta(x_q) \theta\left(1-x_q\right) K_{0}(\Bar{Q}_{R_{4},\xi}|\vec{r} \,|) \frac{\Vec{\epsilon}^{\; \Bar{\lambda} *}_n}{\left[\left(\vec{p}_g - \bar{x} \vec{p}_{\bar{q}}\right)^2 + \bar{x}^2 m_Q^2 \right]}\\  
   & \times  \gamma^+ \left\{ 2 \left(\vec{p}_g - \bar{x} \vec{p}_{\bar{q}} \right)_m  \left[ \delta^{mn} \left( 1 + \frac{\bar{x}}{2}\right) - \frac{\bar{x}}{2} \omega^{nm}  \right] - m_Q  \bar{x}^2 \gamma^n     \right\}  \,,\numberthis 
\end{align*}
for longitudinally polarized photons, and 
\begin{align*}
    &\mathcal{N}_{R_{4}}^{\lambda = \pm 1, \Bar{\lambda}}(p,k,p_g; \vec{r} \,) = g \theta(x_q) \theta\left( 1-x_q \right)  \frac{ \Vec{\epsilon}^{\; \Bar{\lambda} *}_n \Vec{\epsilon}^{\; \lambda}_i }{\left[\left(\vec{p}_g - \bar{x} \vec{p}_{\bar{q}} \right)^2 + \bar{x}^2 m_Q^2 \right]} \\
    &  \times \left\{ - \frac{i \Vec{r}_j}{|\vec{r} \,|} \Bar{Q}_{R_{4},\xi}K_{1}(\Bar{Q}_{R_{4},\xi}|\vec{r} \,|)  \left[ \delta^{ij}(2x_q-1) - \omega^{ij} \right] \gamma^+  -   m_Q K_{0}(\Bar{Q}_{R_{4},\xi}|\vec{r} \,|) \gamma^i \gamma^+    \right\} \\
    & \times \left\{2 \left(\vec{p}_g - \bar{x} \vec{p}_{\bar{q}} \right)_m  \left[ \delta^{mn} \left( 1 + \frac{\bar{x}}{2}\right) - \frac{\bar{x}}{2} \omega^{nm}  \right] - m_Q  \bar{x}^2 \gamma^n  \right\} \numberthis \,,
\end{align*}
for transversely polarized photons. Here we defined the variables
\begin{align*}
   \bar{x} & = \frac{x_g}{x_{\bar{q}}} \\ 
    \Bar{Q}_{R_{4},\xi}^2  & = x_q (1-x_q) Q^2 + m_Q^2 \numberthis[VariablesR2'] \; . 
\end{align*}

\section{Short-distance coefficients for $Q\bar{Q}[\kappa]g$ amplitude in CGC + NRQCD}
\label{sec:projectionQQbar}

Having computed the amplitudes for $Q\bar{Q}g$ production in the previous section, we are now in a position to compute the short-distance coefficients by proper projection to the quantum state $\kappa$.

\subsection{Gluon emission from quark before the shockwave}

We project the amplitude $R_1$ in Eq.\,\eqref{AmplitudeR1QQbar} using Eq.\,\eqref{AmpProjected}, we find 
 \begin{align*}
\mathcal{M}_{R_1}^{\lambda, \Bar{\lambda}, \kappa, J_z}(p,p_g) & = \frac{ee_Q}{2 \pi} \int \der^2 \vec{r}  \; \der^2 \vec{b}  \; \der^2 \vec{z} \;    \Tr\left[\mathcal{C}_{R_1}\left(\vec{b} + \frac{\vec{r} }{2}, \vec{b}  - \frac{\vec{r} }{2}, \vec{z} \, \right) C^{[c]} \right] \; e^{- i \Vec{p} \cdot \Vec{b} }  e^{- i \vec{p}_g \cdot \Vec{z}} \\
& \times   \left( \mathcal{F}_{R_1, reg}^{\lambda, \Bar{\lambda}, \kappa, J_z}(p,p_g,Q; \vec{r} , \vec{b} , \vec{z} \, ) + \mathcal{F}_{R_1, inst}^{\lambda, \Bar{\lambda}, \kappa, J_z}(p,p_g,Q; \vec{r} , \vec{b} , \vec{z} \, )  \right) \,, \numberthis 
 \end{align*}
where we defined the projected perturbative factor:
\begin{align*}
\mathcal{F}^{\lambda,\Bar{\lambda}, \kappa, J_z}_{R_1, reg } (p,p_g,Q; \vec{r} ,\vec{b} , \vec{z} \,) & = \Tr\left[ \Pi^{J J_z}(p,0)\mathcal{N}^{\lambda, \Bar{\lambda}}_{R_1, reg }(p,0,p_g;\vec{r} ,\vec{b} ,\vec{z} \,)\right] \numberthis 
\end{align*}
with a similar expression for the instantaneous counterpart.  
 
First, we evaluate the perturbative factors in Eqs.\,\eqref{eq:N_R1reg_L},\,\eqref{eq:N_R1reg_T} and \eqref{eq:N_R1ins_T} at $k=0$ (corresponding to the $S$-wave), we find
\begin{align*}
& \mathcal{N}_{R_1, reg}^{\lambda = 0 , \Bar{\lambda}}(p,0,p_g; \vec{x} , \vec{y} , \vec{z} \, ) \\*
& = - Q  \frac{ (1+ x_g )}{x_g }  \frac{g}{2\pi}     \Vec{\epsilon}^{\; \Bar{\lambda}*}_n   \gamma^+ \left\{   I_{1m}(\Vec{y}-\Vec{x}, \Vec{z}-\Vec{x}) \left[ \delta^{nm} - x_g\omega^{mn}\right]   - m_Q I_0(\Vec{y}-\Vec{x}, \Vec{z}- \Vec{x}) \frac{2 x_g^2}{1 + x_g } \gamma^n    \right\} \; , \numberthis \\
\end{align*}
\begin{align*}
& \mathcal{N}_{R_1, reg}^{\lambda = \pm 1, \Bar{\lambda}}(p,0,p_g; \vec{x} , \vec{y} , \vec{z} \, )  =  \frac{1}{ x_g (1 -x_g) }  \frac{g}{\pi}   \Vec{\epsilon}^{\; \Bar{\lambda}*}_n \Vec{\epsilon}^{\; \lambda}_i \\ 
&  \times \Bigg\{ I_{2mj}(\Vec{y}- \Vec{x}, \Vec{z}- \vec{x}) \left[ \delta^{nm} + x_g  \omega^{nm}\right]\left[ \delta^{ij} x_g  - \omega^{ij}\right] \gamma^+  - m_Q  I_{1m}(\Vec{y}- \Vec{x}, \Vec{z}- \Vec{x}) \left[\delta^{nm} + x_g  \omega^{nm}\right]  \gamma^i\gamma^+ \nonumber \\
& \left. +  m_Q \Tilde{I}_{1j}(\Vec{y}- \Vec{x}, \Vec{z}- \Vec{x})  \frac{2 x_g^2 }{(1 + x_g )} \gamma^n  \left[ \delta^{ij} x_g - \omega^{ij}\right] \gamma^+  -  m_Q^2  I_0(\Vec{y}- \Vec{x}, \Vec{z}- \Vec{x} ) \frac{2 x_g^2 }{(1 + x_g )} \gamma^n  \gamma^i\gamma^+  \right\} \; ,  \numberthis \\
\end{align*}
\begin{align*}
&\mathcal{N}_{R_1, inst}^{\lambda = \pm 1 , \Bar{\lambda}}(p,0,p_g; \vec{x} , \vec{y} , \vec{z} \, )   =  - \frac{ x_g(1 - x_g )^2 }{2 (1 + x_g )}    \frac{g}{2\pi}  \frac{\bar{Q}_{R_1} K_1(\Bar{Q}_{R_1}X_{R_1}) }{X_{R_1}}  \Vec{\epsilon}^{\; \Bar{\lambda}*}_n \Vec{\epsilon}^{\; \lambda}_i  \gamma^n  \gamma^i \gamma^+ \; , \numberthis
\end{align*}
with 
\begin{align*}
 \bar{Q}_{R_1, \xi = 0}^2 = \Bar{Q}_{R_1}^2 & = Q^2 + m_Q^2 \frac{4}{(1 - x_g  )}  \\
 X_{R_1, \xi = 0}^2  = X_{R_1}^2  & = x_g \left(\frac{1}{2}- \frac{x_g}{2}  \right) \left(\Vec{z} - \Vec{b}- \frac{\Vec{r}}{2}\right)^2 + \left( \frac{1}{2}- \frac{x_g}{2} \right)^2 \Vec{r}^{\; 2} \\
 & +  \left( \frac{1}{2}- \frac{x_g}{2} \right)   x_g \left(\Vec{b}-\frac{\Vec{r}}{2}- \Vec{z}\right)^2 \; . \numberthis[VariablesR1At0] 
\end{align*}
To compute $\mathcal{F}^{\lambda, \Bar{\lambda}, \kappa, J_z}$, we use the projectors in  Eqs.\,\eqref{eq:ProjectorSpin0},\,\eqref{eq:ProjectorSpin1}, and perform the traces over the gamma matrices with the help of the identities in appendix \ref{sec:GammaTrace}. We find 
\begin{align*}
& \mathcal{F}^{\lambda= 0,\Bar{\lambda}, {}^1S_0^{[c]}, J_z}_{R_1, reg} (p,p_g,Q; \vec{r} ,\vec{b} , \vec{z} \,) \\
& =  - i\frac{1}{\sqrt{2m_Q}} Q  (1+ x_g )  \frac{gp^+}{\pi}     \Vec{\epsilon}^{\; \Bar{\lambda}*}_n       \epsilon^{ nm} I_{1m}(\Vec{y}-\Vec{x}, \Vec{z}-\Vec{x})   \,, \numberthis \\
& \mathcal{F}^{\lambda= 0,\Bar{\lambda}, {}^3S_1^{[c]}, J_z}_{R_1, reg} (p,p_g,Q; \vec{r} ,\vec{b} , \vec{z} \,) \\
& =  \varepsilon_\rho^*(J_z) \sqrt{2 m_Q} Q  \frac{ (1+ x_g )}{x_g }  \frac{g}{\pi} \Vec{\epsilon}^{\; \Bar{\lambda}*}_n  \left\{  g^{\rho + } I_{1}^n (\Vec{y}-\Vec{x}, \Vec{z}-\Vec{x})  +   \frac{ x_g^2}{1 + x_g }  I_0(\Vec{y}-\Vec{x}, \Vec{z}- \Vec{x}) \right.  \\
& \left. \times  (  g^{\rho + } \Vec{p}^{\; n} - g^{\rho n} p^+ )  \right\} \,, \numberthis 
\end{align*}
for the photon longitudinal polarization, and 
\begin{align*}
& \mathcal{F}^{\lambda= \pm 1,\Bar{\lambda}, {}^1S_0^{[c]}, J_z}_{R_1, inst} (p,p_g,Q; \vec{r} ,\vec{b} , \vec{z} \,) \\
& = -  i  \frac{1}{\sqrt{8m_Q}} \frac{ x_g(1 - x_g )^2 }{ (1 + x_g )}    \frac{gp^+}{\pi}  \frac{\Bar{Q}_{R_1} K_1(\Bar{Q}_{R_1}X_{R_1}) }{X_{R_1}}  \Vec{\epsilon}^{\; \Bar{\lambda}*}_n \Vec{\epsilon}^{\; \lambda}_i   \epsilon^{n i  } \,, \numberthis \\
& \mathcal{F}^{\lambda= \pm 1,\Bar{\lambda}, {}^3S_1^{[c]}, J_z}_{R_1, inst} (p,p_g,Q; \vec{r} ,\vec{b} , \vec{z} \,)  \\
& = - \varepsilon_\rho^*(J_z) \sqrt{\frac{m_Q}{2}} \frac{ x_g(1 - x_g )^2 }{ (1 + x_g )}    \frac{g}{\pi}  \frac{\Bar{Q}_{R_1} K_1(\Bar{Q}_{R_1}X_{R_1}) }{X_{R_1}}  \Vec{\epsilon}^{\; \Bar{\lambda}*}_n \Vec{\epsilon}^{\; \lambda}_i   g^{\rho +} \delta^{ni}  \,,   \numberthis \\
& \mathcal{F}^{\lambda= \pm 1,\Bar{\lambda}, {}^1S_0^{[c]}, J_z}_{R_1, reg} (p,p_g,Q; \vec{r} ,\vec{b} , \vec{z} \,) \\
& =  -i  \sqrt{\frac{2}{m_Q}}    \frac{1}{ x_g (1 -x_g) }  \frac{g p^+ }{\pi}   \Vec{\epsilon}^{\; \Bar{\lambda}*}_n \Vec{\epsilon}^{\; \lambda}_i  \left\{ - I_{2mj}(\Vec{y}- \Vec{x}, \Vec{z}- \vec{x})  \left[ - \delta^{nm}   \epsilon^{ i j } + x_g^2 \delta^{ij} \epsilon^{ nm  } \right]  \right. \\
& \left. +  \frac{2 x_g^2 }{(1 + x_g )} m_Q^2  I_0(\Vec{y}- \Vec{x}, \Vec{z}- \Vec{x} )   \epsilon^{ni} \right\}  \,, \numberthis \\
& \mathcal{F}^{\lambda= \pm 1,\Bar{\lambda}, {}^3S_1^{[c]}, J_z}_{R_1, reg} (p,p_g,Q; \vec{r} ,\vec{b} , \vec{z} \,) \\
& = \varepsilon^*_\rho(J_z) \sqrt{2m_Q}   \frac{1}{ x_g (1 -x_g) }  \frac{g}{\pi}   \Vec{\epsilon}^{\; \Bar{\lambda}*}_n \Vec{\epsilon}^{\; \lambda}_i  \left\{ - 2 x_g I_{2mj}(\Vec{y}- \Vec{x}, \Vec{z}- \vec{x})  g^{\rho +} \left[\delta^{nm}   \delta^{ij}  +  \epsilon^{ij} \epsilon^{nm} \right]   \right. \\
&   +   I_{1m}(\Vec{y}- \Vec{x}, \Vec{z}- \Vec{x})  \left[ \delta^{nm}  (  g^{\rho + } \Vec{p}^{\; i} - g^{\rho i} p^+ )   -  x_g   \delta^{im}  (   g^{\rho +}  \Vec{p}^{\; n}- p^+  g^{n \rho}) + x_g   \delta^{in}  ( g^{\rho +} \Vec{p}^{\; m}  - p^+  g^{m \rho}     ) \right]   \\
&   - \frac{2 x_g^2 }{(1 + x_g )}  \Tilde{I}_{1j}(\Vec{y}- \Vec{x}, \Vec{z}- \Vec{x})  \left[ \delta^{ij} x_g  ( g^{\rho +} \Vec{p}^{\; n} - g^{\rho n} p^+ ) - \delta^{jn}  (  g^{\rho +} \Vec{p}^{\; i} -  p^+  g^{i \rho } ) + \delta^{in} ( g^{\rho +} \Vec{p}^{\; j}  - p^+  g^{j \rho}    ) \right]  \\ 
& \left.      -  \frac{4 x_g^2 }{(1 + x_g )} m_Q^2  I_0(\Vec{y}- \Vec{x}, \Vec{z}- \Vec{x} )   g^{\rho + } \delta^{ni}   \right\} \,, \numberthis 
\end{align*}
for the transversely polarized photon. 
The integrals $I_0, I_{1m},I_{2mj}, \tilde{I}_{1j}$ (defined in appendix \ref{sec:R1TransverseIntegral}) are evaluated at $ z_1 = \frac{1}{2}- \frac{x_g}{2}$, $z_2 = x_g $ and $z_3= \frac{1}{2}- \frac{x_g}{2}$.

\subsection{Gluon emission from quark after the shockwave}

The projected reduced amplitude reads
\begin{align*}
    \mathcal{M}_{R_2}^{\lambda, \Bar{\lambda}, \kappa, J_z}(p,p_g) & = \frac{e e_Q }{2\pi} \int \der^2 \vec{r} \; \mathcal{F}_{R_2}^{\lambda, \Bar{\lambda}, \kappa, J_z}(p,p_g,Q;\vec{r} \,) \int \der^2 \vec{b} \;   e^{- i \Vec{p} \cdot \Vec{b}} e^{- i \vec{p}_g \cdot \left(\Vec{b} + \frac{\Vec{r}}{2}\right)} \\
    & \times \Tr\left[ \mathcal{C}_{R_2}\left(\vec{b}  + \frac{\vec{r} }{2}; \vec{b}  - \frac{\vec{r}  }{2}\right) C^{[c]} \right]  \,,\numberthis 
\end{align*}
with 
\begin{equation}
\mathcal{F}_{R_2}^{\lambda, \Bar{\lambda}, \kappa, J_z}(p,p_g,Q;\vec{r} \,)  = 
\Tr\left[\Pi^{J J_z}(p,0) \mathcal{N}_{R_2}^{\lambda, \Bar{\lambda}}(p,0,p_g;\vec{r} \,)\right] \\ 
\end{equation}

The computations for $\mathcal{F}_{R_2}^{\lambda, \bar{\lambda}, \kappa, J_z}$ are similar to the ones for $\mathcal{F}_{R_1}$ We will just give the results below.
\begin{align*}
& \mathcal{F}_{R_2}^{\lambda=0,\Bar{\lambda}, {}^1S_0^{[c]}, J_z}(p,p_g,Q; \vec{r} \, )\\
& = i \sqrt{\frac{2}{m_Q}}  g p^+ x_g (1+x_g)   Q K_0(\Bar{Q}_{R_2}|\vec{r} \, | )  \frac{\Vec{\epsilon}^{\; \Bar{\lambda*}}_n}{\left[\left(\vec{p}_g - x \frac{\Vec{p}}{2}\right)^2 + x^2 m_Q^2 \right]} \left(\vec{p}_g - x \frac{\Vec{p}}{2}\right)_m \epsilon^{nm}  \numberthis \\
&   \mathcal{F}_{R_2}^{\lambda=0,\Bar{\lambda}, {}^3S_1^{[c]}, J_z}(p,p_g,Q; \vec{r} \, )\\*
& =   - \varepsilon^*_\rho(J_z) \sqrt{\frac{m_Q}{2}}g(1 + x_g)(1-x_g) Q K_0(\Bar{Q}_{R_2}|\vec{r} \, | )  \frac{\Vec{\epsilon}^{\; \Bar{\lambda*}}_n}{\left[\left(\vec{p}_g - x \frac{\Vec{p}}{2}\right)^2 + x^2 m_Q^2 \right]}  \\ 
& \times  \left\{ 4 \left(\vec{p}_g - x \frac{  \Vec{p}}{2}\right)_m  \delta^{mn} \left(1 + \frac{x}{2}\right)    g^{\rho + }  + x^2   (g^{\rho+} \Vec{p}^{\; n}- g^{\rho n}p^+  )\right\}  \,, \numberthis \\
\end{align*}
for longitudinally polarized photons, and 
\begin{align*}
&   \mathcal{F}_{R_2}^{\lambda=\pm 1,\Bar{\lambda}, {}^1S_0^{[c]}, J_z}(p,p_g,Q; \vec{r} \, )\\
& = \sqrt{\frac{2}{m_Q}} i gp^+ \frac{\Vec{\epsilon}^{\; \Bar{\lambda}*}_n \Vec{\epsilon}_i^{\; \lambda}}{\left[\left(\vec{p}_g - x\frac{\Vec{p}}{2} \right)^2 + x^2 m_Q^2 \right]}  \left\{2i \Bar{Q}_{R_2} K_1(\Bar{Q}_{R_2}|\vec{r} \, |)  \left(\vec{p}_g - x \frac{\Vec{p}}{2}\right)_m   \frac{ \Vec{r}_j}{|\vec{r} \, | } \right. \\
& \left. \times \left[  \left(1 + \frac{x}{2}\right) \delta^{mn}  \epsilon^{ji}  + \frac{x}{2} x_g  \delta^{ij} \epsilon^{nm}\right]  + x^2 m_Q^2 K_0(\Bar{Q}_{R_2}|\vec{r} \, |) \epsilon^{ni} \right\} \,,  \numberthis \\
&   \mathcal{F}_{R_2}^{\lambda=\pm 1,\Bar{\lambda}, {}^3S_1^{[c]}, J_z}(p,p_g,Q; \vec{r} \, )\\
&= -\varepsilon_\rho^*(J_z) \sqrt{2m_Q}   g \frac{\Vec{\epsilon}^{\; \Bar{\lambda}*}_n \Vec{\epsilon}_i^{\; \lambda}}{\left[\left(\vec{p}_g - x\frac{\Vec{p}}{2} \right)^2 + x^2 m_Q^2 \right]}  \left\{4 i \left(\vec{p}_g - x \frac{\Vec{p}}{2}\right)_m   \frac{ \Vec{r}_j}{|\vec{r} \, | } \Bar{Q}_{R_2} K_1(\Bar{Q}_{R_2}|\vec{r} \, |) \right. \\
& \times  g^{\rho + } \left[  \left(1 + \frac{x}{2}\right)x_g  \delta^{ij} \delta^{mn} + \frac{x}{2} \epsilon^{ij}\epsilon^{nm}  \right] + 2 \left(\vec{p}_g - x \frac{\Vec{p}}{2}\right)_m   K_0(\Bar{Q}_{R_2}|\vec{r} \, |) \\
&  \times \left[\delta^{mn}  \left(1 + \frac{x}{2}\right)  \left(\Vec{p}^{\; i} g^{\rho + } -p^+ g^{\rho i}\right)  - \frac{x}{2} \delta^{im}  \left(\Vec{p}^{\; n} g^{\rho + } -p^+ g^{\rho n}\right)  + \frac{x}{2} \delta^{in}  \left(\Vec{p}^{\; m} g^{\rho + } -p^+ g^{\rho m}\right)   \right] \\
& + i  x^2 \frac{ \Vec{r}_j}{|\vec{r} \, | } \Bar{Q}_{R_2} K_1(\Bar{Q}_{R_2}|\vec{r} \, |)   \left[ \delta^{ij} x_g \left(  \Vec{p}^{\; n} g^{\rho + }- g^{\rho n} p^+\right) -  \delta^{jn} \left( \Vec{p}^{\; i} g^{\rho + } - p^+ g^{i\rho} \right) + \delta^{in} \left( \Vec{p}^{\; j} g^{\rho + }  - p^+ g^{j\rho}\right) \right]   \\
&  -2  x^2 m_Q^2 K_0(\Bar{Q}_{R_2}|\vec{r} \, |)  g^{\rho + } \delta^{in} \Bigg\} \;,  \numberthis 
\end{align*}
for transversely polarized photons, where 
\begin{align}
    x &= \frac{2 x_g}{(1-x_g)} \,, \nonumber \\
    \Bar{Q}_{R_2}^2  &= \Bar{Q}_{R_2, \xi = 0}^2  = Q^2 \frac{(1-x_g)(1 + x_g)}{4} + m_Q^2 \,. \label{eq:R2variables}
\end{align}

\subsection{Gluon emission from antiquark}
\paragraph{$R_{3}$ diagram: gluon emission before the shockwave}

\begin{align*}
& \mathcal{F}_{R_{3},reg}^{\lambda= 0, \bar{\lambda}, {}^1S_0^{[8]}, J_z}(p,p_3;\vec{r} , \vec{b} , \vec{z} \, ) \\
& = - i \frac{1}{\sqrt{2m_Q}}  Q    (1+ x_g ) \frac{g p^+ }{\pi} \Vec{\epsilon}^{\; \bar{\lambda}*}_n  I_{1m}(\Vec{x}- \Vec{y}, \Vec{z}- \Vec{y} \, ) \epsilon^{nm} \,, \numberthis \\
& \mathcal{F}_{R_{3},reg}^{\lambda= 0, \bar{\lambda}, {}^3S_1^{[c]}, J_z}(p,p_3;\vec{r} , \vec{b} , \vec{z} \, ) \\
& = - \varepsilon_\rho^*(J_z) \sqrt{2m_Q}   Q    \frac{1+ x_g }{ x_g} \frac{g}{\pi} \Vec{\epsilon}^{\; \bar{\lambda}*}_n   \left\{  g^{\rho +}   I_{1}^n(\Vec{x}- \Vec{y}, \Vec{z}- \Vec{y} \, )  +   \frac{x_g^2}{1 + x_g } I_0(\Vec{x}- \Vec{y}, \Vec{z}- \Vec{y} \, )     \right. \\
& \left. \times \left(  g^{\rho + } \Vec{p}^{\; n}  - g^{\rho n} p^+ \right)\right\} \,, \numberthis 
\end{align*}
for photon longitudinal polarization, and 
\begin{align*}
& \mathcal{F}_{R_{3},inst}^{\lambda= \pm 1, \bar{\lambda}, {}^1S_0^{[8]}, J_z}(p,p_3;\vec{r} , \vec{b} , \vec{z} \, ) \\
& = - i  \frac{1}{\sqrt{8m_Q}}\frac{x_g (1- x_g  )^2}{(1 + x_g )}  \frac{gp^+}{\pi}  \left(\frac{\bar{Q}_{R_{3}}K_{1}(\Bar{Q}_{R_{3}}X_{R_{3}})  }{X_{R_{3}}}\right) \Vec{\epsilon}^{\; \Bar{\lambda}*}_n \Vec{\epsilon}^{\; \lambda}_i   \epsilon^{n i} \,, \numberthis \\ 
& \mathcal{F}_{R_{3},inst}^{\lambda= \pm 1, \bar{\lambda}, {}^3S_1^{[c]}, J_z}(p,p_3;\vec{r} , \vec{b} , \vec{z} \, ) \\
& =   \varepsilon_\rho^*(J_z) \sqrt{\frac{m_Q}{2}} \frac{x_g (1- x_g  )^2}{(1 + x_g )}  \frac{g}{\pi}  \left(\frac{\bar{Q}_{R_{3}}K_{1}(\Bar{Q}_{R_{3}}X_{R_{3}})  }{X_{R_{3}}}\right) \Vec{\epsilon}^{\; \Bar{\lambda}*}_n \Vec{\epsilon}^{\; \lambda}_i  g^{\rho + }\delta^{ in} \,,  \numberthis \\
& \mathcal{F}_{R_{3},reg}^{\lambda= \pm 1, \bar{\lambda}, {}^1S_0^{[8]}, J_z}(p,p_3;\vec{r} , \vec{b} , \vec{z} \, ) \\
& = -i \sqrt{\frac{2}{m_Q}} \frac{1}{x_g (1 - x_g ) } \frac{g p^+}{\pi}    \Vec{\epsilon}^{\; \bar{\lambda}*}_n  \vec{\epsilon}^{\; \lambda }_i  \left\{ I_{2mj}(\Vec{x}-\Vec{y},\Vec{z}- \Vec{y} \, ) \left[ - x_g^2  \delta^{ij} \epsilon^{mn} + \delta^{nm}  \epsilon^{i j }\right]  \right. \\
& \left. -  m_Q^2 \frac{2 x_g^2}{1+ x_g } I_0(\Vec{x}-\Vec{y},\Vec{z}- \Vec{y} \, )\epsilon^{i n } \right\} \,, \numberthis  \\ 
& \mathcal{F}_{R_{3},reg}^{\lambda= \pm 1, \bar{\lambda}, {}^3S_0=1^{[c]}, J_z}(p,p_3;\vec{r} , \vec{b} , \vec{z} \, ) \\
& = \varepsilon_\rho^*(J_z) \sqrt{2 m_Q} \frac{1}{x_g (1 - x_g ) } \frac{g}{\pi} \Vec{\epsilon}^{\; \bar{\lambda}*}_n  \vec{\epsilon}^{\; \lambda }_i \left\{ 2  x_g  I_{2mj}(\Vec{x}-\Vec{y},\Vec{z}- \Vec{y} \, )    g^{\rho +} \left[ \delta^{ij} \delta^{nm} + \epsilon^{ij}\epsilon^{nm}\right] \right.\\
& + \frac{2 x_g^2}{1+ x_g } \Tilde{I}_{1j}(\Vec{x}-\Vec{y},\Vec{z}- \Vec{y} \,  ) \left[  x_g  \delta^{ij}  ( g^{\rho + }\Vec{p}^{\; n}  - g^{\rho n} p^+  )     -   \delta^{jn} \left(  g^{\rho + } \Vec{p}^{\; i}   -   p^+ g^{i \rho } \right)  +     \delta^{in} \left( g^{\rho + }\Vec{p}^{\; j} -  p^+ g^{j \rho }  \right) \right]\\
& - I_{1m}(\Vec{x}-\Vec{y},\Vec{z}- \Vec{y} \, )\left[    \delta^{nm}  (g^{\rho + }\Vec{p}^{\; i}- g^{\rho i } p^+ )  -  x_g  \delta^{im} \left( g^{\rho + }\Vec{p}^{\; n} - p^+  g^{n\rho}  \right) +  x_g \delta^{in} \left(  g^{\rho + }  \vec{p}^{\; m} -p^+  g^{m \rho }  \right) \right] \\
& \left. +   m_Q^2 \frac{4 x_g^2}{1+ x_g } I_0(\Vec{x}-\Vec{y},\Vec{z}- \Vec{y} \, ) g^{ \rho + } \delta^{in} \right\} \,,\numberthis
\end{align*}
for transversely polarized photons. The variables $\bar{Q}_{R_{3}}^2$ and $X_{R_{3}}^2 $ are identical to those in Eq.\,\eqref{VariablesR1At0}. The integrals $I_0, I_{1m},I_{2mj}, \tilde{I}_{1j}$ (defined in appendix \ref{sec:R1TransverseIntegral}) are evaluated at $z_1 = \frac{1}{2}- \frac{x_g}{2}$, $z_2 = x_g $ and $z_3= \frac{1}{2}- \frac{x_g}{2}$.

\paragraph{$R_{4}$ diagram: gluon emission after the shockwave}

\begin{align*}
& \mathcal{F}_{R_{4}}^{\lambda = 0 , \Bar{\lambda}, {}^{1}S_0^{[8]}, J_z}(p,p_3;\vec{r} \,) \\
& =  i  \sqrt{\frac{2}{m_Q}}g p^+  x_g(1 + x_g )  QK_{0}(\Bar{Q}_{R_{4}}|\vec{r} \,|) \frac{\Vec{\epsilon}^{\; \Bar{\lambda} *}_n}{\left[\left(\Vec{p}_3 - \bar{x} \frac{\vec{p}}{2} \right)^2 + \bar{x}^2 m_Q^2 \right]}  \left(\Vec{p}_3 - \bar{x} \frac{\vec{p}}{2}  \right)_m  \epsilon^{nm} \numberthis \\
& \mathcal{F}_{R_{4}}^{\lambda = 0 , \Bar{\lambda}, {}^{3}S_1^{[c]}, J_z}(p,p_3;\vec{r} \,) \\
& =  \varepsilon^*_\rho(J_z) \sqrt{\frac{m_Q}{2}} g (1 - x_g ) (1 + x_g )  QK_{0}(\Bar{Q}_{R_{4}}|\vec{r} \,|) \frac{\Vec{\epsilon}^{\; \Bar{\lambda} *}_n}{\left[\left(\Vec{p}_3 - \bar{x} \frac{\vec{p}}{2} \right)^2 + \bar{x}^2 m_Q^2 \right]} \\    
&  \times \left\{   4 \left(\Vec{p}_3 - \bar{x} \frac{\vec{p}}{2}  \right)_m  \delta^{mn} \left( 1 + \frac{\bar{x}}{2}\right)    g^{\rho +}  +  \bar{x}^2 \left(  g^{\rho + }  \Vec{p}^{\; n} -g^{\rho n} p^+   \right) \right\} \,, \numberthis \\
\end{align*}
for longitudinally polarized photons, and
\begin{align*}
& \mathcal{F}_{R_{4}}^{\lambda = \pm 1 , \Bar{\lambda}, {}^{1}S_0^{[8]}, J_z}(p,p_3;\vec{r} \,) \\ 
& = i \sqrt{\frac{2}{m_Q}} gp^+  \frac{ \Vec{\epsilon}^{\; \Bar{\lambda} *}_n \Vec{\epsilon}^{\; \lambda}_i }{\left[\left(\Vec{p}_3 - \bar{x} \frac{\vec{p}}{2}  \right)^2 + \bar{x}^2 m_Q^2 \right]}   \left\{ - 2 i \Bar{Q}_{R_{4}}K_{1}(\Bar{Q}_{R_{4}}|\vec{r} \,|) \frac{ \Vec{r}_j}{|\vec{r} \,|}  \left(\Vec{p}_3 - \bar{x} \frac{\vec{p}}{2}  \right)_m \right.  \\
& \left.  \times \left[\frac{\bar{x}}{2} x_g \delta^{ij} \epsilon^{nm} +  \left(1 + \frac{\bar{x}}{2}\right) \delta^{mn} \epsilon^{ji} \right]\ + \bar{x}^2 m_Q^2 K_{0}(\Bar{Q}_{R_{4}}|\vec{r} \,|)  \epsilon^{ni} \right\} \,,\numberthis \\
& \mathcal{F}_{R_{4}}^{\lambda=\pm 1,\Bar{\lambda}, {}^3S_1^{[c]}, J_z}(p,p_3; \vec{r} \,)\\ 
& =  \varepsilon_\rho^*(J_z) \sqrt{2m_Q} g  \frac{ \Vec{\epsilon}^{\; \Bar{\lambda} *}_n \Vec{\epsilon}^{\; \lambda}_i }{\left[\left(\Vec{p}_3 - \bar{x} \frac{\vec{p}}{2}  \right)^2 + \bar{x}^2 m_Q^2 \right]} \left\{ - 4 i \Bar{Q}_{R_{4}}K_{1}(\Bar{Q}_{R_{4}}|\vec{r} \,|) \frac{ \Vec{r}_j}{|\vec{r} \,|}  \left(\Vec{p}_3 - \bar{x} \frac{\vec{p}}{2}  \right)_m  g^{\rho + }  \right.\\
&  \times  \left[ x_g \left( 1 + \frac{\bar{x}}{2}\right)  \delta^{ij} \delta^{mn} + \frac{\bar{x}}{2}\epsilon^{ij} \epsilon^{nm}  \right]  - i \bar{x}^2 \Bar{Q}_{R_{4}}K_{1}(\Bar{Q}_{R_{4}}|\vec{r} \,|) \frac{ \Vec{r}_j}{|\vec{r} \,|}    \\
& \times \left[ \delta^{ij} x_g ( g^{\rho + } \Vec{p}^{\; n} - g^{\rho n} p^+ ) -  \delta^{jn} \left( \Vec{p}^{\; i} g^{\rho + } - p^+ g^{i \rho}   \right) +  \delta^{in} \left(  g^{\rho +} \Vec{p}^{\; j} - p^+ g^{j \rho} \right) \right] \\
& - 2 \bar{x}^2 m_Q^2 K_{0}(\Bar{Q}_{R_{4}}|\vec{r} \,|)  g^{\rho + } \delta^{in}    + 2  K_{0}(\Bar{Q}_{R_{4}}|\vec{r} \,|)  \left(\Vec{p}_3 - \bar{x} \frac{\vec{p}}{2}  \right)_m  \\
& \times \left.  \left[  \delta^{mn} \left( 1 + \frac{\bar{x}}{2}\right) ( g^{\rho + } \vec{p}^{\; i} - g^{\rho i } p^+ ) - \frac{\bar{x}}{2} \delta^{im}  (g^{\rho + }  \Vec{p}^{\; n} - p^+ g^{n \rho}) + \frac{\bar{x}}{2} \delta^{in} (  g^{\rho + }  \vec{p}^{\; m} - p^+ g^{m \rho }  ) \right] \right\} \,, \numberthis
\end{align*}
for transversely polarized photons. The variables are defined as $\bar{x}$ and $\Bar{Q}_{R_{4}}^2 $ are given by
\begin{align*}
\bar{x} & = \frac{2x_g}{1 - x_g} \,, \nonumber \\  
\Bar{Q}_{R_{4}}^2  & =  \Bar{Q}_{R_{4}, \xi = 0}^2 = Q^2 \frac{(1 - x_g)(1 + x_g)}{4} + m_Q^2  \,,
\end{align*}
which are the same as in Eq.\,\eqref{eq:R2variables}.

We close this section, by pointing out the symmetry between quark-anti-quark symmetry in the perturbative functions: $\mathcal{F}_{R_i}^{\lambda, \bar{\lambda}, \kappa, J_z}$:
\begin{align}
    \mathcal{F}_{R_{1(2)}}^{\lambda, \bar{\lambda}, {}^1S_0^{[c]}, J_z}(\vec{r}) = \mathcal{F}_{R_{3(4)}}^{\lambda, \bar{\lambda}, {}^1S_0^{[c]}, J_z} (-\vec{r}) \,, \label{eq:symmetry1}
\end{align}
and
\begin{align}
    \mathcal{F}_{R_{1(2)}}^{\lambda, \bar{\lambda}, {}^3S_1^{[c]}, J_z}(\vec{r}) = -\mathcal{F}_{R_{3(4)}}^{\lambda, \bar{\lambda}, {}^3S_1^{[c]}, J_z}(-\vec{r}) \,.
    \label{eq:symmetry2}
\end{align}
The sign difference between the two different spin states is due to their different parities.

\section{Differential cross-section for direct quarkonium + gluon production}
\label{sec:cross-section}
As we discussed in Sec.\,\ref{sec:outline_computation}, it is sufficient to compute the short-distance coefficients for differential cross-section for direct quarkonium + gluon production. We show separately the results for longitudinally and transversely polarized photons, as well as the different $\kappa$ states of the heavy quark pair. As discussed in Sec.\,\ref{sec:outline_computation} there are 16 contributions:
\begin{align}
    \der \hat{\sigma}_{\kappa}^{\lambda} = \sum_{i,j=1}^4 \der \hat{\sigma}_{R_{i} R_{j}, \kappa}^{\lambda} \,,
\end{align}
where
\begin{align}
     \der \hat{\sigma}_{R_{i} R_{j}, \kappa}^{\lambda} = \frac{1}{(2q^+)^2} \frac{\der^2 \Vec{p}}{(2 \pi)^2} \frac{\der^2 \vec{p}_g}{(2 \pi)^2} \frac{d x_g }{4 \pi x_g (1 -x_g)} \frac{1}{N_{\rm color}} \widebar{\sum_{J_z }} \sum_{\Bar{\lambda} = \pm 1} \braket{\mathcal{M}_{R_i}^{\lambda, \Bar{\lambda}, \kappa, J_z}(p,p_g) \mathcal{M}_{R_j}^{\dag ,\lambda, \Bar{\lambda}, \kappa, J_z}(p,p_g)}_Y \,.
\end{align}

The differential cross-section will have the following schematic form:
\begin{align*}
d \hat{\sigma}_{R_{i} R_{j}, \kappa}^{\lambda} 
& = \frac{1}{(2q^+)^2} \frac{\der^2 \Vec{p}}{(2 \pi)^2} \frac{\der^2 \vec{p}_g}{(2 \pi)^2} \frac{d x_g }{ x_g (1 -x_g)} \frac{\alpha_{\mathrm{em}} e_Q^2}{(2\pi)^2} \int \der \Pi_{R_i}(\vec{p} ,\vec{p}_g ;\vec{r},\vec{b},\vec{z}) \der \Pi_{R_j}^\dagger(\vec{p} ,\vec{p}_g ;\vec{r}^{\, \prime},\vec{b}^{\, \prime},\vec{z}^{\, \prime}) \\
& \times \Xi^{[c]}_{R_{i} R_{j},Y}\left(\vec{b}  + \frac{\vec{r} }{2}, \vec{b} - \frac{\vec{r} }{2}, \vec{z} ; \vec{b}^{\, \prime} - \frac{\vec{r}^{\, \prime} }{2}, \vec{b}^{\, \prime}  + \frac{\vec{r}^{\, \prime} }{2}, \vec{z}^{\, \prime} \right)  \Gamma^{\lambda, \kappa}_{R_{i} R_{j}} \left(p,p_g,Q;\vec{r} ,\vec{b} ,\vec{z} , \vec{r}^{\, \prime} ,\vec{b}^{\, \prime} ,\vec{z}^{\, \prime} \right) \,, \numberthis 
\label{eq:xsec-SDC-Quarkonium+gluon}
\end{align*}
where we define the differential element 
\begin{align}
    \der \Pi_{R_i}(\vec{p} ,\vec{p}_g ;\vec{r},\vec{b},\vec{z}) = 
\begin{cases}
& \der^2 \vec{r} \; \der^2 \vec{b}  \;  \der^2 \vec{z} \; e^{- i \Vec{p} \cdot \Vec{b}} e^{- i \vec{p}_g \cdot \Vec{z}}  \quad \text{ if } \quad i = 1,3 \\ 
& \der^2 \vec{r} \; \der^2 \vec{b} \; e^{- i \Vec{p} \cdot \Vec{b}} e^{- i \vec{p}_g \cdot \left(\Vec{b} + \frac{\Vec{r}}{2} \right)}  \quad \text{ if } \quad i = 2 \; \\ 
& \der^2 \vec{r} \; \der^2 \vec{b} \; e^{- i \Vec{p} \cdot \Vec{b}} e^{- i \vec{p}_g \cdot \left(\Vec{b} - \frac{\Vec{r}}{2} \right)}  \quad \text{ if } \quad i = 4 \; , 
\end{cases}
\end{align}
the color correlator 
\begin{equation}
 \Xi^{[c]}_{R_{i} R_{j},Y}\left(\vec{x} , \vec{y} , \vec{z} ; \vec{y}^{\, \prime} , \vec{x}^{\, \prime} , \vec{z}^{\, \prime} \right)  = \frac{1}{N_{\rm color}} \left\langle \Tr\left[\mathcal{C}_{R_{i}}\left(\vec{x} , \vec{y} , \vec{z} \, \right) C^{[c]} \right] \Tr\left[C^{\dag [c]}  \mathcal{C}^\dag_{R_{j}}\left( \vec{x}^{\, \prime} , \vec{y}^{\, \prime} , \vec{z}^{\, \prime}  \right) \right] \right\rangle_Y \,, \label{eq:color_correlators_Quarkonium+g}
\end{equation}
and the perturbative factor
\begin{equation}
\Gamma^{\lambda, \kappa}_{R_{i} R_{j}}\left(p,p_g,Q; \vec{r}, \vec{b}, \vec{z}, \vec{r}^{\, \prime}, \vec{b}^{\, \prime}, \vec{z}^{\, \prime}\right)  =  \sum_{\Bar{\lambda} = \pm 1} 
\widebar{\sum_{J_z}} \mathcal{F}_{R_{i}}^{\lambda, \Bar{\lambda}, \kappa, J_z}\left(p,p_g,Q; \vec{r} , \vec{b} , \vec{z} \, \right) \mathcal{F}_{R_{j}}^{\dag \lambda, \Bar{\lambda}, \kappa, J_z}\left(p,p_g,Q; \vec{r}^{\, \prime} , \vec{b}^{\, \prime} , \vec{z}^{\, \prime}  \right) \,.
\label{eq:perturbative_factor}
\end{equation}

For a $\kappa = {}^{2S+1}L_J^{[c]}$ state, the information on the color state $[c]$ of the heavy-quark pair is fully encoded in the color correlator in Eq.~\eqref{eq:color_correlators_Quarkonium+g}, while the information about the spin, angular momentum and total momentum ${}^{2S+1}L_J$ of the heavy-quark pair, and the polarization $\lambda$ of the photon is completely encompassed in the perturbative factor in Eq.~\eqref{eq:perturbative_factor}.

\subsection{Color Correlators}
\label{sec:color_correlator_summary}
To compute the color correlator in Eq.~\eqref{eq:color_correlators_Quarkonium+g} we recall the color structure $\mathcal{C}_{R_i}$ were given in Eqs.~\eqref{eq:color-structure-R1}, \eqref{eq:color-structure-R2}, \eqref{eq:color-structure-R1c} and \eqref{eq:color-structure-R2c} and the color projector in Eq.~\eqref{eq:ColorProjector}. To simplify our results we repeatedly use the Fierz identity:
\begin{equation}
t^a_{ij} t^a_{kl} = \frac{1}{2} \delta_{kj} \delta_{il} - \frac{1}{2 N_c} \delta_{ij} \delta_{kl} \; . 
\end{equation}
The color correlators will be expressed in terms of the multipole light-like Wilson line correlators:
\begin{align*}
S^{(2)}_Y(\Vec{x}, \Vec{y} \, ) & = \frac{1}{N_c} \left\langle\Tr\left[V(\Vec{x} \, )V^\dag(\Vec{y} \, ) \right]\right\rangle_Y \; , \label{eq:dipole-correlator}\numberthis \\ 
S^{(2,2)}_Y(\Vec{x}, \Vec{y};\Vec{y}^{\, \prime}, \Vec{x}^{\, \prime} ) & = \frac{1}{N_c^2} \left\langle\Tr\left[V(\Vec{x} \, )V^\dag(\Vec{y} \, ) \right]\Tr\left[V(\vec{y}^{\, \prime })V^\dag(\Vec{x}^{\, \prime}) \right]\right\rangle_Y \; , \numberthis \\
S^{(4)}_Y(\Vec{x}, \Vec{y},\Vec{y}^{\, \prime}, \Vec{x}^{\, \prime}) & = \frac{1}{N_c}  \left\langle\Tr\left[V(\Vec{x} \, )V^\dag(\Vec{y}\,)V(\Vec{y}^{\, \prime})V^\dag(\Vec{x}^{\, \prime}) \right]\right\rangle_Y \; \numberthis \\  
S^{(2,4)}(\Vec{x}, \Vec{z}; \Vec{z}, \Vec{y}, \Vec{y}^{\, \prime}, \Vec{x}^{\, \prime})  & = \frac{1}{N_c^2} \left\langle \Tr\left[V(\vec{x} \,) V^\dag(\vec{z} \,)\right] \Tr\left[  V(\vec{z} \,)V^\dag(\vec{y} \,) V(\vec{y}^{\, \prime}) V^\dag (\vec{x}^{\, \prime})  \right] \right\rangle_ Y  \numberthis  \\
S^{(6)}( \vec{z} , \vec{y} , \vec{x} , \vec{z} , \vec{y}^{\, \prime}, \vec{x}^{\, \prime} )  & = \frac{1}{N_c} \left\langle\Tr \left[V(\vec{z} \, )V^\dag(\vec{y} \, ) V(\vec{x} \,) V^\dag(\vec{z} \, )V(\vec{y}^{\, \prime}) V^\dag (\vec{x}^{\, \prime}) \right]\right\rangle_Y \numberthis  \\
S^{(4,4)}(\vec{x} , \vec{z} , \vec{z}^{\, \prime} , \vec{x}^{\, \prime} ; \vec{z} ,\vec{y} , \vec{y}^{\, \prime} , \vec{z}^{\, \prime} )  & = \frac{1}{N_c^2}\left\langle\Tr\left[V(\vec{x} \,)V^\dag(\vec{z} \,) V(\vec{z}^{\, \prime} )V^\dag(\vec{x}^{\, \prime} ) \right]  \Tr\left[V(\vec{z} \,)V^\dag(\vec{y} \,)V(\vec{y}^{\, \prime} )V^\dag(\vec{z}^{\, \prime} ) \right] \right\rangle_Y \; , \numberthis \\
S^{(8)}(\vec{z} , \vec{y} , \vec{x} , \vec{z}^{\, \prime} , \vec{x}^{\, \prime} , \vec{y}^{\, \prime}  ) & = \frac{1}{N_c} \left\langle \Tr\left[V(\vec{z} \,) V^\dag(\vec{y} \,) V(\vec{x} \,) V^\dag(\vec{z} \,) V(\vec{z}^{\, \prime} )V^\dag(\vec{x}^{\, \prime} )V(\vec{y}^{\, \prime} )V^\dag(\vec{z}^{\, \prime} ) \right] \right\rangle_Y \; . \label{eq:octupole-correlator} \numberthis 
\end{align*}
Noting that $\mathcal{C}_{R_1}=\mathcal{C}_{R_3}$, we have the following relations among the color correlators
\begin{align}
    & \Xi^{[c]}_{R_1R_1,Y} = \Xi^{[c]}_{R_1R_3,Y} = \Xi^{[c]}_{R_3R_1,Y} = \Xi^{[c]}_{R_3R_3,Y} \nonumber \\
    & \Xi^{[c]}_{R_1R_2,Y} = \Xi^{[c]}_{R_3R_2,Y} \nonumber \\
    & \Xi^{[c]}_{R_1R_4,Y} = \Xi^{[c]}_{R_3R_4,Y} 
\end{align}
Furthermore, we can relate correlators using hermitian conjugation:
\begin{align}
    \Xi^{[c]}_{R_iR_j,Y}\left(\vec{x} , \vec{y} , \vec{z} ; \vec{y}^{\, \prime} , \vec{x}^{\, \prime} , \vec{z}^{\, \prime} \right) = \Xi^{\dagger [c]}_{R_jR_i,Y}\left(\vec{y}^{\, \prime} , \vec{x}^{\, \prime} , \vec{z}^{\, \prime}; \vec{x} , \vec{y} , \vec{z}  \right) \,.
\end{align}
As a consequence of these relations, it is sufficient to compute 6 color correlators, whose results we give below.

\paragraph{$R_1 R_1$ contribution}
\begin{align*}
 \Xi^{[1]}_{R_1R_1,Y}\left(\vec{x} , \vec{y} , \vec{z} ; \vec{y}^{\, \prime} , \vec{x}^{\, \prime} , \vec{z}^{\, \prime} \right) & =\frac{1}{2} \left( S^{(8)}(\vec{z} , \vec{y} , \vec{x} , \vec{z}^{\, \prime} , \vec{x}^{\, \prime} , \vec{y}^{\, \prime}  ) - S^{(2,2)}(\vec{x} , \vec{y} \,; \vec{y}^{\, \prime} , \vec{x}^{\, \prime} ) \right) \;, \numberthis \\
 \Xi^{[8]}_{R_1R_1,Y}\left(\vec{x} , \vec{y} \,, \vec{z} ; \vec{y}^{\, \prime} , \vec{x}^{\, \prime} , \vec{z}^{\, \prime} \right) 
& =\frac{1}{2(N_c^2 - 1) }  \left\{ N_c^2  S^{(4,4)}(\vec{x} , \vec{z} , \vec{z}^{\, \prime} , \vec{x}^{\, \prime} ; \vec{z} ,\vec{y} , \vec{y}^{\, \prime} , \vec{z}^{\, \prime} )  \right. \\
& -S^{(8)}(\vec{z} , \vec{y} , \vec{x} , \vec{z}^{\, \prime} , \vec{x}^{\, \prime} , \vec{y}^{\, \prime}  ) - S^{(4)}(\vec{x} , \vec{y} , \vec{y}^{\, \prime} , \vec{x}^{\, \prime} ) + S^{(2,2) }(\vec{x} , \vec{y} ; \vec{y}^{\, \prime} , \vec{x}^{\, \prime} ) \\
& - N_c^2 S^{(2,2)}(\vec{x} ,\vec{z} ; \vec{z} , \vec{y} \,) + S^{(2)}(\vec{x} ,\vec{y} \,)  - N_c^2S^{(2,2)}(\vec{y}^{\, \prime} ,\vec{z}^{\, \prime} ;\vec{z}^{\, \prime} ,\vec{x}^{\, \prime} ) \\
& \left.  + S^{(2)}(\vec{y}^{\, \prime} , \vec{x}^{\, \prime} )  + (N_c^2  - 1 ) \right\}   \;. \numberthis 
\end{align*}

\paragraph{$R_2 R_2$ contribution}
\begin{align}
 \Xi^{[1]}_{R_2R_2,Y}\left(\vec{x} , \vec{y}; \vec{y}^{\, \prime} , \vec{x}^{\, \prime}  \right)  &= \frac{1}{2} \left(S^{(4)}(\vec{x} , \vec{y} , \vec{y}^{\, \prime} , \vec{x}^{\, \prime} ) - S^{(2,2)}(\vec{x} , \vec{y} ; \vec{y}^{\, \prime} , \vec{x}^{\, \prime} )\right) \;, \\
\Xi^{[8]}_{R_2R_2,Y}\left(\vec{x} , \vec{y}  ; \vec{y}^{\, \prime} , \vec{x}^{\, \prime}  \right)
& = \frac{1}{2} \left(\frac{N_c^2 - 2}{N_c^2 - 1} S^{(4)}(\vec{x} , \vec{y} , \vec{y}^{\, \prime} , \vec{x}^{\, \prime} )  + \frac{1}{N_c^2 - 1 }S^{(2,2)}(\vec{x} , \vec{y} ; \vec{y}^{\, \prime} , \vec{x}^{\, \prime} )  \right. \nonumber \\  
&  \left. -  S^{(2)}(\vec{x} , \vec{y} \,) - S^{(2)}(\vec{y}^{\, \prime} , \vec{x}^{\, \prime} )   + 1 \right) \; . \numberthis 
\end{align}

\paragraph{$R_4 R_4$ contribution}
\begin{align}
 \Xi^{[1]}_{R_4R_4,Y}\left(\vec{x} , \vec{y}; \vec{y}^{\, \prime} , \vec{x}^{\, \prime}  \right)  &= \frac{1}{2} \left(S^{(4)}(\vec{x} , \vec{y} , \vec{y}^{\, \prime} , \vec{x}^{\, \prime} ) - S^{(2,2)}(\vec{x} , \vec{y} ; \vec{y}^{\, \prime} , \vec{x}^{\, \prime} )\right) \;, \\
\Xi^{[8]}_{R_4R_4,Y}\left(\vec{x} , \vec{y}  ; \vec{y}^{\, \prime} , \vec{x}^{\, \prime}  \right)
& = \frac{1}{2} \left(\frac{N_c^2 - 2}{N_c^2 - 1} S^{(4)}(\vec{x} , \vec{y} , \vec{y}^{\, \prime} , \vec{x}^{\, \prime} )  + \frac{1}{N_c^2 - 1 }S^{(2,2)}(\vec{x} , \vec{y} ; \vec{y}^{\, \prime} , \vec{x}^{\, \prime} )  \right. \nonumber \\  
&  \left. -  S^{(2)}(\vec{x} , \vec{y} \,) - S^{(2)}(\vec{y}^{\, \prime} , \vec{x}^{\, \prime} )   + 1 \right) \;, \numberthis 
\end{align}
which is the same as the $R_2 R_2$ contribution.

\paragraph{$R_1 R_2$ contribution}

\begin{align*}
& \Xi^{[1]}_{R_1R_2}\left(\vec{x} , \vec{y} , \vec{z} ; \vec{y}^{\, \prime} , \vec{x}^{\, \prime}  \right) = \frac{1}{2}\left( S^{(6)}( \vec{z} , \vec{y}, \vec{x} , \vec{z} , \vec{y}^{\, \prime}, \vec{x}^{\, \prime} ) - S^{(2,2)}(\vec{x} , \vec{y}; \vec{y}^{\, \prime}, \vec{x}^{\, \prime} ) \right) \;, \numberthis \\
& \Xi^{[8]}_{R_1R_2}\left(\vec{x} , \vec{y} , \vec{z} ; \vec{y}^{\, \prime} ; \vec{y}^{\, \prime} , \vec{x}^{\, \prime} \right) \\
& = \frac{1}{2(N_c^2 - 1) } \left\{  N_c^2 S^{(2,4)}(\Vec{x}, \Vec{z}; \Vec{z}, \Vec{y}, \Vec{y}^{\, \prime}, \Vec{x}^{\, \prime}) - S^{(6)}(\vec{z} , \vec{y} , \vec{x} , \vec{z} , \vec{y}^{\, \prime}, \vec{x}^{\, \prime} ) - S^{(4)}(\Vec{x}, \Vec{y}, \Vec{y}^{\, \prime}, \Vec{x}^{\, \prime})  \right.  \\
& \left. + S^{(2,2)}(\Vec{x}, \Vec{y}; \Vec{y}^{\, \prime}, \Vec{x}^{\, \prime}) - N_c^2 S^{(2,2)}(\Vec{x}, \Vec{z}; \Vec{z}, \Vec{y})  + S^{(2)}( \Vec{x}, \Vec{y}\,) - (N_c^2 - 1 ) S^{(2)}(\Vec{y}^{\, \prime }, \Vec{x}^{\, \prime}) + (N_c^2 - 1) \right\} \;. \numberthis 
\end{align*}

\paragraph{$R_2R_{4}$ contribution}
\begin{align*}
\Xi_{R_2 R_{4}}^{[1]}\left(\vec{x} , \vec{y}  ; \vec{y}^{\, \prime} , \vec{x}^{\, \prime} \right) & = \frac{1}{2 } \left( S^{(4)}\left( \Vec{x}, \Vec{y}, \Vec{y}^{\, \prime}, \Vec{x}^{\, \prime} \right) - S^{(2,2)}\left( \Vec{x}, \Vec{y}; \Vec{y}^{\, \prime}, \Vec{x}^{\, \prime}\right)\right)  \;,\numberthis \\
\Xi_{R_2 R_{4}}^{[8]}\left(\vec{x} , \vec{y} ; \vec{y}^{\, \prime} , \vec{x}^{\, \prime}  \right) & = \frac{1}{2} \left\{ \frac{N_c^2 + 1 }{N_c^2 - 1 } S^{(2,2)}(\Vec{x}, \Vec{y}; \Vec{y}^{\, \prime}, \Vec{x}^{\, \prime}) - \frac{2}{N_c^2 - 1 } S^{(4)}(\Vec{x}, \Vec{y}, \Vec{y}^{\, \prime}, \Vec{x}^{\, \prime}) -  S^{(2)}\left(\Vec{x}, \Vec{y}\right) \right. \\
& \left. - S^{(2)}(\Vec{y}^{\, \prime}, \Vec{x}^{\, \prime} )  + 1  \right\} \;. \numberthis
\end{align*}

\paragraph{$R_1 R_4$ contribution}
\begin{align*}
&\Xi_{R_1R_4,Y}^{[1]}\left(\vec{x},\vec{y},\vec{z}; \vec{y}^{\, \prime}, \vec{x}^{\, \prime}\right)  = \frac{1}{2}\left\{S^{(6)}\left(\vec{z},\vec{y},\vec{x},\vec{z},\vec{y}^{\, \prime}, \vec{x}^{\, \prime}\right) - S^{(2,2)}\left(\vec{x},\vec{y};\vec{y}^{\, \prime}, \vec{x}^{\, \prime}\right)\right\} \,, \numberthis \\
&\Xi_{R_1R_4,Y}^{[8]}\left(\vec{x},\vec{y},\vec{z}; \vec{y}^{\, \prime}, \vec{x}^{\, \prime}\right) \nonumber \
\nonumber \\
& = \frac{1}{2(N_c^2-1)} \left\{ N_c^2 S^{(2,4)}(\vec{z},\vec{y};\vec{x},\vec{z}, \vec{y}^{\, \prime}, \vec{x}^{\, \prime}) - S^{(6)}(\vec{z} , \vec{y} , \vec{x} , \vec{z} , \vec{y}^{\, \prime}, \vec{x}^{\, \prime} ) - S^{(4)}(\Vec{x}, \Vec{y}, \Vec{y}^{\, \prime}, \Vec{x}^{\, \prime})  \right.  \\
& \left. + S^{(2,2)}(\Vec{x}, \Vec{y}; \Vec{y}^{\, \prime}, \Vec{x}^{\, \prime}) - N_c^2 S^{(2,2)}(\Vec{x}, \Vec{z}; \Vec{z}, \Vec{y})  + S^{(2)}( \Vec{x}, \Vec{y} \,) - (N_c^2 - 1 ) S^{(2)}(\Vec{y}^{\, \prime }, \Vec{x}^{\, \prime}) + (N_c^2 - 1) \right\} \,. \numberthis 
\end{align*}

\subsection{Perturbative factors}
\label{sec:perturbative_factor_summary}
Exploiting the symmetry found in Eqs.\,\eqref{eq:symmetry1} and \eqref{eq:symmetry2}, we have the following relations among the perturbative factors
\begin{align}
    \Gamma^{\lambda,{}^1S_0^{[c]} }_{R_1 R_1}(\vec{r},\vec{r}^{\, \prime}) & = \Gamma^{\lambda,{}^1S_0^{[c]} }_{R_3 R_3} (-\vec{r},-\vec{r}^{\, \prime}) = \Gamma^{\lambda,{}^1S_0^{[c]} }_{R_1 R_3}(\vec{r},-\vec{r}^{\, \prime}) \,, \nonumber \\
    \Gamma^{\lambda,{}^1S_0^{[c]} }_{R_2 R_2}(\vec{r},\vec{r}^{\, \prime}) &= \Gamma^{\lambda,{}^1S_0^{[c]} }_{R_4 R_4}(-\vec{r},-\vec{r}^{\, \prime}) = \Gamma^{\lambda,{}^1S_0^{[c]} }_{R_2 R_4}(\vec{r},-\vec{r}^{\, \prime}) \,, \nonumber \\
    \Gamma^{\lambda,{}^1S_0^{[c]} }_{R_1 R_2}(\vec{r},\vec{r}^{\, \prime}) & = \Gamma^{\lambda,{}^1S_0^{[c]} }_{R_3 R_4}(-\vec{r},-\vec{r}^{\, \prime}) = \Gamma^{\lambda,{}^1S_0^{[c]} }_{R_1 R_4}(\vec{r},-\vec{r}^{\, \prime}) = \Gamma^{\lambda,{}^1S_0^{[c]} }_{R_3 R_2}(-\vec{r},\vec{r}^{\, \prime}) \,,
\end{align}
and
\begin{align}
    \Gamma^{\lambda,{}^3S_1^{[c]} }_{R_1 R_1}(\vec{r},\vec{r}^{\, \prime}) & = \Gamma^{\lambda,{}^3S_1^{[c]} }_{R_3 R_3} (-\vec{r},-\vec{r}^{\, \prime}) = -\Gamma^{\lambda,{}^3S_1^{[c]} }_{R_1 R_3}(\vec{r},-\vec{r}^{\, \prime}) \,,\nonumber \\
    \Gamma^{\lambda,{}^3S_1^{[c]} }_{R_2 R_2}(\vec{r},\vec{r}^{\, \prime}) &= \Gamma^{\lambda,{}^3S_1^{[c]} }_{R_4 R_4}(-\vec{r},-\vec{r}^{\, \prime}) = -\Gamma^{\lambda,{}^3S_1^{[c]} }_{R_2 R_4}(\vec{r},-\vec{r}^{\, \prime}) \,, \nonumber \\
    \Gamma^{\lambda,{}^3S_1^{[c]} }_{R_1 R_2}(\vec{r},\vec{r}^{\, \prime}) & = \Gamma^{\lambda,{}^3S_1^{[c]} }_{R_3 R_4}(-\vec{r},-\vec{r}^{\, \prime}) = -\Gamma^{\lambda,{}^3S_1^{[c]} }_{R_1 R_4}(\vec{r},-\vec{r}^{\, \prime}) = -\Gamma^{\lambda,{}^3S_1^{[c]} }_{R_3 R_2}(-\vec{r},\vec{r}^{\, \prime}) \,.
\end{align}
In the relations above all other arguments of the functions $\Gamma^{\lambda,\kappa}_{R_i R_j}$ are kept fixed, and for the sake of compactness, they are not shown explicitly.

Furthermore, hermitian conjugation implies
\begin{align}
    \Gamma^{\lambda,\kappa}_{R_i R_j}(p,p_g,Q;\vec{r},\vec{b},\vec{z},\vec{r}^{\, \prime},\vec{b}^{\, \prime},\vec{z}^{\, \prime}) = \Gamma^{\dagger\lambda,\kappa}_{R_j R_i} (p,p_g,Q;\vec{r}^{\, \prime},\vec{b}^{\, \prime},\vec{z}^{\, \prime},\vec{r},\vec{b},\vec{z}) \,.
\end{align}
As a consequence of these relations, it is sufficient to compute the contributions for $\Gamma^{\lambda,\kappa}_{R_1 R_1}$, $\Gamma^{\lambda,\kappa}_{R_2 R_2}$ and $\Gamma^{\lambda,\kappa}_{R_1 R_2}$. 

In order to compute the perturbative factors, we use repeatedly:
\begin{equation}
    \sum_{\lambda = \pm 1} \vec{\epsilon}^{\; \lambda}_i \vec{\epsilon}^{\; *  \lambda}_j = - g_{\perp ij} = \delta_{ij} \; , 
\end{equation}
when summing over transverse polarizations for the gluon (and photon). We also use
 \begin{equation}
      \epsilon^{ji} \epsilon^{mn} = \left(\delta^{jm}\delta^{in}- \delta^{jn}\delta^{im}\right) \; . 
 \end{equation}
For the particular case of $\Gamma^{\lambda, {}^3S_1^{[c]}}$ state, we will use the completeness relation in Eq.\,\eqref{SumTensorSWave} and the identities in Appendix\,\ref{sec:Lorentz-Contractions}. 

We present the results separately for longitudinally and transversely polarized photons. They will depend on transverse integrals that are defined in appendix~\ref{sec:R1TransverseIntegral} and that are to be evaluated at $z_1 = \frac{1}{2}- \frac{x_g}{2}$, $z_2 = x_g $ and $z_3= \frac{1}{2}- \frac{x_g}{2}$.
and on different variables defined in eqs.~\eqref{VariablesR1At0} and \eqref{eq:R2variables}.

\subsubsection{Longitudinally polarized photon}
\paragraph{$R_1R_1$ contribution}
\begin{align}
    \Gamma_{R_1R_1}^{\lambda= 0, \kappa} = \Gamma_{R_1regR_1reg}^{\lambda= 0, \kappa} \,,
\end{align}
where
\begin{align*}
& \Gamma^{\lambda = 0, {}^1S_0^{[c]} }_{R_1regR_1reg} \left(p,p_g,Q;\vec{r} ,\vec{b} ,\vec{z} , \vec{r}^{\, \prime} ,\vec{b}^{\, \prime} ,\vec{z}^{\, \prime} \right) \\
& = \frac{Q^2}{2m_Q}  (1+ x_g )^2  \, \frac{g^2 (p^+)^2}{\pi^2 }  I_{1}\left(\Vec{y}-\Vec{x}, \Vec{z}-\Vec{x} \, \right)  \cdot I_{1}^*\left(\vec{y}^{\, \prime}-\vec{x}^{\, \prime}, \vec{z}^{\, \prime}-\vec{x}^{\, \prime}\right) \;, \numberthis
\end{align*}
for the $\kappa={}^{1} S_0$ state, and
\begin{align*}
& \Gamma^{\lambda= 0, {}^3S_1^{[c]} }_{R_1regR_1reg}\left(p,p_g,Q;\vec{r} ,\vec{b} ,\vec{z} ,\vec{r}^{\, \prime} ,\vec{b}^{\, \prime} ,\vec{z}^{\, \prime} \right)  \\ 
& = \frac{2m_Q}{3} Q^2 \frac{(1 + x_g)^2}{x_g^2} \frac{g^2(p^+)^2}{\pi^2} \left\{  \frac{1}{4m_Q^2} I_{1} \left(\Vec{y}-\Vec{x}, \Vec{z}-\Vec{x} \, \right) \cdot  I_{1}^{*}\left(\Vec{y}^{\, \prime}-\Vec{x}^{\, \prime}, \Vec{z}^{\, \prime} -\vec{x}^{\, \prime}\right) + \frac{ 2 x_g^4}{(1+ x_g)^2} \right. \\
& \left. \times I_0 \left(\Vec{y}-\Vec{x}, \Vec{z}- \Vec{x}\, \right)  I_0^*\left(\Vec{y}^{\, \prime}-\Vec{x}^{\, \prime}, \Vec{z}^{\, \prime}- \Vec{x}^{\, \prime}\right)   \right\} \;, \numberthis
\end{align*}
for the $\kappa={}^{3} S_1$ state.

\paragraph{$R_2R_2$ contribution}
\begin{align*}
& \Gamma^{\lambda= 0, {}^1S_0^{[c]}}_{R_2R_2} (p,p_g,Q;\vec{r} , \vec{r}^{\, \prime} ) \\*
& = \frac{2}{m_Q}  g^2 (p^+)^2 x_g^2 (1+x_g)^2  Q K_0(\Bar{Q}_{R_2}|\vec{r} \,| )   Q K_0(\Bar{Q}_{R_2}|\vec{r}^{\, \prime} | ) \frac{ \left(\vec{p}_g - x \frac{\Vec{p}}{2}\right)^2}{\left[\left(\vec{p}_g - x \frac{\Vec{p}}{2}\right)^2 + x^2 m_Q^2 \right]^2} \;, \numberthis 
\end{align*}
for the $\kappa={}^{1} S_0$ state, and
\begin{align*}
& \Gamma^{\lambda= 0, {}^3S_1^{[c]} }_{R_2R_2}(p,p_g,Q;\vec{r} , \vec{r}^{\, \prime} ) \\
& = \frac{1}{3m_Q} (p^+)^2 g^2(1 + x_g)^2(1-x_g)^2 Q K_0(\Bar{Q}_{R_2}|\vec{r} \,| )  Q K_0(\Bar{Q}_{R_2}|\vec{r}^{\, \prime} | ) \frac{1 }{\left[\left(\vec{p}_g - x \frac{\Vec{p}}{2}\right)^2 + x^2 m_Q^2 \right]^2} \\
& \times  \left\{ 2 \left(1 + \frac{x}{2}\right)^2  \left(\vec{p}_g - x \frac{  \Vec{p}}{2}\right)^2   +   x^4 m_Q^2 \right\} \;, \numberthis
\end{align*}
for the $\kappa={}^{3} S_1$ state.

\paragraph{$R_1 R_2$ contribution}
\begin{align}
    \Gamma_{R_1R_2}^{\lambda= 0, \kappa} = \Gamma_{R_1regR_2}^{\lambda= 0, \kappa} \,,
\end{align}
where
\begin{align*}
& \Gamma_{R_1regR_2}^{\lambda = 0, {}^1S_0^{[c] }} \left(p,p_g,Q; \vec{r} ,\vec{b} ,\vec{z}  , \vec{r}^{\, \prime}\right)  \\
& = - \frac{1}{m_Q} x_g (1+ x_g )^2  \frac{g^2(p^+)^2}{\pi}   Q^2 K_0(\Bar{Q}_{R_2}|\vec{r}^{\, \prime}|)   \frac{\left(\vec{p}_g - x \frac{\Vec{p}}{2}\right) \cdot I_{1}\left(\Vec{y}-\Vec{x}, \Vec{z}-\Vec{x} \, \right) }{\left[\left(\vec{p}_g - x \frac{\Vec{p}}{2}\right)^2 + x^2 m_Q^2 \right]} \;, \numberthis
\end{align*}
for the $\kappa={}^{1} S_0$ state, and
\begin{align*}
& \Gamma_{R_1regR_2}^{\lambda = 0, {}^3S_1^{[c] }} \left(p,p_g,Q; \vec{r} ,\vec{b} ,\vec{z}  , \vec{r}^{\, \prime}\right)  \\
& = - \frac{2m_Q}{3}  \frac{(1 + x_g)^2(1-x_g) }{x_g }  \frac{g^2(p^+)^2 }{\pi} Q^2 K_0(\Bar{Q}_{R_2}|\vec{r}^{\, \prime}| )  \frac{1}{\left[\left(\vec{p}_g - x \frac{\Vec{p}}{2}\right)^2 + x^2 m_Q^2 \right]}  \\
& \times  \left\{  \frac{1}{2m_Q^2} \left(1 + \frac{x}{2}\right)  I_{1}\left(\Vec{y}-\Vec{x}, \Vec{z}-\Vec{x} \, \right)  \cdot  \left(\vec{p}_g - x \frac{  \Vec{p}}{2}\right)      + \frac{ x^2   x_g^2}{1 + x_g }  I_0 \left(\Vec{y}-\Vec{x}, \Vec{z}- \Vec{x}\, \right)   \right\} \;, \numberthis 
\end{align*}
for the $\kappa={}^{3} S_1$ state.

\subsubsection{Transversely polarized photon}
Here we present the results averaged over the two transverse polarizations. 
\paragraph{$R_1R_1$ contribution}
\begin{align}
    \Gamma_{R_1R_1}^{\lambda= \pm 1, \kappa} = \Gamma_{R_1instR_1inst}^{\lambda= \pm 1, \kappa} + \Gamma_{R_1regR_1inst}^{\lambda= \pm 1, \kappa} + \Gamma_{R_1instR_1reg}^{\lambda= \pm 1, \kappa} + \Gamma_{R_1regR_1reg}^{\lambda= \pm 1, \kappa} \,,
\end{align}
where
\begin{align*}
& \sum_{\lambda = \pm 1 }\Gamma_{R_1instR_1inst}^{\lambda= \pm 1, {}^1S_0^{[c]}}\left(p,p_g,Q; \vec{r},\vec{b}, \vec{z}, \vec{r}^{\, \prime}, \vec{b}^{\, \prime}, \vec{z}^{\, \prime} \right) \\ 
& =  \frac{1}{4m_Q} \frac{ x_g^2 (1 - x_g )^4 }{ (1 + x_g )^2}  \frac{g^2 (p^+)^2}{\pi^2} \frac{\Bar{Q}_{R_1} K_1(\Bar{Q}_{R_1}X_{R_1}) }{X_{R_1}} \frac{\Bar{Q}_{R_1} K_1(\Bar{Q}_{R_1}X_{R_1}) }{X_{R_1}} \;, \numberthis \\
& \sum_{\lambda = \pm 1} \Gamma_{R_1regR_1inst}^{\lambda= \pm 1, {}^1S_0^{[c]}}\left(p,p_g,Q; \vec{r},\vec{b}, \vec{z}, \vec{r}^{\, \prime}, \vec{b}^{\, \prime}, \vec{z}^{\, \prime} \right) \\ 
& =  \frac{1}{2m_Q}    \frac{(1 - x_g )}{ (1 + x_g ) }  \frac{g^2 (p^+)^2 }{\pi^2}     \frac{\Bar{Q}_{R_1} K_1(\Bar{Q}_{R_1}X_{R_1}) }{X_{R_1}}    \left\{ - I_{2mj}\left(\Vec{y}- \Vec{x}, \Vec{z}- \vec{x} \, \right)    (1 + x_g^2) \delta^{jm} \right. \\
& \left. +  \frac{4 x_g^2 }{(1 + x_g )} m_Q^2  I_0\left(\Vec{y}- \Vec{x}, \Vec{z}- \Vec{x} \, \right)    \right\}  \;,  \numberthis \\
& \sum_{\lambda = \pm 1} \Gamma_{R_1instR_1reg}^{\lambda= \pm 1, {}^1S_0^{[c]}}\left(p,p_g,Q; \vec{r},\vec{b}, \vec{z}, \vec{r}^{\, \prime}, \vec{b}^{\, \prime}, \vec{z}^{\, \prime} \right) \\ 
& =  \frac{1}{2m_Q}    \frac{(1 - x_g )}{ (1 + x_g ) }  \frac{g^2 (p^+)^2 }{\pi^2}     \frac{\Bar{Q}_{R_1} K_1(\Bar{Q}_{R_1}X_{R_1}) }{X_{R_1}}    \left\{ - I^*_{2mj}\left(\Vec{y}^{\, \prime}- \Vec{x}^{\, \prime}, \Vec{z}^{\, \prime}- \vec{x}^{\, \prime}\right)    (1 + x_g^2) \delta^{jm} \right. \\
& \left. +  \frac{4 x_g^2 }{(1 + x_g )} m_Q^2  I_0^*\left(\Vec{y}^{\, \prime}- \Vec{x}^{\, \prime}, \Vec{z}^{\, \prime}- \Vec{x}^{\, \prime} \right)    \right\}    \numberthis \\
& \sum_{\lambda = \pm 1} \Gamma_{R_1regR_1reg}^{\lambda= \pm 1, {}^1S_0^{[c]}} \left(p,p_g,Q; \vec{r},\vec{b}, \vec{z}, \vec{r}^{\, \prime}, \vec{b}^{\, \prime}, \vec{z}^{\, \prime} \right) \\ 
& = \frac{2}{m_Q} \frac{1}{ x_g^2 (1 -x_g)^2 }  \frac{g^2 (p^+)^2 }{\pi^2}  \left\{   I_{2mj}\left(\Vec{y}- \Vec{x}, \Vec{z}- \vec{x}\, \right)   I_{2rl}^*\left(\Vec{y}^{\, \prime}- \Vec{x}^{\, \prime}, \Vec{z}^{\, \prime}- \vec{x}^{\, \prime}\right)     \right. \\
&  \times \left[ (1 + x_g^4)  \delta^{rm} \delta^{jl} - 2 x_g^2   (\delta^{lm} \delta^{jr} - \delta^{lr} \delta^{jm} )   \right] - \frac{2 x_g^2 }{(1 + x_g )} m_Q^2   I_{2mj}\left(\Vec{y}- \Vec{x}, \Vec{z}- \vec{x}\, \right) \\
& \times I_0^*\left(\Vec{y}^{\, \prime}- \Vec{x}^{\, \prime}, \Vec{z}^{\, \prime} - \Vec{x}^{\, \prime} \right)   (1 + x_g^2 ) \delta^{mj}  -\frac{2 x_g^2 }{(1 + x_g )} m_Q^2  I_0\left(\Vec{y}- \Vec{x}, \Vec{z}- \Vec{x}\, \right) \\
& \times  I_{2rl}^*\left(\Vec{y}^{\, \prime}- \Vec{x}^{\, \prime}, \Vec{z}^{\, \prime}- \vec{x}^{\, \prime}\right) (1 + x_g^2) \delta^{rl} + \frac{8 x_g^4 }{(1 + x_g )^2} m_Q^4  I_0\left(\Vec{y}- \Vec{x}, \Vec{z}- \Vec{x} \, \right) \\
& \left.  \times I_0^*\left(\Vec{y}^{\, \prime}- \Vec{x}^{\, \prime}, \Vec{z}^{\, \prime} - \Vec{x}^{\, \prime} \right)   \right\} \;, \numberthis \\
\end{align*}
for the $\kappa={}^{1} S_0$ state, and
\begin{align*}
    & \sum_{\lambda = \pm 1 }\Gamma_{R_1instR_1inst}^{\lambda= \pm 1, {}^3S_1^{[c]}}\left(p,p_g,Q; \vec{r},\vec{b}, \vec{z}, \vec{r}^{\, \prime}, \vec{b}^{\, \prime}, \vec{z}^{\, \prime} \right) \\ 
&= \frac{1}{12 m_Q} \frac{ x_g^2 (1 - x_g )^4 }{ (1 + x_g )^2 } \frac{g^2 (p^+)^2}{\pi^2} \frac{\Bar{Q}_{R_1} K_1(\Bar{Q}_{R_1}X_{R_1}) }{X_{R_1}} \frac{\Bar{Q}_{R_1} K_1(\Bar{Q}_{R_1}X_{R_1}) }{X_{R_1}} \;, \numberthis \\
& \sum_{\lambda = \pm 1} \Gamma_{R_1regR_1inst}^{\lambda= \pm 1, {}^3S_1^{[c]}}\left(p,p_g,Q; \vec{r},\vec{b}, \vec{z}, \vec{r}^{\, \prime}, \vec{b}^{\, \prime}, \vec{z}^{\, \prime} \right) \\ 
& =  \frac{1 }{3m_Q}    \frac{ x_g (1 - x_g ) }{ (1 + x_g )}   \frac{g^2(p^+)^2}{\pi^2}   \frac{\Bar{Q}_{R_1} K_1(\Bar{Q}_{R_1}X_{R_1}) }{X_{R_1}}   \left\{   I_{2mj}\left(\Vec{y}- \Vec{x}, \Vec{z}- \vec{x}\, \right)   \delta^{mj}      +  \frac{2 x_g }{(1 + x_g )} m_Q^2 \right. \\
& \left. \times I_0\left(\Vec{y}- \Vec{x}, \Vec{z}- \Vec{x} \, \right)   \right\} \;, \numberthis \\
& \sum_{\lambda = \pm 1} \Gamma_{R_1regR_1reg}^{\lambda= \pm 1, {}^3S_1^{[c]}}\left(p,p_g,Q; \vec{r},\vec{b}, \vec{z}, \vec{r}^{\, \prime}, \vec{b}^{\, \prime}, \vec{z}^{\, \prime} \right) \\ 
&  = \frac{4m_Q}{ 3}   \frac{1}{ x_g^2 (1 -x_g)^2 } \frac{g^2  (p^+)^2  }{\pi^2}   \left\{  \frac{ x_g^2}{m_Q^2}  I_{2mj}\left(\Vec{y}- \Vec{x}, \Vec{z}- \vec{x}\, \right)   I^*_{2rl}\left(\Vec{y}^{\, \prime}- \Vec{x}^{\, \prime}, \Vec{z}^{\, \prime}- \vec{x}^{\, \prime}\right)   \right.  \\ 
&  \times \left[  \delta^{rm}   \delta^{lj} - \delta^{lm}   \delta^{rj}   +  \delta^{jm} \delta^{lr} \right] +    \frac{2 x_g^3 }{(1 + x_g )} I_{2mj}\left(\Vec{y}- \Vec{x}, \Vec{z}- \vec{x}\, \right) I_0^*\left(\Vec{y}^{\, \prime}- \Vec{x}^{\, \prime}, \Vec{z}^{\, \prime}- \Vec{x}^{\, \prime} \right) \delta^{jm}       \\
&  +    I_{1m}\left(\Vec{y}- \Vec{x}, \Vec{z}- \Vec{x}\, \right) I^*_{1r}\left(\Vec{y}^{\, \prime}- \Vec{x}^{\, \prime}, \Vec{z}^{\, \prime}- \Vec{x}^{\, \prime}\right) \delta^{rm}  \left[  1     +   x_g^2         \right]\\
&  + \frac{ x_g^2 }{(1 + x_g )}   I_{1m}\left(\Vec{y}- \Vec{x}, \Vec{z}- \Vec{x}\, \right) \Tilde{I}^*_{1l}\left(\Vec{y}^{\, \prime}- \Vec{x}^{\, \prime}, \Vec{z}^{\, \prime}- \Vec{x}^{\, \prime}\right) \delta^{lm} \left[  1-x_g   \right]^2 \\ 
& + \frac{ x_g^2 }{(1 + x_g )}  \Tilde{I}_{1j}\left(\Vec{y}- \Vec{x}, \Vec{z}- \Vec{x}\, \right)    I^*_{1r}\left(\Vec{y}^{\, \prime}- \Vec{x}^{\, \prime}, \Vec{z}^{\, \prime}- \Vec{x}^{\, \prime}\right)  \delta^{jr}  \left[1- x_g\right]^2 \\ 
& + \frac{4 x_g^4 }{(1 + x_g )^2}   \Tilde{I}_{1j}\left(\Vec{y}- \Vec{x}, \Vec{z}- \Vec{x}\,\right)  \Tilde{I}^*_{1l}\left(\Vec{y}^{\, \prime}- \Vec{x}^{\, \prime}, \Vec{z}^{\, \prime}- \Vec{x}^{\, \prime}\right)   \delta^{jl} \left[  x_g^2   +  1 \right] \\ 
&  +   \frac{2x_g^3 }{(1 + x_g )}   I_0\left(\Vec{y}- \Vec{x}, \Vec{z}- \Vec{x} \, \right)  I^*_{2rl}\left(\Vec{y}^{\, \prime}- \Vec{x}^{\, \prime}, \Vec{z}^{\, \prime}- \vec{x}^{\, \prime}\right)       \delta^{lr}  \\
&  \left. +  m_Q^2 \frac{4 x_g^4 }{(1 + x_g )^2}  I_0\left(\Vec{y}- \Vec{x}, \Vec{z}- \Vec{x} \, \right)   I_0^*\left(\Vec{y}^{\, \prime}- \Vec{x}^{\, \prime}, \Vec{z}^{\, \prime}- \Vec{x}^{\, \prime} \right)  \right\}  \;. \numberthis \\
& \sum_{\lambda = \pm 1} \Gamma_{R_1instR_1reg}^{\lambda= \pm 1, {}^3S_1^{[c]}}\left(p,p_g,Q; \vec{r},\vec{b}, \vec{z}, \vec{r}^{\, \prime}, \vec{b}^{\, \prime}, \vec{z}^{\, \prime} \right) \\ 
& =  \frac{1 }{3m_Q}    \frac{ x_g (1 - x_g ) }{ (1 + x_g )}   \frac{g^2(p^+)^2}{\pi^2}   \frac{\Bar{Q}_{R_1} K_1(\Bar{Q}_{R_1}X_{R_1}) }{X_{R_1}}   \left\{   I^*_{2mj}(\Vec{y}^{\, \prime}- \Vec{x}^{\, \prime}, \Vec{z}^{\, \prime}- \vec{x}^{\, \prime})   \delta^{mj}      +  \frac{2 x_g }{(1 + x_g )} m_Q^2 \right. \\
&  \times I_0^*(\Vec{y}^{\, \prime}- \Vec{x}^{\, \prime}, \Vec{z}^{\, \prime}- \Vec{x}^{\, \prime} )   \Bigg\}  \numberthis \;, 
\end{align*}
for the $\kappa={}^{3} S_1$ state.

\paragraph{$R_2R_2$ contribution}
\begin{align*}
& \sum_{\lambda = \pm 1} \Gamma^{\lambda = \pm 1, {}^1S_0^{[c]}}_{R_2R_2}(p,p_g,Q;\vec{r} , \vec{r}^{\, \prime} )  \\
&= \frac{2}{m_Q}  g^2 (p^+)^2   \frac{1}{\left[\left(\vec{p}_g - x\frac{\Vec{p}}{2} \right)^2 + x^2 m_Q^2 \right]^2}  \\ 
& \times \left\{ 4 \left[\left(1 + \frac{x}{2}\right)^2 + \frac{x^2}{4} x_g^2 \right]  \left(\vec{p}_g - x \frac{\Vec{p}}{2}\right)^2   \frac{ \Vec{r} \cdot \Vec{r}^{\, \prime} }{|\vec{r} \,|  |\vec{r}^{\, \prime} |  }  \Bar{Q}_{R_2} K_1(\Bar{Q}_{R_2}|\vec{r} \,|) \Bar{Q}_{R_2} K_1(\Bar{Q}_{R_2}|\vec{r}^{\, \prime} |)  \right. \\
& +  i 2 x^2  \left[ \left(1 + \frac{x}{2}\right) + \frac{x}{2} x_g \right] m_Q^2  \Vec{r} \cdot \left(\vec{p}_g - x \frac{\Vec{p}}{2}\right)    \frac{\Bar{Q}_{R_2} K_1(\Bar{Q}_{R_2}|\vec{r} \,|)   }{|\vec{r} \,| } K_0(\Bar{Q}_{R_2}|\vec{r}^{\, \prime} |) \\ 
& - i     2 x^2 \left[\left(1 + \frac{x}{2}\right)    + \frac{x}{2} x_g \right] m_Q^2   \Vec{r}^{\, \prime} \cdot   \left(\vec{p}_g - x \frac{\Vec{p}}{2}\right) \frac{\Bar{Q}_{R_2} K_1(\Bar{Q}_{R_2}|\vec{r}^{\, \prime} |)  }{|\vec{r}^{\, \prime} | }   K_0(\Bar{Q}_{R_2}|\vec{r} \,|)  \\ 
&   + 2 x^4 m_Q^4 K_0(\Bar{Q}_{R_2}|\vec{r} \,|) K_0(\Bar{Q}_{R_2}|\vec{r}^{\, \prime} |)    \Bigg\} \,,\numberthis
\end{align*}
for the $\kappa={}^{1} S_0$ state, and
\begin{align*}
& \sum_{\lambda = \pm 1} \Gamma^{\lambda = \pm 1, {}^3S_1^{[c]}}_{R_2R_2}( p, p_g, Q; \vec{r}  , \vec{r}^{\, \prime} ) \\
& = \frac{2m_Q }{3} g^2 (p^+)^2  \frac{1}{\left[\left(\vec{p}_g - x\frac{\Vec{p}}{2} \right)^2 + x^2 m_Q^2 \right]^2} \\
& \left\{  \frac{4}{m_Q^2} \left(\vec{p}_g - x \frac{\Vec{p}}{2}\right)^2     \frac{  \Vec{r}^{\, \prime }\cdot \Vec{r} }{|\vec{r}^{\, \prime} | |\vec{r} \, | } \Bar{Q}_{R_2} K_1(\Bar{Q}_{R_2}|\vec{r} \, |) \Bar{Q}_{R_2} K_1(\Bar{Q}_{R_2}|\vec{r}^{\, \prime} |)  \right.  \left[    \left(1 + \frac{x}{2}\right)^2 x_g^2   + \left( \frac{x}{2} \right)^2   \right] \\
& -  i 2  x^2   \left(\vec{p}_g - x \frac{\Vec{p}}{2}\right)  \cdot   \frac{ \Vec{r}}{|\vec{r} \, | } \Bar{Q}_{R_2} K_1(\Bar{Q}_{R_2}|\vec{r} \, |) K_0(\Bar{Q}_{R_2}|\vec{r}^{\, \prime} |)    \left[  \left(1 + \frac{x}{2}\right)x_g   + \frac{x}{2}  \right]     \\ 
& + 8 \left(\vec{p}_g - x \frac{\Vec{p}}{2}\right)^2 K_0(\Bar{Q}_{R_2}|\vec{r} \, |)     K_0(\Bar{Q}_{R_2}|\vec{r}^{\, \prime} |)   \left[   \left(1 + \frac{x}{2}\right)^2 + \left( \frac{x}{2} \right)^2  \right] \\
& +  i2 x^2  \left(\vec{p}_g - x \frac{\Vec{p}}{2}\right) \cdot   \frac{ \Vec{r}^{\, \prime }}{|\vec{r}^{\, \prime} | } K_0(\Bar{Q}_{R_2}|\vec{r} \, |)  \Bar{Q}_{R_2} K_1(\Bar{Q}_{R_2}|\vec{r}^{\, \prime} |)    \left[ 1  -  x_g    \right] \\
&- i  2 x^2  \left(\vec{p}_g - x \frac{\Vec{p}}{2}\right) \cdot    \frac{ \Vec{r}}{|\vec{r} \, | } K_0(\Bar{Q}_{R_2}|\vec{r}^{\, \prime} |) \Bar{Q}_{R_2} K_1(\Bar{Q}_{R_2}|\vec{r} \, |)  \left[ 1  - x_g            \right] \\
& + 2  x^4  \frac{ \Vec{r} \cdot \Vec{r}^{\, \prime }}{|\vec{r} \, |  |\vec{r}^{\, \prime} | } \Bar{Q}_{R_2} K_1(\Bar{Q}_{R_2}|\vec{r} \, |)  \Bar{Q}_{R_2} K_1(\Bar{Q}_{R_2}|\vec{r}^{\, \prime} |)  \left[  x_g^2       +  1    \right] \\ 
&  + i 2   x^2     \left(\vec{p}_g - x \frac{\Vec{p}}{2}\right)    \cdot \frac{  \Vec{r}^{\, \prime }}{|\vec{r}^{\, \prime} | } \Bar{Q}_{R_2} K_1(\Bar{Q}_{R_2}|\vec{r}^{\, \prime} |) K_0(\Bar{Q}_{R_2}|\vec{r} \, |) \left[  \left(1 + \frac{x}{2}\right)x_g  + \frac{x}{2}   \right]  \\
&   +  2   x^4  m_Q^2 K_0(\Bar{Q}_{R_2}|\vec{r} \, |)  K_0(\Bar{Q}_{R_2}|\vec{r}^{\, \prime} |)   \Bigg\}  \,, \numberthis   
\end{align*}
for the $\kappa={}^{3} S_1$ state.

\paragraph{$R_1 R_2$ contribution}
\begin{align}
    \Gamma_{R_1R_2}^{\lambda= \pm 1, \kappa} =  \Gamma_{R_1insR_2reg}^{\lambda= \pm 1, \kappa} + \Gamma_{R_1regR_2reg}^{\lambda= \pm 1, \kappa} \,,
\end{align}
where
\begin{align*}
& \sum_{\lambda = \pm 1}\Gamma_{R_1instR_2}^{\lambda = \pm 1, {}^1S_0^{[c] }}\left(p,p_g,Q; \vec{r} ,\vec{b} ,\vec{z}  , \vec{r}^{\, \prime}\right)  \\
& =  - \frac{1}{m_Q} \frac{ x_g(1 - x_g )^2 }{ (1 + x_g )}    \frac{g^2 (p^+)^2}{\pi}  \frac{\Bar{Q}_{R_1} K_1(\Bar{Q}_{R_1}X_{R_1}) }{X_{R_1}}  \frac{1}{\left[\left(\vec{p}_g - x\frac{\Vec{p}}{2} \right)^2 + x^2 m_Q^2 \right]}  \\
& \times     \left\{ - i  \left(\vec{p}_g - x \frac{\Vec{p}}{2}\right) \cdot   \frac{  \Vec{r}^{\, \prime}}{|\vec{r}^{\, \prime}| } \Bar{Q}_{R_2} K_1(\Bar{Q}_{R_2}|\vec{r}^{\, \prime}|)   \left[\left(1 + \frac{x}{2}\right)    + \frac{x}{2} x_g    \right]  +  x^2 m_Q^2 K_0(\Bar{Q}_{R_2}|\vec{r}^{\, \prime}|)    \right\} \,, \numberthis \\
& \sum_{\lambda = \pm 1}\Gamma_{R_1regR_2}^{\lambda = \pm 1, {}^1S_0^{[c] }}\left(p,p_g,Q; \vec{r} ,\vec{b} ,\vec{z}  , \vec{r}^{\, \prime}\right)  \\
& = -   \frac{2}{m_Q}    \frac{1}{ x_g (1 -x_g) }  \frac{g^2 (p^+)^2 }{\pi}  \frac{1}{\left[\left(\vec{p}_g - x\frac{\Vec{p}}{2} \right)^2 + x^2 m_Q^2 \right]}  \\
& \times \left\{ 2i   \Bar{Q}_{R_2} K_1(\Bar{Q}_{R_2}|\vec{r}^{\, \prime}|) I_{2mj}\left(\Vec{y}- \Vec{x}, \Vec{z}- \vec{x}\,\right) \left(\vec{p}_g - x \frac{\Vec{p}}{2}\right)_r  \frac{ \Vec{r}^{\, \prime}_l}{|\vec{r}^{\, \prime}| } \right. \\ 
& \times \left[  \left(1 + \frac{x}{2}  + x_g^3  \frac{x}{2}\right) \delta^{rm}   \delta^{jl} +  \left( x_g^2 \left(1 + \frac{x}{2}\right) + \frac{x}{2} x_g  \right)
\left(\delta^{rl}\delta^{mj} - \delta^{rj} \delta^{ml}\right)  
\right] \\ 
& - x^2 \left(   1 + x_g^2    \right) m_Q^2 K_0(\Bar{Q}_{R_2}|\vec{r}^{\, \prime}|)  I_{2mj}\left(\Vec{y}- \Vec{x}, \Vec{z}- \vec{x}\,\right) \delta^{jm}      \\
& -i \frac{4 x_g^2 }{(1 + x_g )} m_Q^2 \Bar{Q}_{R_2} K_1(\Bar{Q}_{R_2}|\vec{r}^{\, \prime}|)   I_0\left(\Vec{y}- \Vec{x}, \Vec{z}- \Vec{x} \, \right)   \left(\vec{p}_g - x \frac{\Vec{p}}{2}\right) \cdot \frac{ \Vec{r}^{\, \prime}}{|\vec{r}^{\, \prime}| }   \left[ \left(1 + \frac{x}{2}\right)    + \frac{x}{2} x_g    \right] \\ 
& \left. +  \frac{4 x_g^2 x^2 }{(1 + x_g )} m_Q^4   K_0(\Bar{Q}_{R_2}|\vec{r}^{\, \prime}|)  I_0(\Vec{y}- \Vec{x}, \Vec{z}- \Vec{x} )  \right\} \,, \numberthis 
\end{align*}
for the $\kappa={}^{1} S_0$ state, and 
\begin{align*}
& \sum_{\lambda = \pm 1}\Gamma_{R_1instR_2}^{\lambda = \pm 1, {}^3S_1^{[c] }}\left(p,p_g,Q; \vec{r} ,\vec{b} ,\vec{z}  , \vec{r}^{\, \prime}\right)  \\
& =  \frac{1 }{3m_Q}  \frac{ x_g(1 - x_g )^2 }{ (1 + x_g )}    \frac{g^2 (p^+)^2}{\pi}  \frac{\Bar{Q}_{R_1} K_1(\Bar{Q}_{R_1}X_{R_1}) }{X_{R_1}} \frac{1}{\left[\left(\vec{p}_g - x\frac{\Vec{p}}{2} \right)^2 + x^2 m_Q^2 \right]}    \\
& \times       \left\{- i  \left(\vec{p}_g - x \frac{\Vec{p}}{2}\right)\cdot    \frac{ \Vec{r}^{\, \prime}}{|\vec{r}^{\, \prime}| } \Bar{Q}_{R_2} K_1(\Bar{Q}_{R_2}|\vec{r}^{\, \prime}|)   \left[  \left(1 + \frac{x}{2}\right)x_g   + \frac{x}{2} \right]  -  x^2 m_Q^2 K_0(\Bar{Q}_{R_2}|\vec{r}^{\, \prime}|)   \right\}  \numberthis \\
& \sum_{\lambda = \pm 1}\Gamma_{R_1regR_2}^{\lambda = \pm 1, {}^1S_0^{[c] }}\left(p,p_g,Q; \vec{r} ,\vec{b} ,\vec{z}  , \vec{r}^{\, \prime}\right)  \\
& = -   \frac{2}{m_Q}    \frac{1}{ x_g (1 -x_g) }  \frac{g^2 (p^+)^2 }{\pi}  \frac{1}{\left[\left(\vec{p}_g - x\frac{\Vec{p}}{2} \right)^2 + x^2 m_Q^2 \right]}  \\
& \times \left\{ 2i   \Bar{Q}_{R_2} K_1(\Bar{Q}_{R_2}|\vec{r}^{\, \prime}|) I_{2mj}\left(\Vec{y}- \Vec{x}, \Vec{z}- \vec{x}\,\right) \left(\vec{p}_g - x \frac{\Vec{p}}{2}\right)_r  \frac{ \Vec{r}^{\, \prime}_l}{|\vec{r}^{\, \prime}| } \right. \\ 
& \times \left[  \left(1 + \frac{x}{2}  + x_g^3  \frac{x}{2}\right) \delta^{rm}   \delta^{jl} +  \left( x_g^2 \left(1 + \frac{x}{2}\right) + \frac{x}{2} x_g  \right)
\left(\delta^{rl}\delta^{mj} - \delta^{rj} \delta^{ml}\right)  
\right] \\ 
& - x^2 \left(   1 + x_g^2    \right) m_Q^2 K_0(\Bar{Q}_{R_2}|\vec{r}^{\, \prime}|)  I_{2mj}\left(\Vec{y}- \Vec{x}, \Vec{z}- \vec{x}\,\right) \delta^{jm}      \\
& -i \frac{4 x_g^2 }{(1 + x_g )} m_Q^2 \Bar{Q}_{R_2} K_1(\Bar{Q}_{R_2}|\vec{r}^{\, \prime}|)   I_0\left(\Vec{y}- \Vec{x}, \Vec{z}- \Vec{x} \, \right)   \left(\vec{p}_g - x \frac{\Vec{p}}{2}\right) \cdot \frac{ \Vec{r}^{\, \prime}}{|\vec{r}^{\, \prime}| }   \left[ \left(1 + \frac{x}{2}\right)    + \frac{x}{2} x_g    \right] \\ 
& \left. +  \frac{4 x_g^2 x^2 }{(1 + x_g )} m_Q^4   K_0(\Bar{Q}_{R_2}|\vec{r}^{\, \prime}|)  I_0(\Vec{y}- \Vec{x}, \Vec{z}- \Vec{x} )  \right\} \numberthis \\
& \sum_{\lambda = \pm 1}\Gamma_{R_1regR_2}^{\lambda = \pm 1, {}^3S_1^{[c] }}\left(p,p_g,Q; \vec{r} ,\vec{b} ,\vec{z}  , \vec{r}^{\, \prime}\right)  \\
& =  -  \frac{2m_Q }{3}   \frac{1}{ x_g (1 -x_g) }  \frac{g^2 (p^+)^2}{\pi} \frac{1}{\left[\left(\vec{p}_g - x\frac{\Vec{p}}{2} \right)^2 + x^2 m_Q^2 \right]}  \\
& \times \left\{ - 2 x_g \Bar{Q}_{R_2} K_1(\Bar{Q}_{R_2}|\vec{r}^{\, \prime}|)   I_{2mj}\left(\Vec{y}- \Vec{x}, \Vec{z}- \vec{x}\,\right) \left(\vec{p}_g - x \frac{\Vec{p}}{2}\right)_r   \frac{-i \Vec{r}^{\, \prime}_l}{|\vec{r}^{\, \prime}| }    \frac{1}{ m_Q^2}  
\left[x_g \left(1 + \frac{x}{2}\right) +  \frac{x}{2} \right] \right. \\
& \times \left( \delta^{rm}   \delta^{lj}   + \delta^{lr} \delta^{jm}  -  \delta^{lm}   \delta^{rj}  \right)  +   2 x_g x^2  K_0(\Bar{Q}_{R_2}|\vec{r}^{\, \prime}|)  I_{2mj}\left(\Vec{y}- \Vec{x}, \Vec{z}- \vec{x}\,\right)   \delta^{jm}  \\ 
& +   4 K_0(\Bar{Q}_{R_2}|\vec{r}^{\, \prime}|)  I_{1}\left(\Vec{y}- \Vec{x}, \Vec{z}- \Vec{x}\,\right)    \cdot \left(\vec{p}_g - x \frac{\Vec{p}}{2}\right)     \left[ \left(1 + \frac{x}{2}\right)    +  x_g \frac{x}{2}    \right] \\
& + i  x^2 \left( 1-x_g \right)^2  \Bar{Q}_{R_2} K_1(\Bar{Q}_{R_2}|\vec{r}^{\, \prime}|)    I_{1}\left(\Vec{y}- \Vec{x}, \Vec{z}- \Vec{x}\,\right)  \cdot  \frac{ \Vec{r}^{\, \prime}}{|\vec{r}^{\, \prime}| }  \\ 
&   + \frac{4 x_g^2 \left(1 - x_g      \right)  }{(1 + x_g )} K_0(\Bar{Q}_{R_2}|\vec{r}^{\, \prime}|) \Tilde{I}_{1}\left(\Vec{y}- \Vec{x}, \Vec{z}- \Vec{x}\,\right)  \cdot \left(\vec{p}_g - x \frac{\Vec{p}}{2}\right)    \\
&   + i  \frac{4 x_g^2 x^2  \left(x_g^2  + 1   \right) }{(1 + x_g )} \Bar{Q}_{R_2} K_1(\Bar{Q}_{R_2}|\vec{r}^{\, \prime}|) \Tilde{I}_{1}\left(\Vec{y}- \Vec{x}, \Vec{z}- \Vec{x}\,\right) \cdot \frac{ \Vec{r}^{\, \prime}}{|\vec{r}^{\, \prime}| }     \\ 
&   + i  \frac{4 x_g^2 }{(1 + x_g )} \Bar{Q}_{R_2} K_1(\Bar{Q}_{R_2}|\vec{r}^{\, \prime}|)    I_0\left(\Vec{y}- \Vec{x}, \Vec{z}- \Vec{x}\, \right)     \left(\vec{p}_g - x \frac{\Vec{p}}{2}\right) \cdot    \frac{ \Vec{r}^{\, \prime}}{|\vec{r}^{\, \prime}| } \left[  x_g \left(1 + \frac{x}{2}\right)     +   \frac{x}{2}    \right] \\
&   \left. +  \frac{4 x_g^2 x^2 }{(1 + x_g )} m_Q^2  K_0(\Bar{Q}_{R_2}|\vec{r}^{\, \prime}|)  I_0\left(\Vec{y}- \Vec{x}, \Vec{z}- \Vec{x}\, \right)  \right\}  \;, .  \numberthis 
\end{align*}
for the $\kappa={}^{3} S_1$ state.

\section{Summary and Outlook}
\label{sec:outlook}

In this work, we performed the first computation of the short-distance coefficients for direct quarkonium + gluon production in electron-nucleus collisions at small $x$ within the joint CGC + NRQCD framework. We study both color octet and singlet contributions and focus on the $S$-wave states. We revisited the computation of $Q\bar{Q}g$ production in the CGC within covariant perturbation theory with momentum space Feynman rules, and CGC effective vertices which resum coherent multiple interactions of the heavy quark pair and the gluon with the gluon background field of the nucleus to all orders. Then, we projected the amplitudes to the specific quantum state $\kappa$ for the heavy quark pair. 

Our final results for the short distance coefficients for the differential cross-section are presented in Sec.\,\ref{sec:cross-section}. They are expressed as a sum over 16 contributions $\der \sigma_{R_i R_j, \kappa}^{\lambda}$ corresponding to all possible locations for the emission of the gluon. We show the results for 6 of these contributions, as the others can be easily obtained by quark-antiquark exchange or complex conjugation. Each of these contributions is written as a convolution of perturbatively calculable impact factors $\Gamma_{R_i R_j, \kappa}^{\lambda}$ with multi-point lightlike
Wilson line correlators $\Xi_{R_i R_j, \kappa}$. The color correlators are collected in Sec.\,\ref{sec:color_correlator_summary} for both singlet and octet states. The perturbative factors depend both on the polarization of the virtual photon and the spin state $\kappa$ and are presented in Sec.\,\ref{sec:perturbative_factor_summary}. The calculation for the $P$-wave can be carried out in a similar fashion, and it will be presented in an upcoming paper. 

In the future, we plan to implement the fragmentation of the gluon into a jet (hadron), to study the azimuthal correlations of direct quarkonium + jet (hadron) at the future Electron-Ion Collider. It would be interesting to compare our results with those obtained within the TMD formalism \cite{DAlesio:2019qpk}. In particular, we expect a matching between both formalisms in the limit in which the transverse momentum imbalance between the quarkonium and the jet is small compared to the invariant mass of the quarkonium-jet system, as it has been observed to occur for other processes \cite{Dominguez:2011wm,Xiao:2017yya,Taels:2022tza,Caucal:2022ulg,Caucal:2023nci,Caucal:2023fsf}. 

Furthermore, by integrating over the phase space of the gluon, we can obtain the next-to-leading order real corrections for direct quarkonium production in DIS at small-$x$. Last, we could focus on the regime where the final state gluon is soft and derive expressions for the associated Sudakov logarithms for direct quarkonium production in the limit of small transverse momentum \cite{Mueller:2012uf,Qiu:2013qka,Hatta:2021jcd}. 

\section*{Acknowledgements}
We are grateful to Philip Velie for valuable discussions and contributions at the early stages of this work.
F.S. is supported by the National Science Foundation under grant No. PHY-1945471, the UC Southern California Hub, with funding from the UC National Laboratories division of the University of California Office of the President, and by the U.S. Department of Energy under Contract No. DE-AC02-05CH11231, by NSF under Grant No. OAC-2004571 within the X-SCAPE Collaboration. This work is also supported by the U.S. Department of Energy, Office of Science, Office of Nuclear Physics, within the framework of the Saturated Glue (SURGE) Topical Theory Collaboration.

\newpage

\appendix

\section{Feynman rules}
\label{sec:Feynmanrules}

In this section, we gather the elementary Feynman rules needed for our computation. We remind the reader that we work in light-cone gauge $A^+=0$. We label $\sigma,\sigma'$ the spinor indices and $i,j$, $a,b$ the color indices in the $SU(N_c)$ fundamental and adjoint representations respectively. 

The free massive quark and gluon propagator are 
\begin{align*}
S_0^{\sigma \sigma', ij}(l) & = \frac{i(\slashed{l}+ m)^{\sigma \sigma'}}{l^2 - m^2 + i \varepsilon } \delta^{ij} \numberthis \,, \\ 
G^{\mu \nu , ab}_0 (l) & = \frac{i}{l^2 + i \varepsilon} \Pi^{\mu \nu }(l) \delta^{ab} \,, \numberthis 
\end{align*}
where  
\begin{equation}
\Pi^{\mu \nu }(l)  = - g_{\mu \nu} + \frac{l^\mu n_2^\nu + l^\nu n_2^\mu}{l^+} \,,
\end{equation}
is the gluon tensor. It has the following properties: 
\begin{align}
\Pi_{\alpha \beta}(l) & = \sum_{\bar{\lambda} = \pm 1} \varepsilon_\alpha(l,\bar{\lambda})  \varepsilon^*_\beta(l,\bar{\lambda}) + \frac{l^2}{(l^+)^2} n_{2 \alpha} n_{2 \beta} \,, \\ 
\Pi_{\alpha \beta}(l) \Pi^{\beta \delta}(l') & = - \sum_{\bar{\lambda}=\pm 1} \varepsilon_\alpha(l,\bar{\lambda}) \varepsilon^{*\delta}(l',\bar{\lambda}) \,, \\ 
\varepsilon_\beta^*(l,\Bar{\lambda}) \Pi^{\beta \delta}(l') & = - \varepsilon^{*\delta}(l',\bar{\lambda}) \,,\label{eq:GluonPts}
\end{align}
where $\Bar{\lambda}$ is the gluon polarization. 

\section{Gamma traces}
\label{sec:GammaTrace}

This section is dedicated to collecting some useful traces of gamma matrices. Here,  $ \omega_{\mu \nu} = \frac{1}{2} \left[\gamma^\mu, \gamma^\nu\right]$. These results are used in particular to compute $\mathcal{F}_{R_i}^{\lambda, \Bar{\lambda}, \kappa, J_z}$ which is given by Eq.~\eqref{eq:Ffunction}. They are: 

\begin{align*}
\Tr(\text{odd nb of } \gamma) & = 0 \numberthis\\ 
\Tr(\gamma^\mu \gamma^\nu) & = 4 g^{\mu \nu} \numberthis\\
\Tr(\gamma^\mu \gamma^\nu \gamma^\rho \gamma^\sigma) &= 4(g^{\mu \nu} g^{\rho \sigma} - g^{\mu \rho} g^{\nu \sigma} + g^{\mu \sigma }g^{\nu \rho}) \numberthis \\  
\Tr(\gamma^\alpha \gamma^\beta \gamma^\mu \gamma^\nu \gamma^\rho \gamma^\sigma) & = 4g^{\alpha \beta}(g^{\mu \nu}g^{\rho \sigma} - g^{\mu \rho}g^{\nu \sigma} + g^{\mu \sigma} g^{\nu \rho}) - 4 g^{\alpha \mu} (g^{\beta \nu}g^{\rho \sigma} - g^{\beta \rho}g^{\nu \sigma} + g^{\beta \sigma} g^{\nu \rho}) \\ 
& + 4g^{\alpha \nu} (g^{\beta \mu}g^{\rho \sigma} - g^{\beta \rho}g^{\mu \sigma} + g^{\beta \sigma} g^{\mu \rho}) -4g^{\alpha \rho } (g^{\beta \mu}g^{\nu \sigma} - g^{\beta \nu}g^{\mu \sigma} + g^{\beta \sigma} g^{\mu \nu}) \\ 
& + 4g^{\alpha \sigma} (g^{\beta \mu}g^{\nu \rho} - g^{\beta \nu}g^{\mu \rho} + g^{\beta \rho} g^{\mu \nu}) \numberthis \\
\Tr\left(\gamma^5 \gamma^\mu \gamma^\nu\right) & = 0 \numberthis \\ 
\Tr \left( \gamma^5 \gamma^\mu \gamma^\nu \gamma^\alpha \gamma^\beta \right) & = - 4 i \epsilon^{\mu \nu \alpha \beta } \numberthis \\ 
\Tr \left(\gamma^5 \gamma^\alpha \omega^{ij}\gamma^\beta \right) & = \frac{1}{2} \left[  \Tr \left(\gamma^5 \gamma^\alpha \gamma^i \gamma^j \gamma^+ \right) -  \Tr \left(\gamma^5 \gamma^\alpha \gamma^j \gamma^i \gamma^+ \right)\right] \nonumber \\ 
& =  \frac{1}{2} \left[ - 4 i \epsilon^{\alpha i j \beta } - (- 4 i \epsilon^{\alpha j i  \beta }) \right] \nonumber \\
& = - 4 i \epsilon^{\alpha i j \beta } \numberthis \\
\Tr(\gamma^\alpha \gamma^\beta \omega^{ij}) &= \frac{1}{2}\left[\Tr(\gamma^\alpha \gamma^\beta \gamma^i \gamma^j)-\Tr(\gamma^\alpha \gamma^\beta \gamma^j \gamma^i)\right] \\ 
& = \frac{1}{2}\left[4(g^{\alpha \beta} g^{ij}-g^{\alpha i}g^{\beta j}+ g^{\alpha j}g^{\beta i} ) -4(g^{\alpha \beta} g^{ij}-g^{\alpha j}g^{\beta i}+ g^{\alpha i}g^{\beta j} )\right] \\
& = 4(g^{\alpha j}g^{\beta i} -g^{\alpha i}g^{\beta j} ) \numberthis 
\end{align*}
Hence, for the last trace, if either $\alpha$ or $\beta$ is $+,-$, the trace is equal to 0.

\section{Dirac structure manipulation for gluon emission }
\label{sec:Dirac-identities}
We use the Ward identity $l_2 \cdot \slashed{\varepsilon}^* (l_2,\Bar{\lambda}) $, Eq.~\eqref{eq:polarization_vector_gluonTransverse}, the properties of the Gamma matrices and the Dirac equations $\Bar{u}_s(p_q)(\slashed{p}_q - m_Q) = 0$ and $(\slashed{p}_{\bar{q}} + m_Q)v_{\bar{s}}(p_{\Bar{q}})$ in the below manipulations. 

\paragraph{Emission of a gluon by the quark before the shockwave }

\begin{align*}
& \gamma^+ \left[(\slashed{l}_1- \slashed{l}_2) + m_Q\right] \slashed{\varepsilon}^* (l_2,\Bar{\lambda})(\slashed{l}_1 + m_Q) \\
& = \gamma^+ \left[ 2(l_1 - l_2) \cdot \varepsilon^*(l_2,\Bar{\lambda}) - \slashed{\varepsilon}^*(l_2,\Bar{\lambda}) (\slashed{l}_1- \slashed{l}_2) + m_Q \slashed{\varepsilon}^*(l_2,\Bar{\lambda})\right] (\slashed{l}_1 + m_Q) \\
& = \gamma^+ \left[ 2 \left( l_1^+ \frac{\Vec{l}_2 \cdot \Vec{\epsilon}^{\; \Bar{\lambda} *}}{l_2^+ } - \Vec{l}_1 \cdot \Vec{\epsilon}^{\; \Bar{\lambda}*}\right) + \Vec{\epsilon}^{\; \Bar{\lambda}*}_n \gamma^n (\slashed{l}_1- \slashed{l}_2) - m_Q \Vec{\epsilon}^{\; \Bar{\lambda}*}_n \gamma^n \right](\slashed{l}_1 + m_Q) \\
& = \gamma^+ \left[ 2 \frac{l_1^+}{l_2^+}\left( \Vec{l}_2 - \frac{l_2^+}{l_1^+} \Vec{l}_1  \right)\cdot \Vec{\epsilon}^{\; \Bar{\lambda}*} + \Vec{\epsilon}^{\; \Bar{\lambda}*}_n \gamma^n \left( \Vec{l}_2 - \frac{l_2^+ }{l_1^+} \Vec{l}_1 \right)_m \gamma^m + \Vec{\epsilon}^{\; \Bar{\lambda}*}_n \gamma^n \left(1 - \frac{l_2^+}{l_1^+} \right) (\slashed{l}_1 - m_Q + m_Q ) \right. \\
& \left. - m_Q \Vec{\epsilon}^{\; \Bar{\lambda}*}_n \gamma^n     \right] (\slashed{l}_1 + m_Q) \\ 
& = \gamma^+ \left[2 \frac{l_1^+ }{l_2^+} \left( \Vec{l}_2 - \frac{l_2^+ }{l_1^+} \Vec{l}_1 \right)_m \Vec{\epsilon}^{\; \Bar{\lambda}*}_n \left[\delta^{nm} + \frac{l_2^+}{2 l_1^+ } \gamma^n \gamma^m \right] + \Vec{\epsilon}^{\; \Bar{ \lambda}*}_n \gamma^n \left( 1 - \frac{l_2^+ }{l_1^+} \right)(\slashed{l}_1 - m_Q) \right. \\
& \left. - \frac{l_2^+ }{l_1^+} m_Q \Vec{\epsilon}^{\; \Bar{\lambda}*}_n \gamma^n  \right] (\slashed{l}_1 + m_Q) \numberthis[Dirac_Gluon_Q_before]
\end{align*}

\paragraph{Emission of a gluon by the quark after the shockwave  }

\begin{align*}
& \Bar{u}_s(p_q) \slashed{\varepsilon}^*(p_g,\Bar{\lambda})(\slashed{p}_q+ \slashed{p}_g+ m_Q) \gamma^+ \\ 
& = \Bar{u}_s(p_q)  \left[ 2 \varepsilon^*(p_g,\Bar{\lambda})\cdot (p_q + p_g) - (\slashed{p}_q + \slashed{p}_g) \slashed{\varepsilon}^*(p_g,\Bar{\lambda}) + m_Q \slashed{\varepsilon}^*(p_g,\Bar{\lambda})  \right] \gamma^+ \\ 
& = \Bar{u}_s(p_q) \left[2\left(\frac{\vec{p}_g \cdot \Vec{\epsilon}^{\; \Bar{\lambda}*}}{p_g^+}p_q^+ - \Vec{\epsilon}^{\; \Bar{\lambda}*} \cdot \Vec{p}_q\right) - (\slashed{p}_q - m_Q)  \slashed{\varepsilon}^*(p_g,\Bar{\lambda}) - \slashed{p}_g \slashed{\varepsilon}^*(p_g,\Bar{\lambda})\right] \gamma^+ \\ 
& = \Bar{u}_s(p_q) \left[\frac{2}{x} \left(\vec{p}_g - x\Vec{p}_q \right) \cdot \Vec{\epsilon}^{\; \Bar{\lambda}*} + \left(\slashed{p}_g- x\slashed{p}_q + x \slashed{p}_q \right) \Vec{\epsilon}^{\; \Bar{\lambda}*}_n \gamma^n\right] \gamma^+ \\ 
& = \Bar{u}_s(p_q) \left[ \frac{2}{x} \left(\vec{p}_g - x\Vec{p}_q \right) \cdot \Vec{\epsilon}^{\; \Bar{\lambda}*} - (\vec{p}_g - x \Vec{p}_q)_m \gamma^m \Vec{\epsilon}^{\; \Bar{\lambda}*}_n \gamma^n + x m_Q \Vec{\epsilon}^{\; \Bar{\lambda}*}_n   \gamma^n   \right] \gamma^+ \\ 
& = \Bar{u}_s(p_q) \Vec{\epsilon}^{\; \Bar{\lambda}*}_n  \left[  \frac{2}{x} (\vec{p}_g - x \Vec{p}_q)_m \left( \delta^{mn} - \frac{x}{2} \gamma^m \gamma^n \right) + xm_Q \gamma^n \right] \gamma^+ \\ 
& = \Bar{u}_s(p_q) \Vec{\epsilon}^{\; \Bar{\lambda}*}_n  \left\{  \frac{2}{x} (\vec{p}_g - x \Vec{p}_q)_m \left[ \delta^{mn}\left(1 + \frac{x}{2}\right) + \frac{x}{2} \omega^{nm} \right] + xm_Q \gamma^n \right\} \gamma^+ \numberthis[Dirac_Gluon_Q_after]
\end{align*}
We have defined $x = p_g^+ /p_q^+$. 

\paragraph{Emission of a gluon by the antiquark before the shockwave}

\begin{align*}
& \left[ - \slashed{l}_1 + m_Q\right]  \slashed{\varepsilon}^*(l_2,\bar{\lambda}) \left[ (\slashed{l}_2 - \slashed{l}_1 )+ m_Q \right]   \gamma^+ \\
& = \left[ - \slashed{l}_1 + m_Q\right]  \left[2 \varepsilon^*(l_2,\bar{\lambda}) \cdot (l_2 - l_1) - (\slashed{l}_2 - \slashed{l}_1) \slashed{\varepsilon}^*(l_2,\Bar{\lambda}) + m_Q \slashed{\varepsilon}^*(l_2,\bar{\lambda})  \right] \gamma^+ \\
& = \left[ - \slashed{l}_1 + m_Q\right]  \left[- 2 \left(\frac{\Vec{\epsilon}^{\; \bar{\lambda}*} \cdot \Vec{l}_2}{l_2^+} l_1^+ - \Vec{\epsilon}^{\; \Bar{\lambda}*} \cdot \Vec{l}_1\right) + (\slashed{l}_2 - \slashed{l}_1) \Vec{\epsilon}^{\; \Bar{\lambda}*}_n \gamma^n   - m_Q \Vec{\epsilon}^{\; \Bar{\lambda}*}_n \gamma^n  \right] \gamma^+ \\
& = \left[ - \slashed{l}_1 + m_Q\right]  \left[ - 2 \frac{l_1^+}{l_2^+} \left(\Vec{l}_2 - \frac{l_2^+}{l_1^+} \Vec{l}_1\right)\cdot \Vec{\epsilon}^{\; \bar{\lambda}*} + \left(\slashed{l}_2 - \frac{l_2^+}{l_1^+} \slashed{l}_1 +  \frac{l_2^+}{l_1^+} \slashed{l}_1   - \slashed{l}_1 \right) \Vec{\epsilon}^{\; \Bar{\lambda}*}_n \gamma^n   - m_Q \Vec{\epsilon}^{\; \Bar{\lambda}*}_n \gamma^n  \right]\gamma^+ \\
& = \left[ - \slashed{l}_1 + m_Q\right]  \left[ - 2 \frac{l_1^+}{l_2^+} \left(\Vec{l}_2 - \frac{l_2^+}{l_1^+} \Vec{l}_1\right)\cdot \Vec{\epsilon}^{\; \bar{\lambda}*} - \left(\Vec{l}_2- \frac{l_2^+}{l_1^+} \Vec{l}_1 \right)_m \gamma^m   \Vec{\epsilon}^{\; \Bar{\lambda}*}_n \gamma^n \right. \\
& \left. +  \left( \frac{l_2^+}{l_1^+} - 1 \right) \left( \slashed{l}_1  + m_Q - m_Q \right)    \Vec{\epsilon}^{\; \Bar{\lambda}*}_n \gamma^n   - m_Q \Vec{\epsilon}^{\; \Bar{\lambda}*}_n \gamma^n  \right]\gamma^+ \\
& = \left[ - \slashed{l}_1 + m_Q\right]  \left\{  - 2 \frac{l_1^+}{l_2^+} 
\left(\Vec{l}_2- \frac{l_2^+}{l_1^+} \Vec{l}_1 \right)_m  \Vec{\epsilon}^{\; \Bar{\lambda}*}_n  \left[\delta^{nm} + \frac{l_2^+}{2l_1^+} \gamma^m \gamma^n \right] - \frac{l_2^+}{l_1^+} m_Q  \Vec{\epsilon}^{\; \Bar{\lambda}*}_n \gamma^n     \right. \\
& \left.  + \left( \frac{l_2^+}{l_1^+} - 1 \right) \left(\slashed{l}_1 + m_Q\right) \Vec{\epsilon}^{\; \Bar{\lambda}*}_n \gamma^n  \right\} \gamma^+ \\
& = \left[ - \slashed{l}_1 + m_Q\right]  \left\{  - 2 \frac{l_1^+}{l_2^+} 
\left(\Vec{l}_2- \frac{l_2^+}{l_1^+} \Vec{l}_1 \right)_m  \Vec{\epsilon}^{\; \Bar{\lambda}*}_n  \left[\delta^{nm} \left(1 - \frac{l_2^+}{2l_1^+} \right) + \frac{l_2^+}{2l_1^+} \omega^{mn} \right] - \frac{l_2^+}{l_1^+} m_Q  \Vec{\epsilon}^{\; \Bar{\lambda}*}_n \gamma^n    \right\} \gamma^+ \\
& + \left(1 - \frac{l_2^+}{l_1^+} \right)(l_1^2 - m_Q^2) \Vec{\epsilon}^{\; \Bar{\lambda}*}_n \gamma^n \gamma^+ \\
& = \left[ - \slashed{l}_1 + m_Q\right]  \left\{  - 2 \frac{p_g^+ + p_{\bar{q}}^+}{p_g^+} \left(\Vec{l}_2- \frac{p_g^+}{q^+ - p_q^+} \Vec{l}_1 \right)_m  \Vec{\epsilon}^{\; \Bar{\lambda}*}_n  \left[\delta^{nm} \left(\frac{p_g^+ + 2 p_{\bar{q}}^+ }{2(p_g^+ + p_{\bar{q}}^+)} \right) + \frac{p_g^+}{2(p_g^+ + p_{\bar{q}}^+)} \omega^{mn} \right] \right. \\
& \left. - \frac{p_g^+}{p_g^+ + p_{\bar{q}}^+} m_Q  \Vec{\epsilon}^{\; \Bar{\lambda}*}_n \gamma^n    \right\} \gamma^+ + \left(1 - \frac{p_g^+}{p_g^+ + p_{\bar{q}}^+ } \right)(l_1^2 - m_Q^2) \Vec{\epsilon}^{\; \Bar{\lambda}*}_n \gamma^n \gamma^+ \\ 
& = \left[ - \slashed{l}_1 + m_Q\right]  \left\{  - 2 \frac{p_{\bar{q}}^+}{p_g^+} \left(\Vec{l}_2- \frac{x_g}{1-x_q} \Vec{l}_1 \right)_m  \Vec{\epsilon}^{\; \Bar{\lambda}*}_n  \left[\delta^{nm} \left(\frac{p_g^+ + 2 p_{\bar{q}}^+ }{2p_{\bar{q}}^+} \right) + \frac{p_g^+}{2p_{\bar{q}}^+} \omega^{mn} \right] \right. \\
& \left. - \frac{p_{\bar{q}}^+}{p_g^+ + p_{\bar{q}}^+} \frac{p_g^+}{p_{\Bar{q}}^+} m_Q  \Vec{\epsilon}^{\; \Bar{\lambda}*}_n \gamma^n    \right\} \gamma^+  +  \frac{p_{\Bar{q}}^+}{q^+ - p_q^+}(l_1^2 - m_Q^2) \Vec{\epsilon}^{\; \Bar{\lambda}*}_n \gamma^n \gamma^+ \\ 
& = 2 \frac{p_{\bar{q}}^+}{p_g^+} \Vec{\epsilon}^{\; \Bar{\lambda}*}_n  \left[ - \slashed{l}_1 + m_Q\right]  \gamma^+  \left\{  -  \left(\Vec{l}_2- \frac{x_g}{1-x_q} \Vec{l}_1 \right)_m   \left[\delta^{nm} \left(1 + \frac{p_g^+  }{2p_{\bar{q}}^+} \right) + \frac{p_g^+}{2p_{\bar{q}}^+} \omega^{mn} \right] \right. \\
& \left. + \frac{m_Q}{2}  \frac{p_{\bar{q}}^+}{p_g^+ + p_{\bar{q}}^+} \left( \frac{p_g^+}{p_{\Bar{q}}^+} \right)^2   \gamma^n    \right\}  +  \frac{x_{\bar{q}}}{1-x_q}(l_1^2 - m_Q^2) \Vec{\epsilon}^{\; \Bar{\lambda}*}_n \gamma^n \gamma^+ \numberthis[Dirac_Gluon_Qbar_before]
\end{align*}

\paragraph{Emission of a gluon by the antiquark after the shockwave}

\begin{align*}
   &   \gamma^+ (- \slashed{p}_g - \slashed{p}_{\Bar{q}} + m_Q) \slashed{\varepsilon}^*(p_g,\Bar{\lambda}) v_{\Bar{s}}(p_{\Bar{q}}) \\
   & = \gamma^+ \left[ 2(-p_g -p_{\Bar{q}}) \cdot \varepsilon^*(p_g,\Bar{\lambda}) - \slashed{\varepsilon}^*(p_g,\Bar{\lambda})(-\slashed{p}_g- \slashed{p}_{\Bar{q}}) + m_Q \slashed{\varepsilon}^*(p_g,\Bar{\lambda})\right] v_{\Bar{s}}(p_{\Bar{q}}) \\
   & = \gamma^+ \left[ - 2 p_{\Bar{q}} \cdot \varepsilon^*( p_g,\Bar{\lambda}) + \slashed{\varepsilon}^*(p_g,\Bar{\lambda})\slashed{p}_g+ \slashed{\varepsilon}^*(p_g,\Bar{\lambda}) (\slashed{p}_{\Bar{q}} + m_Q ) \right] v_{\Bar{s}}(p_{\Bar{q}}) \\
   & = \gamma^+ \left[ - 2 \left( p_{\Bar{q}}^+ \frac{\vec{p}_g \cdot \Vec{\epsilon}^{\; \Bar{\lambda}*}}{p_g^+} - \Vec{p}_{\Bar{q}} \cdot \Vec{\epsilon}^{\; \Bar{\lambda} *  } \right)- \Vec{\epsilon}^{\; \Bar{\lambda} *}_n \gamma^n \slashed{p}_g  \right]  v_{\Bar{s}}(p_{\Bar{q}}) \\
   & = \gamma^+ \left[ - \frac{2}{\bar{x}} (\vec{p}_g - \bar{x} \Vec{p}_{\Bar{q}}) \cdot \Vec{\epsilon}^{\; \Bar{\lambda }*} - \Vec{\epsilon}^{\; \Bar{\lambda}*}_n \gamma^n (\slashed{p}_g- \bar{x} \slashed{p}_{\Bar{q}}) - \bar{x} \Vec{\epsilon}^{\; \Bar{\lambda}*}_n \gamma^n \slashed{p}_{\Bar{q}}       \right] v_{\Bar{s}}(p_{\Bar{q}}) \\
   & = \gamma^+ \left[ - \frac{2}{\bar{x}} (\vec{p}_g - \bar{x} \Vec{p}_{\Bar{q}})_m  \Vec{\epsilon}^{\; \Bar{\lambda }*}_n \delta^{mn} + (\vec{p}_g - \bar{x} \Vec{p}_{\Bar{q}})_m \Vec{\epsilon}^{\; \Bar{\lambda}*}_n \gamma^n \gamma^m + \bar{x} m_Q \Vec{\epsilon}^{\; \Bar{\lambda}*}_n \gamma^n   \right] v_{\Bar{s}}(p_{\Bar{q}}) \\
   & = \gamma^+ \left[- \frac{2}{\bar{x}} (\vec{p}_g - \bar{x} \Vec{p}_{\Bar{q}})_m  \Vec{\epsilon}^{\; \Bar{\lambda }*}_n \left( \delta^{mn} - \frac{\bar{x}}{2}\gamma^n \gamma^m \right)  + \bar{x} m_Q \Vec{\epsilon}^{\; \Bar{\lambda}*}_n \gamma^n \right]v_{\Bar{s}}(p_{\Bar{q}}) \\ 
   & = \Vec{\epsilon}^{\; \Bar{\lambda}*}_n  \gamma^+ \left\{ - \frac{2}{\bar{x}} (\vec{p}_g - \bar{x} \Vec{p}_{\Bar{q}})_m  \left[ \delta^{mn} \left( 1 + \frac{\bar{x}}{2}\right) - \frac{\bar{x}}{2} \omega^{nm}  \right] + \bar{x} m_Q \gamma^n     \right\} v_{\Bar{s}}(p_{\Bar{q}}) \numberthis[Dirac_Gluon_Qbar_after]  \; . 
\end{align*}
where $\bar{x} = p_g^+ /p_{\Bar{q}}^+$.

\section{Lorentz contraction with $\mathbb{P}_{\alpha \rho}$}
\label{sec:Lorentz-Contractions}
Here, we list a few useful identities that will aid in the calculation of the perturbative factor defined in Eq.\,\eqref{eq:perturbative_factor} when $\kappa= {}^{3} S_1^{[c]}$. The following contractions are encountered when summing over $J_z$ and using Eq.\,\eqref{SumTensorSWave}:
\begin{align}
\left(-g_{\rho \alpha} + \frac{p_\rho p_\alpha }{p^2}\right) g^{\rho + }   g^{\alpha + }   & = \frac{(p^+)^2 }{p^2} \; , \\
\left(-g_{\rho \alpha} + \frac{p_\rho p_\alpha }{p^2}\right) g^{\rho + }    (g^{\alpha+} \Vec{p}^{\; s}- g^{\alpha s}p^+  ) & = \Vec{p}^{\; s} \frac{(p^+)^2}{p^2} - p^+ \frac{\Vec{p}^{\; s} p^+ }{p^2} = 0 \; , \\ 
\left(-g_{\rho \alpha} + \frac{p_\rho p_\alpha }{p^2}\right)  (g^{\rho+} \Vec{p}^{\; n}- g^{\rho n}p^+  )   (g^{\alpha+} \Vec{p}^{\; s}- g^{\alpha s}p^+  ) & =  (p^+)^2 (-g^{ns}) = (p^+)^2 \delta^{ns} \; . 
\label{3S1Identities}
\end{align}

\section{Transverse momentum integrals for the $R_1$ diagrams  }

\label{sec:R1TransverseIntegral}

We calculate in this appendix the transverse integrals that appear in diagrams $R_1$ and $R_{3}$.

\begin{align}
I(\vec{r}_1 , \vec{r}_2)  & = \int \frac{\der^2 \Vec{l}_1 \der^2 \Vec{l}_2}{(2\pi)^2}  \frac{e^{i \Vec{l}_1 \cdot \Vec{r}_1} e^{i \Vec{l}_2 \cdot \Vec{r}_2} }{Q^2 + \frac{\Vec{l}_1^{\; 2} + m_Q^2}{z_1} + \frac{\Vec{l}_2^{\; 2}}{z_2} + \frac{(\Vec{l}_1 + \Vec{l}_2)^2 + m_Q^2}{z_3}} \\ 
I_0(\vec{r}_1, \vec{r}_2)  & = \int \frac{\der^2 \Vec{l}_1 \der^2 \Vec{l}_2}{(2\pi)^2} \frac{e^{i \Vec{l}_1 \cdot \Vec{r}_1} e^{i \Vec{l}_2 \cdot \Vec{r}_2} }{\left[Q^2 z_1 (1 - z_1) + \Vec{l}_1^{\; 2} + m_Q^2\right] \left[Q^2 + \frac{\Vec{l}_1^{\; 2} + m_Q^2}{z_1} + \frac{\Vec{l}_2^{\; 2}}{z_2} + \frac{(\Vec{l}_1 + \Vec{l}_2)^2 + m_Q^2}{z_3}\right]}
\end{align}

\begin{align*}
I_{1m}(\vec{r}_1, \vec{r}_2)  
& =  \int \frac{\der^2 \Vec{l}_1 \der^2 \Vec{l}_2}{(2\pi)^2} \frac{ e^{i \Vec{l}_1 \cdot \Vec{r}_1} e^{i \Vec{l}_2 \cdot \Vec{r}_2}  \left(\Vec{l}_{2m} + \frac{z_2}{1-z_1}\Vec{l}_{1m}\right)}{\left[Q^2 z_1 (1 - z_1) + \Vec{l}_1^{\; 2} + m_Q^2\right] \left[Q^2 + \frac{\Vec{l}_1^{\; 2} + m_Q^2}{z_1} + \frac{\Vec{l}_2^{\; 2}}{z_2} + \frac{(\Vec{l}_1 + \Vec{l}_2)^2 + m_Q^2}{z_3}\right]}   \\
& = - i \left( \frac{\partial}{\partial \Vec{r}_{2}^{\; m}} +  \frac{z_2}{1-z_1} \frac{\partial}{\partial \Vec{r}_{1}^{\; m}} \right) I_0(\vec{r}_1, \vec{r}_2) \numberthis  
\end{align*}

\begin{align*}
I_{2mj}(\vec{r}_1, \vec{r}_2) &=  \int \frac{\der^2 \Vec{l}_1 \der^2 \vec{l}_2}{(2\pi)^2} \frac{ e^{i \Vec{l}_1 \cdot \Vec{r}_1} e^{i \Vec{l}_2 \cdot \Vec{r}_2} \; \Vec{l}_{1j} \left(\Vec{l}_{2m} + \frac{z_2}{1-z_1}\Vec{l}_{1m}\right) }{\left[Q^2 z_1 (1 - z_1) + \Vec{l}_1^{\; 2} + m_Q^2\right] \left[Q^2 + \frac{\Vec{l}_1^{\; 2} + m_Q^2}{z_1} + \frac{\Vec{l}_2^{\; 2}}{z_2} + \frac{(\Vec{l}_1 + \Vec{l}_2)^2 + m_Q^2}{z_3}\right]} \\*
& = - i \frac{\partial}{\partial\Vec{r}_{1}^{\; j}}   I_{1m}(\vec{r}_1, \vec{r}_2)  \numberthis 
\end{align*}

\begin{align*}
\Tilde{I}_{1j}(\vec{r}_1,r_{2_\perp}) & = \int \frac{\der^2 \Vec{l}_1 \der^2 \Vec{l}_2}{(2\pi)^2} \frac{e^{i \Vec{l}_1 \cdot \Vec{r}_1} e^{i \Vec{l}_2 \cdot \Vec{r}_2} \; \Vec{l}_{1j} }{\left[Q^2 z_1 (1 - z_1) + \Vec{l}_1^{\; 2} + m_Q^2\right] \left[Q^2 + \frac{\Vec{l}_1^{\; 2} + m_Q^2}{z_1} + \frac{\Vec{l}_2^{\; 2}}{z_2} + \frac{(\Vec{l}_1 + \Vec{l}_2)^2 + m_Q^2}{z_3}\right]} \\ 
& =  - i \frac{\partial}{\partial\Vec{r}_{1}^{\; j}}   I_{0}(\vec{r}_1, \vec{r}_2)  \numberthis 
\end{align*}

$z_1, z_2, z_3$ such that $ z_1 + z_2 + z_3 = 1$. 

I will use the following relation to compute them 
\begin{equation}
\int_0^{\infty} d s\; s^{\nu-1} e^{-s A^2}  e^{-B^2 / s} = 2\left(\frac{A}{B}\right)^{-\nu} K_{-\nu}\left(2AB\right) 
\end{equation}
with $A^2 > 0, B^2 > 0$.

\subsection{Schwinger’s parametrization}

Schwinger’s parametrization 
\begin{equation}
    \frac{1}{D^\beta} = \frac{1}{\Gamma(\beta)} \int_{0}^{ \infty} d s s^{\beta - 1} e^{- s D }
\end{equation}
for $\operatorname{Re}(\beta) > 0$.

\begin{align*}
& I(\vec{r}_1, \vec{r}_2)  \\ 
& =  \int_0^\infty ds \int \frac{\der^2 \Vec{l}_1 \der^2 \Vec{l}_2}{(2\pi)^2} \;  e^{i \Vec{l}_1 \cdot \Vec{r}_1} \; e^{i \Vec{l}_2 \cdot \Vec{r}_2}  \; e^{- s \left[Q^2 + \Vec{l}_2^{\; 2} \left(\frac{1}{z_2}+ \frac{1}{z_3}\right) +  \Vec{l}_1^{\; 2} \left(\frac{1}{z_1}+ \frac{1}{z_3}\right)+ m_Q^2 \left(\frac{1}{z_1}+ \frac{1}{z_3}\right) + \Vec{l}_2 \cdot \frac{2}{z_3}\Vec{l}_1\right]} \\ 
& =  \int_0^\infty d s \;  e^{- s \left(Q^2 + m_Q^2 \frac{1-z_2}{z_1 z_3}\right)} \int \frac{\der^2 \Vec{l}_1}{2\pi} \; e^{i \Vec{l}_1 \cdot \Vec{r}_1} \; e^{- \Vec{l}_1^{\; 2} \frac{s(1-z_2)}{z_1z_3}} \int \frac{\der^2 \Vec{l}_2}{2\pi} \; e^{\Vec{l}_2 \cdot \left(i\Vec{r}_2 - \frac{2s}{z_3}\Vec{l}_1\right)} \; e^{- \Vec{l}_2^{\; 2} \frac{s (1-z_1)}{z_2 z_3}} \\ 
& =  \int_0^\infty d s \;  e^{- s \left(Q^2 + m_Q^2 \frac{1-z_2}{z_1 z_3}\right)} \int \frac{\der^2 \Vec{l}_1}{2\pi} \; e^{i \Vec{l}_1 \cdot \Vec{r}_1} \; e^{- \Vec{l}_1^{\; 2} \frac{s(1-z_2)}{z_1z_3}} \frac{1}{2\pi} \left(\frac{\pi z_2 z_3}{s(1-z_1)}\right) e^{\frac{\left(i\Vec{r}_2 - \frac{2s}{z_3}\Vec{l}_1\right)^2}{4s(1-z_1)}z_2z_3} \\ 
& = \frac{  z_2z_3}{2(1-z_1) } \int_0^\infty \frac{ds}{s}  \; e^{- s \left(Q^2 + m_Q^2 \frac{1-z_2}{z_1 z_3}\right)} \; e^{- \Vec{r}_2^{\; 2} \frac{z_2 z_3}{4s(1-z_1)}} \int \frac{\der^2 \Vec{l}_1}{2\pi} \;  e^{\Vec{l}_1 \cdot \left( i \Vec{r}_1 - i \Vec{r}_2 \frac{z_2}{1-z_1}\right)} \; e^{-\Vec{l}_1^{\; 2} \frac{s}{z_1 (1-z_1)}} \\ 
& =\frac{  z_2z_3}{2 (1-z_1) } \int_0^\infty \frac{ds}{s} \; e^{- s \left(Q^2 + m_Q^2 \frac{1-z_2}{z_1 z_3}\right)} \; e^{- \Vec{r}_2^{\; 2} \frac{z_2 z_3}{4s(1-z_1)}} \frac{1}{2\pi} \left(\frac{\pi z_1(1-z_1)}{s}\right) \;  e^{- \left(\Vec{r}_1 - \frac{z_2}{1-z_1}\Vec{r}_2\right)^2 \frac{z_1(1-z_1)}{4s}} \\ 
& = \frac{z_1z_2z_3}{4} \int_0^\infty \frac{ds}{s^2} \; e^{- s\left(Q^2 + m_Q^2 \frac{1-z_2}{z_1 z_3}\right)} \; e^{- \frac{1}{4s} \left[\Vec{r}_2^{\; 2} \frac{z_2 z_3}{1-z_1} + \left(\Vec{r}_1 - \frac{z_2}{1-z_1}\Vec{r}_2\right)^2 z_1(1-z_1)\right]} \\ 
& = \frac{z_1z_2z_3}{4} \int_0^\infty ds \;  s^{(-1)-1} \;  e^{- s \Bar{Q}^2} \; e^{- \frac{X^2}{4s}} \\ 
& =  \frac{z_1z_2z_3}{4} 2\left(\frac{2\Bar{Q}}{X}\right) K_{1}(\Bar{Q}X) \\ 
& = z_1z_2z_3 \frac{\bar{Q}K_{1}(\Bar{Q}X)}{X} \numberthis
\end{align*}

where 
\begin{align*}
    \bar{Q}^2 & = Q^2 + m_Q^2 \frac{1-z_2}{z_1 z_3} \\
    X^2 & = \Vec{r}_{2}^{\; 2} \frac{z_2 z_3}{1-z_1} + \left(\Vec{r}_1 - \frac{z_2}{1-z_1} \Vec{r}_2\right)^2 z_1 (1-z_1) \\ 
    & = \Vec{r}_{2}^{\; 2} \frac{z_2 z_3}{1-z_1}  + \frac{z_2^2 z_1}{1-z_1} \Vec{r}_2^{\; 2} + z_1(z_2+z_3)\Vec{r}_1^{\; 2} - 2z_2z_1\Vec{r}_1 \cdot \Vec{r}_2 \\ 
    & = z_1 z_2(\Vec{r}_1 - \Vec{r}_2)^2 + z_1 z_3 \Vec{r}_1^{\; 2} + \Vec{r}_2^{\; 2 } \left(\frac{z_2 z_3}{1-z_1} + \frac{z_2^2 z_3}{1-z_1}  -z_1 z_2 \right) \Vec{r}_2^{\; 2} \\ 
    & =  z_1 z_2(\Vec{r}_1 - \Vec{r}_2)^2 + z_1 z_3 \Vec{r}_1^{\; 2} + \Vec{r}_2^{\; 2 } \left(\frac{z_2 z_3 + z_2^2 z_1- z_2 z_1 (z_2 + z_3)}{1-z_1}\right) \Vec{r}_2^{\; 2} \\
    & = z_1 z_2(\Vec{r}_1- \Vec{r}_2)^2 + z_1 z_3 \Vec{r}_1^{\; 2} + z_2 z_3 \Vec{r}_2^{\; 2} \numberthis[X2] \\
\end{align*}

\begin{align*}
& I_0(\vec{r}_1, \vec{r}_2)  \\ 
& =  \int \frac{\der^2 \Vec{l}_1 \der^2  \Vec{l}_2}{(2\pi)^2} \frac{e^{i \Vec{l}_1 \cdot \Vec{r}_1} e^{i \Vec{l}_2 \cdot \Vec{r}_2} }{\left[Q^2 z_1 (1 - z_1) + \Vec{l}_1^{\; 2} + m_Q^2\right] \left[Q^2 + \frac{\Vec{l}_1^{\; 2} + m_Q^2}{z_1} + \frac{\Vec{l}_2^{\; 2}}{z_2} + \frac{(\Vec{l}_1 + \Vec{l}_2)^2 + m_Q^2}{z_3}\right]} \\ 
& =  \frac{1}{z_1 (1-z_1)}\int_{0}^{ \infty} ds \int_{0}^{ \infty} dt \;  \int \frac{\der^2 \Vec{l}_1 \der^2 \Vec{l}_2}{(2\pi)^2} \;  e^{- s \left[Q^2 + \frac{\Vec{l}_1^{\; 2 } + m_Q^2 }{z_1 (1-z_1)}\right]} \; e^{- t \left[ Q^2 + \frac{\Vec{l}_1^{\; 2} + m_Q^2}{z_1} + \frac{\Vec{l}_2^{\; 2}}{z_2} + \frac{(\Vec{l}_1 + \Vec{l}_2)^2 + m_Q^2}{z_3} \right]} \\ 
& =  \frac{1}{z_1 (1-z_1)}\int_{0}^{ \infty} ds \int_{0}^{ \infty} dt \; e^{- (s+ t) Q^2} \; e^{- m_Q^2 \left(\frac{s}{z_1 (1-z_1)} + \frac{(1-z_2)t}{z_1 z_3}\right)} \\
& \times \int \frac{\der^2 \Vec{l}_1}{2\pi} \; e^{- \Vec{l}_1^{\; 2} \left(\frac{s}{z_1 (1-z_1)} + \frac{(1-z_2)t}{z_1 z_3}\right)} \; e^{i \Vec{l}_1 \cdot \Vec{r}_1} \int \frac{\der^2 \Vec{l}_2}{2\pi} \; e^{-  \frac{t(1-z_1)}{z_2 z_3} \Vec{l}_2^{\; 2}} \; e^{\Vec{l}_2 \cdot \left(i \Vec{r}_2 - \frac{2t}{z_3} \Vec{l}_1\right)} \\ 
& = \frac{1}{z_1 (1-z_1)}\int_{0}^{ \infty} ds \int_{0}^{ \infty} dt \; e^{- (s+ t) Q^2} \; e^{- m_Q^2 \left(\frac{s}{z_1 (1-z_1)} + \frac{(1-z_2)t}{z_1 z_3}\right)} \\ 
& \times  \int \frac{\der^2 \Vec{l}_1}{2\pi} \; e^{- \Vec{l}_1^{\; 2} \left(\frac{s}{z_1 (1-z_1)} + \frac{(1-z_2)t}{z_1 z_3}\right)} \; e^{i \Vec{l}_1 \cdot \Vec{r}_1} 
\frac{1}{2\pi} \left(\frac{\pi z_2 z_3 }{(1-z_1)t}\right) e^{\frac{\left(i\Vec{r}_2 - \frac{2t}{z_3}\Vec{l}_1\right)^2}{4t(1-z_1)}z_2 z_3} \\ 
& =  \frac{ z_2 z_3}{2 z_1 (1-z_1)^{2}} \int_{0}^{ \infty} ds \int_{0}^{\infty} \frac{dt}{t} \; e^{- (s+ t) Q^2} \;  e^{- m_Q^2 \left(\frac{s}{z_1 (1-z_1)} + \frac{(1-z_2)t}{z_1 z_3}\right)} \;  e^{- \Vec{r}_2^{\; 2} \frac{z_2 z_3}{4 t (1-z_1)}} \\ 
& \times \int \frac{\der^2 \Vec{l}_1}{2\pi} \; e^{-\Vec{l}_1^{\; 2} \left(\frac{s+t}{z_1(1-z_1)}\right)} \; e^{\Vec{l}_1 \cdot \left(i \Vec{r}_1 - i \frac{z_2}{1-z_1} \Vec{r}_2 \right)} \\ 
& =  \frac{ z_2 z_3}{2 z_1 (1-z_1)^{2}} \int_{0}^{ \infty} ds \int_{0}^{ \infty} \frac{dt}{t} \; e^{- (s+ t) Q^2} \; e^{- m_Q^2 \left(\frac{s}{z_1 (1-z_1)} + \frac{(1-z_2)t}{z_1 z_3}\right)} \; e^{- \Vec{r}_2^{\; 2} \frac{z_2 z_3}{4 t (1-z_1)}} \\  
& \times \frac{1}{2\pi} \left(\frac{\pi z_1 (1-z_1)}{s+t}\right) e^{-\frac{\left(\Vec{r}_1 - \frac{z_2}{1-z_1} \Vec{r}_2\right)^2}{4(s+t)}z_1 (1-z_1)} \\
& = \frac{z_2 z_3}{4 (1-z_1)}  \int_{0}^{ \infty} ds \int_{0}^{ \infty} \frac{dt}{t} \; \frac{1}{s+t} \;  e^{- (s+ t) Q^2} \; e^{- m_Q^2 \left(\frac{s}{z_1 (1-z_1)} + \frac{(1-z_2)t}{z_1 z_3}\right)}  \\
& \times e^{- \Vec{r}_2^{\; 2} \frac{z_2 z_3}{4 t (1-z_1)}}  \; e^{-\frac{\left(\Vec{r}_1 - \frac{z_2}{1-z_1} \Vec{r}_2\right)^2}{4(s+t)}z_1 (1-z_1)} 
\end{align*}

We do the following change of variables $u = s + t$ and $v = t $ with $s = u-v > 0, t = v >0 \Rightarrow u > v > 0 $ to obtain finally  
\begin{align*}
 I_0(\vec{r}_1, \vec{r}_2)  
& = \frac{z_2 z_3}{4 (1-z_1)}  \int_{0}^{ \infty} \frac{du}{u}  \int_{0}^{u} \frac{dv}{v} \;   e^{- u Q^2} \; e^{- m_Q^2 \left(\frac{u-v}{z_1 (1-z_1)} + \frac{(1-z_2)v}{z_1 z_3}\right)}  \\
& \times e^{- \Vec{r}_2^{\; 2} \frac{z_2 z_3}{4 v (1-z_1)}}  \; e^{-\frac{\left(\Vec{r}_1 - \frac{z_2}{1-z_1} \Vec{r}_2\right)^2}{4u }z_1 (1-z_1)}  \\ 
& =  \frac{z_2 z_3}{4 (1-z_1)}  \int_{0}^{ \infty}  \frac{du}{u} \; e^{- u Q^2} \;  e^{- m_Q^2  \frac{u}{z_1 (1-z_1)}} e^{-\frac{\left(\Vec{r}_1 - \frac{z_2}{1-z_1} \Vec{r}_2\right)^2}{4u }z_1 (1-z_1)} \\
& \times \int_{0}^{u} \frac{dv}{v} \;  e^{- m_Q^2 v  \left(\frac{z_2}{z_3 (1-z_1)}\right)} \; e^{- \Vec{r}_2^{\; 2} \frac{z_2 z_3}{4 v (1-z_1)}} \numberthis \; . 
\end{align*}

Using 
\begin{align*}
& - i \left(  \frac{\partial}{\partial \Vec{r}_{2}^{\; m}} +  \frac{z_2}{1-z_1} \frac{\partial}{\partial \Vec{r}_{1}^{\; m}} \right)  \left[- \frac{\left(\Vec{r}_1 - \frac{z_2}{1-z_1} \Vec{r}_2\right)^2}{4u }z_1 (1-z_1)  -\Vec{r}_2^{\; 2} \frac{z_2 z_3}{4 v (1-z_1)}  \right] \\ 
& = i \left\{ \frac{z_2}{1-z_1} \frac{2 \left(\Vec{r}_1 - \frac{z_2}{1-z_1} \Vec{r}_2 \right)_m z_1 (1-z_1)}{4u} + \left(- \frac{z_2}{1-z_1} \right)  \frac{2 \left(\Vec{r}_1 - \frac{z_2}{1-z_1} \Vec{r}_2 \right)_m z_1 (1-z_1)}{4u}  \right. \\ 
& \left. + \frac{z_2 z_3 }{4 v (1-z_1)} 2 \Vec{r}_{2 m }\right\} \\ 
& = i \frac{z_2z_3}{2 v (1-z_1)} \Vec{r}_{2 m } \; , 
\end{align*}
we obtain 
\begin{align*}
I_{1m}(\vec{r}_1, \vec{r}_2)  
& = \frac{ (z_2 z_3)^{2}}{8 (1- z_1)^2 } i \Vec{r}_{2m} \int_{0}^{ \infty} \frac{du}{u} \; e^{- u Q^2} \;  e^{-m_Q^2 \frac{u}{z_1 (1-z_1)}} \; e^{- \left(\Vec{r}_1 - \frac{z_2}{1-z_1} \Vec{r}_2\right)^2\frac{z_1(1-z_1)}{4u}} \\ 
& \times \int_0^u \frac{dv}{v^2} \; e^{- m_Q^2 v \frac{ z_2}{z_3(1-z_1)}} \; e^{- \Vec{r}_2^{\; 2} \frac{z_2z_3}{4v(1-z_1)}} \numberthis[I1mSchwinger] \;  .  
\end{align*}

This leads to 
\begin{align*}
I_{2mj}(\vec{r}_1,\vec{r}_2) & = \frac{(z_2 z_3)^{2}}{8 (1-z_1)^2} i \Vec{r}_{2m} \int_0^\infty \frac{d u}{u} \; e^{- u Q^2} \; e^{- m_Q^2 \frac{u}{z_1(1-z_1)}} \;  e^{- \left(\Vec{r}_1 - \frac{z_2}{1-z_1} \Vec{r}_2\right)^2\frac{z_1(1-z_1)}{4u}} \\ 
& \times \left(i \frac{z_1(1-z_1)}{4u} 2\left(\Vec{r}_1 - \frac{z_2}{1-z_1}\Vec{r}_2\right)_j\right)  \int_0^u \frac{dv}{v^{2}} \; e^{- m_Q^2 v \frac{ z_2}{z_3(1-z_1)}} \; e^{- \Vec{r}_2^{\; 2} \frac{z_2z_2}{4v(1-z_1)}} \\ 
& = - \frac{ (z_2 z_3)^{2} z_1}{16 (1-z_1)} \Vec{r}_{2m} \left(\Vec{r}_1 - \frac{z_2}{1-z_1}\Vec{r}_2\right)_j \int_0^\infty \frac{du}{u^{2}} \; e^{-uQ^2} \; e^{- m_Q^2 \frac{u}{z_1(1-z_1)}} \\
& \times e^{- \left(\Vec{r}_1 - \frac{z_2}{1-z_1} \Vec{r}_2\right)^2\frac{z_1(1-z_1)}{4u}} \; \int_0^u \frac{dv}{v^{2}} e^{- m_Q^2 v \frac{ z_2}{z_3(1-z_1)}} \; e^{- \Vec{r}_2^{\; 2} \frac{z_2z_3}{4v(1-z_1)}} \; . \numberthis
\end{align*}

Finally, the last integral  has the following expression 

\begin{align*}
\Tilde{I}_{1j}(\vec{r}_1,\vec{r}_2) & = \frac{z_2z_3}{4 (1-z_1)} \int_0^\infty \frac{du}{u} \; e^{-u Q^2} \; e^{-m_Q^2 \frac{u}{z_1(1-z_1)}} \left(i \frac{z_1(1-z_1)}{4u} 2 \left(\Vec{r}_1 - \frac{z_2}{1-z_1}\Vec{r}_2\right)_j\right) \\
& \times e^{- \left(\Vec{r}_1 - \frac{z_2}{1-z_1} \Vec{r}_2\right)^2\frac{z_1(1-z_1)}{4u}}  \int_0^u \frac{dv}{v} \; e^{- m_Q^2 v \frac{ z_2}{z_3(1-z_1)}} \; e^{- \Vec{r}_2^{\; 2} \frac{z_2z_3}{4v(1-z_1)}} \\ 
& =  \frac{z_1z_2z_3}{8} i \left(\Vec{r}_1 - \frac{z_2}{1-z_1}\Vec{r}_2\right)_j \int_0^\infty \frac{du}{u^{2}} \;  e^{- u Q^2} \; e^{- m_Q^2 \frac{u}{z_1(1-z_1)}} \\
& \times e^{- \left(\Vec{r}_1 - \frac{z_2}{1-z_1} \Vec{r}_2\right)^2\frac{z_1(1-z_1)}{4u}} \;  \int_0^u \frac{dv}{v} \; e^{- m_Q^2 v \frac{ z_2}{z_3(1-z_1)}} \;  e^{- \Vec{r}_2^{\; 2} \frac{z_2z_3}{4v(1-z_1)}} \numberthis
\end{align*}

\subsection{Feynman parametrization}

\begin{align*}
I_0(\vec{r}_1,\vec{r}_2) & =  \int \frac{\der^2 \Vec{l}_1 \der^2 \Vec{l}_2}{(2\pi)^2} \; \frac{e^{i \Vec{l}_1 \cdot \Vec{r}_1} \; e^{i \Vec{l}_2 \cdot \Vec{r}_2} }{\left[Q^2 z_1 (1 - z_1) + \Vec{l}_1^{\; 2} + m_Q^2\right] \left[Q^2 + \frac{\Vec{l}_1^{\; 2} + m_Q^2}{z_1} + \frac{\Vec{l}_2^{\; 2}}{z_2} + \frac{(\Vec{l}_1 + \Vec{l}_2)^2 + m_Q^2}{z_3}\right]} \\ 
& = \frac{1}{z_1(1-z_1)} \int_0^1 dx \int \frac{\der^2 \Vec{l}_1 \der^2 \Vec{l}_2}{(2\pi)^2} \frac{e^{i \Vec{l}_1 \cdot \Vec{r}_1} e^{i \Vec{l}_2 \cdot \Vec{r}_2} }{D^2}
\end{align*}
with 
\begin{align*}
D & = xQ^2 + (1-x)Q^2 + x \left( \frac{\Vec{l}_1^{\; 2} + m_Q^2}{z_1} + \frac{\Vec{l}_2^{\; 2}}{z_2} + \frac{(\Vec{l}_1 + \Vec{l}_2)^2 + m_Q^2}{z_3} \right) + (1-x) \frac{\Vec{l}_1^{\; 2} + m_Q^2}{z_1(1-z_1)} \\ 
& = Q^2 + m_Q^2\frac{x(1-z_2)(1-z_1)+ (1-x)z_3}{z_1z_3(1-z_1)} + \Vec{l}_1^{\; 2} \frac{x(1-z_2)(1-z_1)+ (1-x)z_3}{z_1z_3(1-z_1)} \\
& + \Vec{l}_2^{\; 2} \frac{x(1-z_1)}{z_2z_3} + \Vec{l}_2 \cdot \frac{2x}{z_3} \Vec{l}_1 \\ 
& = Q^2 + m_Q^2 \frac{xz_1 z_2 + z_3 }{z_1 z_3 (1-z_1)} + \Vec{l}_1^{\; 2 } \frac{xz_1 z_2 + z_3 }{z_1 z_3 (1-z_1)} + \Vec{l}_2^{\; 2} \frac{x(1-z_1)}{z_2z_3} + \Vec{l}_2 \cdot \frac{2x}{z_3} \Vec{l}_1 \; . 
\end{align*}

Using the Schwinger parametrization again, we have
\begin{align*}
I_0(\vec{r}_1,\vec{r}_2) & =  \frac{1}{z_1(1-z_1)} \int_0^1 dx \int_0^\infty ds \;  s \; e^{-s \left(Q^2 + m_Q^2 \frac{xz_1 z_2 + z_3 }{z_1 z_3 (1-z_1)}\right)} \int \frac{\der^2 \Vec{l}_1 }{2\pi} \; e^{i \Vec{l}_1 \cdot \vec{r}_1} \; e^{- \Vec{l}_1^{\; 2} s  \frac{xz_1 z_2 + z_3 }{z_1 z_3 (1-z_1)}} \\
& \times \int \frac{\der^2 \Vec{l}_2 }{2\pi} \; e^{\Vec{l}_2 \cdot \left(i \Vec{r}_2 - \frac{2xs}{z_3}\Vec{l}_1\right)} \; e^{-\Vec{l}_2^{\; 2} \frac{sx(1-z_1)}{z_2z_3}} \\ 
&= \frac{1}{z_1(1-z_1)} \int_0^1 dx \int_0^\infty ds \;  s \; e^{-s \left(Q^2 + m_Q^2 \frac{xz_1 z_2 + z_3 }{z_1 z_3 (1-z_1)}\right)} \int \frac{\der^2 \Vec{l}_1 }{2\pi} \; e^{i \Vec{l}_1 \cdot \vec{r}_1} \; e^{- \Vec{l}_1^{\; 2} s  \frac{xz_1 z_2 + z_3 }{z_1 z_3 (1-z_1)}} \\
& \times \frac{1}{2\pi} \left(\frac{\pi z_2z_3}{sx(1-z_1)}\right) e^{\left(i\Vec{r}_2 - \frac{2xs}{z_3}\Vec{l}_1\right)^2 \frac{z_2z_3}{4sx(1-z_1)}} \\ 
&=  \frac{ z_2z_3}{2 z_1(1-z_1)^{2} } \int_0^1 \frac{dx}{x} \int_0^\infty ds \;  e^{-s \left(Q^2 + m_Q^2 \frac{xz_1 z_2 + z_3 }{z_1 z_3 (1-z_1)}\right)} \; e^{- \Vec{r}_2^{\; 2} \frac{z_2z_3}{4sx(1-z_1)}} \\
& \times \int \frac{\der^2 \Vec{l}_1 }{2\pi} \;  e^{\Vec{l}_1 \cdot \left(i\Vec{r}_1 - i \frac{z_2}{1-z_1}\Vec{r}_2\right)} \; e^{- \Vec{l}_1^{\; 2} \frac{s}{z_1(1-z_1)}} \\
& = \frac{z_2z_3}{2 z_1(1-z_1)^{2}} \int_0^1 \frac{dx}{x} \int_0^\infty ds \;   e^{-s \left(Q^2 + m_Q^2 \frac{xz_1 z_2 + z_3 }{z_1 z_3 (1-z_1)}\right)} \; e^{- \Vec{r}_2^{\; 2} \frac{z_2z_3}{4sx(1-z_1)}} \\ 
& \times \frac{1}{2\pi} \left(\frac{\pi z_1 (1-z_1)}{s}\right) \;  e^{-\left(\Vec{r}_1-\frac{z_2}{1-z_1}\Vec{r}_2\right)^2 \frac{z_1(1-z_1)}{4s}}  \\
&  = \frac{ z_2z_3 }{4(1-z_1)} \int_0^1 \frac{dx}{x} \int_0^\infty ds \; s^{-1} \; e^{-s \left(Q^2 + m_Q^2 \frac{xz_1 z_2 + z_3 }{z_1 z_3 (1-z_1)}\right)} \; e^{- \frac{1}{4s} \left[\vec{r}_2^{\; 2} \frac{z_2z_3}{(1-z_1)x}+ \left(\Vec{r}_1-\frac{z_2}{1-z_1}\Vec{r}_2\right)^2  z_1 (1-z_1)\right]} \\ 
& =  \frac{ z_2z_3 }{4(1-z_1)} \int_0^1 \frac{dx}{x} \int_0^\infty ds \;  s^{0-1} \;  e^{-sQ_F^2(x)} \; e^{- \frac{X_F^2(x)}{4s}}  \\ 
& = \frac{z_2z_3}{2  (1-z_1)} \int_0^1 \frac{dx}{x} \;  K_{0}(Q_F(x)X_F(x)) \numberthis 
\end{align*}
where I define 
\begin{align}
    Q_F^2(x) & = Q^2 + m_Q^2 \frac{xz_1 z_2 + z_3 }{z_1 z_3 (1-z_1)} \; ,\\ 
    X_F^2(x) & = \vec{r}_2^{\; 2} \frac{z_2z_3}{(1-z_1)x}+ \left(\Vec{r}_1-\frac{z_2}{1-z_1}\Vec{r}_2\right)^2  z_1 (1-z_1) \; . 
\end{align}

Using 
\begin{align*}
    & - i \left(\frac{\partial}{\partial \Vec{r}_2^{\; m}} + \frac{z_2}{1-z_1}\frac{\partial}{\partial \Vec{r}_1^{\; m}} \right) \left[- \frac{X_F^2(x)}{4s}  \right] \\ 
    & = i \left\{ \frac{2\Vec{r}_{2m}z_2z_3}{4(1-z_1)xs} + \frac{z_1(1-z_1)}{4s}\left(- \frac{z_2}{1-z_1}\right)2 \left(\Vec{r}_1 - \frac{z_2}{1-z_1}\Vec{r}_2\right)_m  \right. \\
    & \left. + \frac{z_2}{1-z_1}  \frac{z_1(1-z_1)}{4s}  2 \left(\Vec{r}_1 - \frac{z_2}{1-z_1}\Vec{r}_2\right)_m \right\} \\ 
    & = \frac{z_2z_3}{2xs(1-z_1)} i\Vec{r}_{2m} \; , 
\end{align*}
we have 
\begin{align*}
I_{1m}(\vec{r}_1, \vec{r}_2) & =  - i \left(\frac{\partial}{\partial \Vec{r}_2^{\; m}} + \frac{z_2}{1-z_1}\frac{\partial}{\partial \Vec{r}_1^{\; m}} \right) I_0(\vec{r}_1, \vec{r}_2) \\ 
& = \frac{(z_2z_3)^{2}}{8(1-z_1)^2} i\Vec{r}_{2m} \int_0^1 \frac{dx}{x^{2}} \int_0^\infty ds \; s^{(-1)-1} \; e^{-sQ_F^2(x)} \;  e^{- \frac{X_F^2(x)}{4s}} \\ 
& = \frac{(z_2z_3)^{2}}{2(1-z_1)^2 } i \Vec{r}_{2m} \int_0^1 \frac{dx}{x^{2}} \frac{Q_F(x)}{X_F(x)} K_{1}(Q_F(x)X_F(x)) \; . \numberthis[I1m] 
\end{align*}

Using 
\begin{equation*}
- i \frac{\partial}{\partial \Vec{r}_1^{\; j}}\left(- \frac{X_F^2(x)}{4s}\right) = i \frac{z_1(1-z_1)}{2s} \left(\Vec{r}_1 - \frac{z_2}{1-z_1}\Vec{r}_2\right)_j \; , 
\end{equation*}
we obtain 
\begin{align*}
I_{2mj}(\vec{r}_1, \vec{r}_2) & = \frac{\partial}{\partial \Vec{r}_1^{\; j}} I_{1m}(\vec{r}_1,\vec{r}_2) \\ 
& = - \frac{ (z_2z_3)^{2}z_1}{(1-z_1)16}  \left(\Vec{r}_1 - \frac{z_2}{1-z_1}\Vec{r}_2\right)_j \Vec{r}_{2m} \int_0^1 \frac{dx}{x^{2}} \int_0^\infty ds \; s^{-2-1} \; e^{-sQ_F^2(x)} \;  e^{- \frac{X_F^2(x)}{4s}} \\ 
& = - \frac{(z_2z_3)^{2}z_1}{(1-z_1)2}  \left(\Vec{r}_1 - \frac{z_2}{1-z_1}\Vec{r}_2\right)_j \Vec{r}_{2m} \int_0^1 \frac{dx}{x^{2}} \left(\frac{Q_F(x)}{X_F(x)}\right)^2 K_2(Q_F(x)X_F(x) ) \; , \numberthis 
\end{align*}
and 
\begin{align*}
\Tilde{I}_{1j}(\vec{r}_1,\vec{r}_2) & = - i  \frac{\partial}{\partial \Vec{r}_1^{\; j}} I_0(\vec{r}_1,\vec{r}_2) \\ 
& = \frac{z_1z_2z_3}{8} i \left(\Vec{r}_{1} - \frac{z_2}{1-z_1} \Vec{r}_2 \right)_j \int_0^1 \frac{dx}{x} \int_0^\infty ds \; s^{(-1)-1} \; e^{-sQ_F^2(x)} \; e^{- \frac{X_F^2(x)}{4s}} \\ 
& = \frac{ z_1z_2z_3}{2} i \left(\Vec{r}_{1} - \frac{z_2}{1-z_1} \Vec{r}_2 \right)_j  \int_0^1 \frac{dx}{x}  \frac{Q_F(x)}{X_F(x)}  K_{1}(Q_F(x)X_F(x) ) \; . \numberthis  
\end{align*}

\bibliography{ref}
\bibliographystyle{JHEP}

\end{document}